\documentclass[epj]{svjour}
\usepackage{graphicx}
\newcommand{\al}{$\alpha$}

\newcommand{\raa}{($\alpha$,$\alpha$)}

\newcommand{\rag}{($\alpha$,$\gamma$)}
\newcommand{\ran}{($\alpha$,$n$)}
\newcommand{\rap}{($\alpha$,$p$)}
\newcommand{\rapo}{($\alpha$,$p_0$)}
\newcommand{\raX}{($\alpha$,$X$)}
\newcommand{\rang}{($\alpha$,$n\gamma$)}

\newcommand{\spro}{$s$-process}
\newcommand{\rppro}{$rp$-process}
\newcommand{\Nsv}{$N_A$$\left< \sigma v \right>$}
\newcommand{\sreac}{$\sigma_{\rm{reac}}$}
\newcommand{\sred}{$\sigma_{\rm{red}}$}
\newcommand{\Ered}{$E_{\rm{red}}$}
\begin{document}
\title{
Cross sections of $\alpha$-induced reactions for targets with masses
$A \approx 20-50$ at low energies 
}
%
%
\author{Peter Mohr\inst{1,2}
\thanks{\emph{Email: WidmaierMohr@t-online.de; mohr@atomki.mta.hu}} 
}                     
%
%
\institute{
Diakonie-Klinikum, D-74523 Schw\"abisch Hall, Germany
 \and 
Institute for Nuclear Research ATOMKI, H-4001 Debrecen, Hungary}
\date{Received: date / Revised version: date}
%
\abstract{
A simple reduction scheme using so-called reduced energies $E_{\rm{red}}$ and
reduced cross sections $\sigma_{\rm{red}}$ allows the comparison of heavy-ion
induced reaction cross sections for a broad range of masses of projectile and
target and over a wide energy range. A global behavior has been found for
strongly bound projectiles whereas much larger reduced cross sections have
been observed for weakly bound and halo projectiles. It has been shown that
this simple reduction scheme works also well for $\alpha$-particle induced
reactions on heavy target nuclei, but very recently significant deviations
have been seen for $\alpha$+$^{33}$S and $\alpha$+$^{23}$Na. Motivated by
these unexpected 
discrepancies, the present study analyses $\alpha$-induced reaction
cross sections for targets with masses $A \approx 20-50$. The study shows that
the experimental data for $\alpha$-induced reactions on nuclei with $A \approx
20-50$ deviate slightly from the global behavior of reduced cross
sections. However, in general the deviations evolve smoothly towards lower
masses. The only significant outliers are the recent 
data for $^{33}$S and $^{23}$Na which are far above the general
systematics, and some very old data may indicate that $^{36}$Ar and $^{40}$Ar
are below the general trend. As expected, also the doubly-magic $^{40}$Ca
nucleus lies slightly below the results for its neighboring
nuclei. Overall, the experimental data are nicely reproduced by a
statistical model calculation utilizing the simple $\alpha$-nucleus potential
by McFadden and Satchler. Simultaneously with the deviation of reduced cross
sections $\sigma_{\rm{red}}$ from the general behavior, the outliers
$^{23}$Na, $^{33}$S, $^{36}$Ar, and $^{40}$Ar also show significant
disagreement between experiment and statistical model calculation.
} 
\maketitle
\section{Introduction}
\label{sec:intro}
The cross sections of \al -induced reactions play an important role in nuclear
astrophysics. Stellar evolution and nucleosynthesis depend on the
Maxwellian-averaged cross sections or reaction rates \Nsv . Some prominent
examples in the mass range under study are the $^{18}$Ne\rap $^{21}$Na
reaction which is important for the break-out from hot CNO-cycles to the
so-called rapid proton capture process (\rppro ) \cite{Sch06,Par08}, the
$^{22}$Ne\ran $^{25}$Mg reaction which is an important neutron source in the
slow neutron capture process (\spro ) \cite{Kaepp11,Kaepp06,Stra06}, the
$^{23}$Na\rap $^{26}$Mg reaction which affects the production of galactic
$^{26}$Al \cite{Ili11}, and the $^{40}$Ca\rag $^{44}$Ti and $^{44}$Ti\rap
$^{47}$V reactions which govern the production and destruction of the tracer
radionuclide $^{44}$Ti in core-collapse supernovae \cite{The98,Mag10}. Beyond
stellar evolution and nucleosynthesis, \al -induced reactions are also
relevant for radionuclide production by energetic solar particles. It has been
shown recently that \al -induced reactions may significantly contribute to the
production of positron emitters \cite{Mur14}, and the abundance of the
radionuclide $^{36}$Cl may even be dominated by this scenario instead of
stellar nucleosynthesis \cite{Guan06}.

In addition to the astrophysical motivation, \al -induced reactions can also
be used for analytical purposes. The thin-layer activation analysis technique
has been suggested for the measurement of vanadium and chromium contents by
\raX\ reactions on $^{\rm{nat}}$V and $^{\rm{nat}}$Cr
\cite{Peng99,Chow95}. The concentration of sulfur which is an important
element for material deterioration can be measured using the $^{32}$S\rap
$^{35}$Cl reaction \cite{Solt96}. At slightly higher energies \al -induced
reactions are used for the production of important tracer elements. For
biological and medical studies $^{43}$K can be produced by $^{40}$Ar\rap
$^{43}$K \cite{Feny95}, and $^{30}$P can be made from $^{27}$Al\ran $^{30}$P
\cite{Saha79}. The behavior of aluminum in bio- and eco-systems can be traced
by $^{29}$Al which is produced by the $^{26}$Mg\rap $^{29}$Al reaction
\cite{Minai90}.

For heavy nuclei it has been found that the total reaction cross section
\sreac\ follows a general trend in the mass range around $A \approx 90 - 150$
\cite{Mohr10,Mohr13}. This trend becomes nicely visible when so-called reduced
cross sections \sred\ are plotted versus the reduced energy \Ered\ as
suggested by \cite{Gom05}. However, very recently huge discrepancies have been
found for the light nuclei $^{23}$Na and $^{33}$S where much larger values for
\sred\ have been observed; a detailed discussion of \al -induced reactions of
$^{33}$S was provided recently in \cite{Mohr14}.

Cross sections for intermediate mass and heavy targets are usually calculated
within the statistical model (StM). The basic prerequisite of the StM is a
sufficiently high level density in the compound nucleus at the excitation
energy $E^\ast = E_{\rm{c.m.}} + S_\alpha$ where $E_{\rm{c.m.}}$ is the
energy in the center-of-mass system and $S_\alpha$ is the separation energy of
the \al -particle in the compound nucleus. This prerequisite is certainly
fulfilled for heavy nuclei, but it will be shown that the experimental cross
sections in the lower mass range under study can also be nicely reproduced by
StM calculations.

For heavy nuclei \al -induced cross sections are typically very well
reproduced by StM calculations at energies above the Coulomb barrier. This
finding is almost independent of the chosen parameterization of the underlying
\al -nucleus potential. At low energies the application of the widely used \al
-nucleus potential by McFadden and Satchler \cite{McF66} typically
overestimates the experimental cross sections. This general behavior of the
McFadden/Satchler potential extends down to at least $^{64}$Zn \cite{Gyu12}
and $^{58}$Ni \cite{Quinn14}. Much efforts have been done in the last decade
to provide improved \al -nucleus potentials (e.g.,
\cite{Mohr13,Dem02,Avr09,Avr10,Avr14}).

Contrary to this general behavior for
heavy nuclei, StM calculations underestimate the experimental results for
$^{23}$Na and $^{33}$S. However, for the even lighter target $^{18}$Ne and the
$^{18}$Ne\rap $^{21}$Na reaction it was found that StM calculations are -- at
least on average -- in reasonable agreement with experimental results
although the excitation function is governed by many resonances which cannot
be reproduced by the StM \cite{Mohr14b}. It is the main scope of the present
study to analyze the mass region around $A \approx 20-50$ and search for
systematic trends for the cross sections of \al -induced reactions. For this
purpose all available reaction data in the EXFOR database \cite{EXFOR} are
reviewed and compared to StM calculations. It will be shown that there is a
smooth trend of increasing reduced cross sections \sred\ with decreasing mass
$A$ with two exceptionally large \sred\ values for $^{23}$Na and
$^{33}$S and perhaps exceptionally low \sred\ values for $^{36}$Ar and
$^{40}$Ar (based on very few data points from an experiment in the 1950s).

The paper is organized as follows. Sec.~\ref{sec:gen} provides general
information on reduced energies \Ered\ and reduced cross sections
\sred\ (Sec.~\ref{sec:red}), 
on the \al -nucleus potential which is the essential ingredient for the StM
(Sec.~\ref{sec:pot}), and on the statistical model itself
(Sec.~\ref{sec:statmod}).
Sec.~\ref{sec:exp} briefly discusses experimental
techniques for the determination of the relevant cross sections. Results for
nuclei with $A \approx 20-50$ are presented in Sec.~\ref{sec:res}. For each
nucleus under study the available experimental data are reviewed and compared
to a theoretical prediction from the StM, and the reduced cross section is
shown. The results are discussed in Sec.~\ref{sec:disc}, and finally
conclusions are drawn in Sec.~\ref{sec:conc}. A list of all nuclei under study
is provided in Table \ref{tab:Q} which includes the residual nuclei and the
reaction $Q$-values for the \rag , \ran , and \rap\ reactions. The $Q$-value
data are taken from nuclear masses in the latest AME2012 evaluation
\cite{Audi12a,Audi12b}.
\begin{table*}
\caption{$Q$-values of \al -induced reactions (taken from AME2012
  \cite{Audi12a,Audi12b}). Stable residual nuclei are underlined.}
\label{tab:Q}
\begin{center}
\begin{tabular}{|c|crr|cr|cr|cr|}
\hline
& & & &
\multicolumn{2}{c|}{\rag} &
\multicolumn{2}{c|}{\ran} &
\multicolumn{2}{c|}{\rap} \\
Section
& target & $Z$ & $N$ 
& residual & $Q$ (MeV)
& residual & $Q$ (MeV)
& residual & $Q$ (MeV) \\
%
\hline
%
\ref{sec:cr50} &
$^{50}$Cr & 24 & 26 &
{\underline{$^{54}$Fe}} & 
$+8.417$ &
$^{53}$Fe & 
$-4.961$ &
$^{53}$Mn & 
$-0.437$ \\
\ref{sec:v51} &
$^{51}$V & 23 & 28 &
{\underline{$^{55}$Mn}} & 
$+7.933$ &
$^{54}$Mn & 
$-2.294$ &
{\underline{$^{54}$Cr}} & 
$-0.134$ \\
\ref{sec:v50} &
$^{50}$V & 23 & 27 &
$^{54}$Mn & 
$+8.758$ &
$^{53}$Mn & 
$-0.181$ &
{\underline{$^{53}$Cr}} & 
$+1.198$ \\
\ref{sec:ti48} &
$^{48}$Ti & 22 & 26 &
{\underline{$^{52}$Cr}} & 
$+9.351$ &
$^{51}$Cr & 
$-2.687$ &
{\underline{$^{51}$V}} & 
$-1.152$ \\
\ref{sec:ti46} &
$^{46}$Ti & 22 & 24 &
{\underline{$^{50}$Cr}} & 
$+8.560$ &
$^{49}$Cr & 
$-4.441$ &
$^{49}$V & 
$-1.030$ \\
\ref{sec:ti44} &
$^{44}$Ti & 22 & 22 &
$^{48}$Cr & 
$+7.698$ &
$^{47}$Cr & 
$-8.634$ &
$^{47}$V & 
$-0.407$ \\
\ref{sec:sc45} &
$^{45}$Sc & 21 & 24 &
$^{49}$V & 
$+9.315$ &
$^{48}$V & 
$-2.241$ &
{\underline{$^{48}$Ti}} & 
$+2.257$ \\
\ref{sec:ca48} &
$^{48}$Ca & 20 & 28 &
$^{52}$Ti & 
$+7.669$ &
$^{51}$Ti & 
$-0.139$ &
$^{49}$Sc & 
$-5.860$ \\
\ref{sec:ca42} &
$^{42}$Ca & 20 & 22 &
{\underline{$^{46}$Ti}} & 
$+8.005$ &
$^{45}$Ti & 
$-5.185$ &
{\underline{$^{45}$Sc}} & 
$-2.340$ \\
\ref{sec:ca40} &
$^{40}$Ca & 20 & 20 &
$^{44}$Ti & 
$+5.127$ &
$^{43}$Ti & 
$-11.172$ &
$^{43}$Sc & 
$-3.522$ \\
\ref{sec:k41} &
$^{41}$K & 19 & 22 &
{\underline{$^{45}$Sc}} & 
$+7.937$ &
$^{44}$Sc & 
$-3.390$ &
{\underline{$^{44}$Ca}} & 
$+1.045$ \\
\ref{sec:k40} &
$^{40}$K & 19 & 21 &
$^{44}$Sc & 
$+6.705$ &
$^{43}$Sc & 
$-2.994$ &
{\underline{$^{43}$Ca}} & 
$+0.009$ \\
\ref{sec:k39} &
$^{39}$K & 19 & 20 &
$^{43}$Sc & 
$+4.806$ &
$^{42}$Sc & 
$-7.332$ &
{\underline{$^{42}$Ca}} & 
$-0.124$ \\
\ref{sec:ar40} &
$^{40}$Ar & 18 & 22 &
{\underline{$^{44}$Ca}} & 
$+8.854$ &
{\underline{$^{43}$Ca}} & 
$-2.277$ &
$^{43}$K & 
$-3.329$ \\
\ref{sec:ar36} &
$^{36}$Ar & 18 & 18 &
{\underline{$^{40}$Ca}} & 
$+7.040$ &
$^{39}$Ca & 
$-8.595$ &
{\underline{$^{39}$K}} & 
$-1.288$ \\
\ref{sec:cl37} &
$^{37}$Cl & 17 & 20 &
{\underline{$^{41}$K}} & 
$+6.223$ &
$^{40}$K & 
$-3.872$ &
{\underline{$^{40}$Ar}} & 
$-1.586$ \\
\ref{sec:cl35} &
$^{35}$Cl & 17 & 18 &
{\underline{$^{39}$K}} & 
$+7.219$ &
$^{38}$K & 
$-5.859$ &
{\underline{$^{38}$Ar}} & 
$+0.837$ \\
\ref{sec:s34} &
$^{34}$S & 16 & 18 &
{\underline{$^{38}$Ar}} & 
$+7.208$ &
$^{37}$Ar & 
$-4.630$ &
{\underline{$^{37}$Cl}} & 
$-3.034$ \\
\ref{sec:s33} &
$^{33}$S & 16 & 17 &
$^{37}$Ar & 
$+6.787$ &
{\underline{$^{36}$Ar}} & 
$-2.001$ &
$^{36}$Cl & 
$-1.928$ \\
\ref{sec:s32} &
$^{32}$S & 16 & 16 &
{\underline{$^{36}$Ar}} & 
$+6.641$ &
$^{35}$Ar & 
$-8.615$ &
{\underline{$^{35}$Cl}} & 
$-1.866$ \\
\ref{sec:p31} &
$^{31}$P & 15 & 16 &
{\underline{$^{35}$Cl}} & 
$+6.998$ &
$^{34}$Cl & 
$-5.647$ &
{\underline{$^{34}$S}} & 
$+0.627$ \\
\ref{sec:si30} &
$^{30}$Si & 14 & 16 &
{\underline{$^{34}$S}} & 
$+7.924$ &
{\underline{$^{33}$S}} & 
$-3.494$ &
$^{33}$P & 
$-2.960$ \\
\ref{sec:si29} &
$^{29}$Si & 14 & 15 &
{\underline{$^{33}$S}} & 
$+7.116$ &
{\underline{$^{32}$S}} & 
$-1.526$ &
$^{32}$P & 
$-2.454$ \\
\ref{sec:si28} &
$^{28}$Si & 14 & 14 &
{\underline{$^{32}$S}} & 
$+6.948$ &
$^{31}$S & 
$-8.097$ &
{\underline{$^{31}$P}} & 
$-1.916$ \\
\ref{sec:al27} &
$^{27}$Al & 13 & 14 &
{\underline{$^{31}$P}} & 
$+9.669$ &
$^{30}$P & 
$-2.643$ &
{\underline{$^{30}$Si}} & 
$+2.372$ \\
\ref{sec:mg26} &
$^{26}$Mg & 12 & 14 &
{\underline{$^{30}$Si}} & 
$+10.643$ &
{\underline{$^{29}$Si}} & 
$+0.034$ &
$^{29}$Al & 
$-2.874$ \\
\ref{sec:mg25} &
$^{25}$Mg & 12 & 13 &
{\underline{$^{29}$Si}} & 
$+11.127$ &
{\underline{$^{28}$Si}} & 
$+2.654$ &
$^{28}$Al & 
$-1.206$ \\
\ref{sec:mg24} &
$^{24}$Mg & 12 & 12 &
{\underline{$^{28}$Si}} & 
$+9.984$ &
$^{27}$Si & 
$-7.196$ &
{\underline{$^{27}$Al}} & 
$-1.601$ \\
\ref{sec:na23} &
$^{23}$Na & 11 & 12 &
{\underline{$^{27}$Al}} & 
$+10.092$ &
$^{26}$Al & 
$-2.966$ &
{\underline{$^{26}$Mg}} & 
$+1.821$ \\
\ref{sec:ne22} &
$^{22}$Ne & 10 & 12 &
{\underline{$^{26}$Mg}} & 
$+10.615$ &
{\underline{$^{25}$Mg}} & 
$-0.478$ &
$^{25}$Na & 
$-3.531$ \\
\ref{sec:ne21} &
$^{21}$Ne & 10 & 11 &
{\underline{$^{25}$Mg}} & 
$+9.886$ &
{\underline{$^{24}$Mg}} & 
$+2.555$ &
$^{24}$Na & 
$-2.178$ \\
\ref{sec:ne20} &
$^{20}$Ne & 10 & 10 &
{\underline{$^{24}$Mg}} & 
$+9.317$ &
$^{23}$Mg & 
$-7.215$ &
{\underline{$^{23}$Na}} & 
$-2.376$ \\
\ref{sec:ne18} &
$^{18}$Ne & 10 & 8 &
$^{22}$Mg & 
$+8.142$ &
$^{21}$Mg & 
$-11.242$ &
$^{21}$Na & 
$+2.638$ \\
\ref{sec:f19} &
$^{19}$F &  9 & 10 &
{\underline{$^{23}$Na}} & 
$+10.467$ &
$^{22}$Na & 
$-1.952$ &
{\underline{$^{22}$Ne}} & 
$+1.673$ \\
%
%
\hline
\end{tabular}
\end{center}
\end{table*}

\section{General considerations}
\label{sec:gen}

\subsection{Reduced energy \Ered\ and reduced cross section \sred }
\label{sec:red}
\begin{figure}[htb]
  \includegraphics{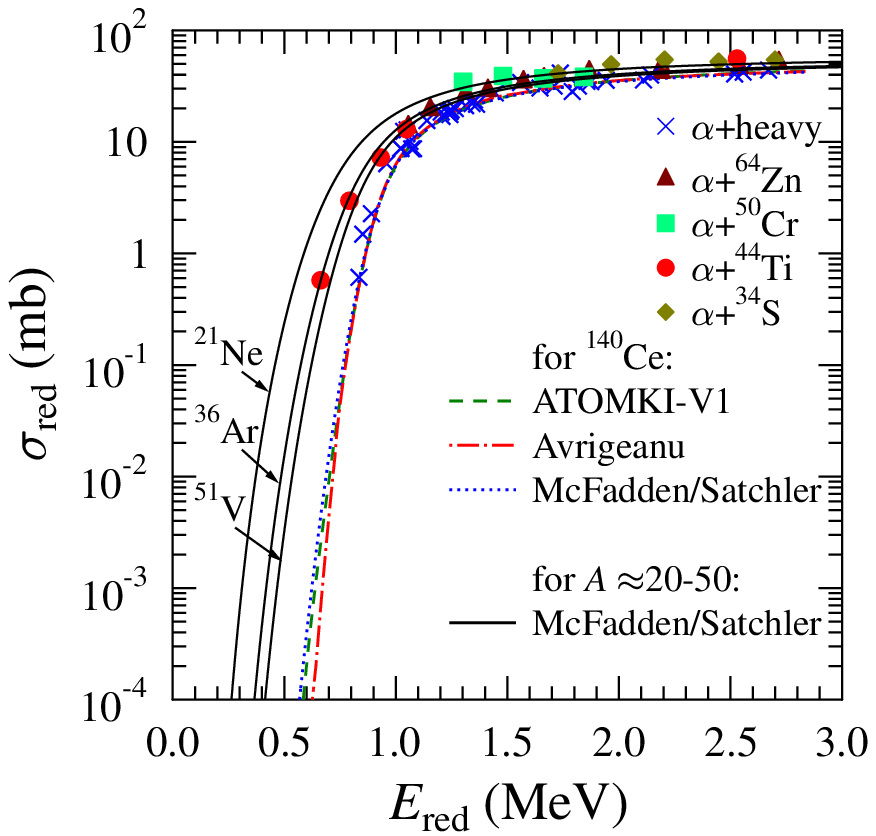}
\caption{
Reduced cross section \sred\ versus reduced energy \Ered\ for \al -induced
reactions on heavy
nuclei. Data from elastic \raa\ scattering of heavy target nuclei have been
taken from \cite{Mohr10,Mohr13} and are shown as blue crosses. Slightly higher
values for \sred\ are found for $^{64}$Zn and for the mass range of this work
($^{50}$Cr, $^{44}$Ti and $^{34}$S). The dashed, dash-dotted, and dotted lines
show calculations for a heavy target nucleus ($^{140}$Ce) using three
different \al -nucleus potentials (data taken from \cite{Mohr13b}). The full
lines show calculations for 
$^{21}$Ne, $^{36}$Ar, and $^{51}$V, i.e.\ in the full mass range of this
study, using the \al -nucleus potential of McFadden and Satchler \cite{McF66}.
}
\label{fig:sred_heavy}
\end{figure}

A simple reduction scheme for the comparison of heavy-ion induced reactions
has been suggested by Gomes {\it et al.}\ \cite{Gom05}. The so-called reduced
cross sections \sred\ and reduced energies \Ered\ are defined by:
\begin{eqnarray}
E_{\rm{red}} & = & \frac{\bigl(A_P^{1/3}+A_T^{1/3}\bigr) E_{\rm{c.m.}}}{Z_P Z_T} \\
\sigma_{\rm{red}} & = & \frac{\sigma_{\rm{reac}}}{\bigl(A_P^{1/3}+A_T^{1/3}\bigr)^2}
\label{eq:red}
\end{eqnarray}
The reduced energy \Ered\ takes into account the different heights of
the Coulomb barrier in the systems under consideration, whereas the reduced
reaction cross section \sred\ scales the measured total reaction cross section
\sreac\ according to the geometrical size of the projectile-plus-target
system. It is found that the reduced cross sections \sred\ show a very similar
behavior for a broad range of projectiles and targets over a wide energy
range. Significantly higher values of  \sred\ are found for weakly bound
projectiles (like e.g.\ $^{6,7}$Li) or halo projectiles (e.g.\ $^6$He), see
\cite{Far10,Lep14}. Results for \al -induced reactions on heavy target nuclei
fit into the systematics of heavy-ion induced reactions \cite{Mohr10,Mohr13}.
Results are shown in Fig.~\ref{fig:sred_heavy}.

\begin{figure}[htb]
\includegraphics[bbllx=40pt,bblly=20pt,bburx=480pt,bbury=530pt,width=0.98\columnwidth]{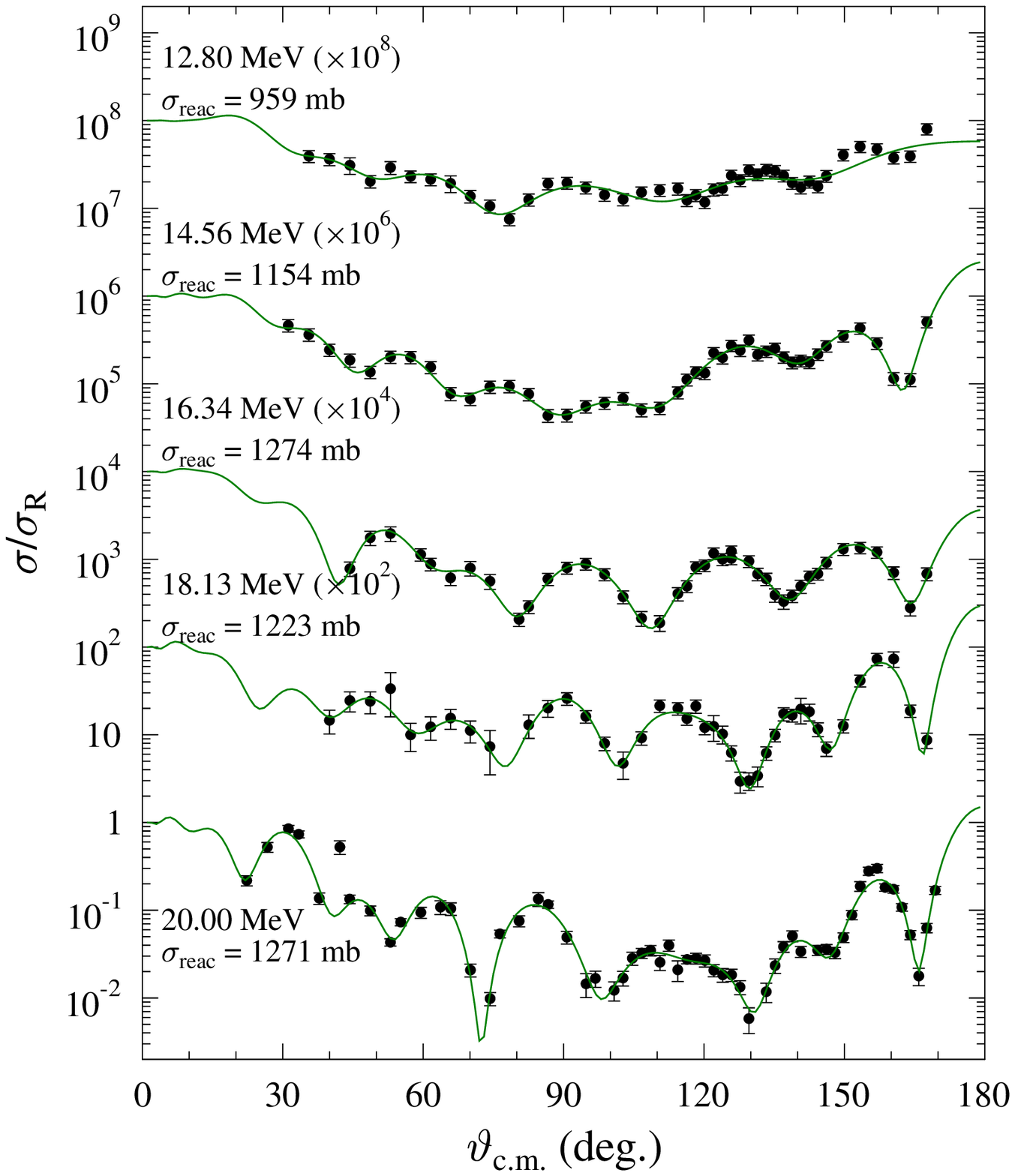}
\caption{
$^{34}$S\raa $^{34}$S elastic scattering: experimental data \cite{Bredbecka94}
  and a phase shift fit to determine the total reaction cross section
  \sreac\ using the method from \cite{Chi96}.
}
\label{fig:s34scat}
\end{figure}

\begin{figure}[htb]
\includegraphics[bbllx=40pt,bblly=20pt,bburx=480pt,bbury=465pt,width=0.98\columnwidth]{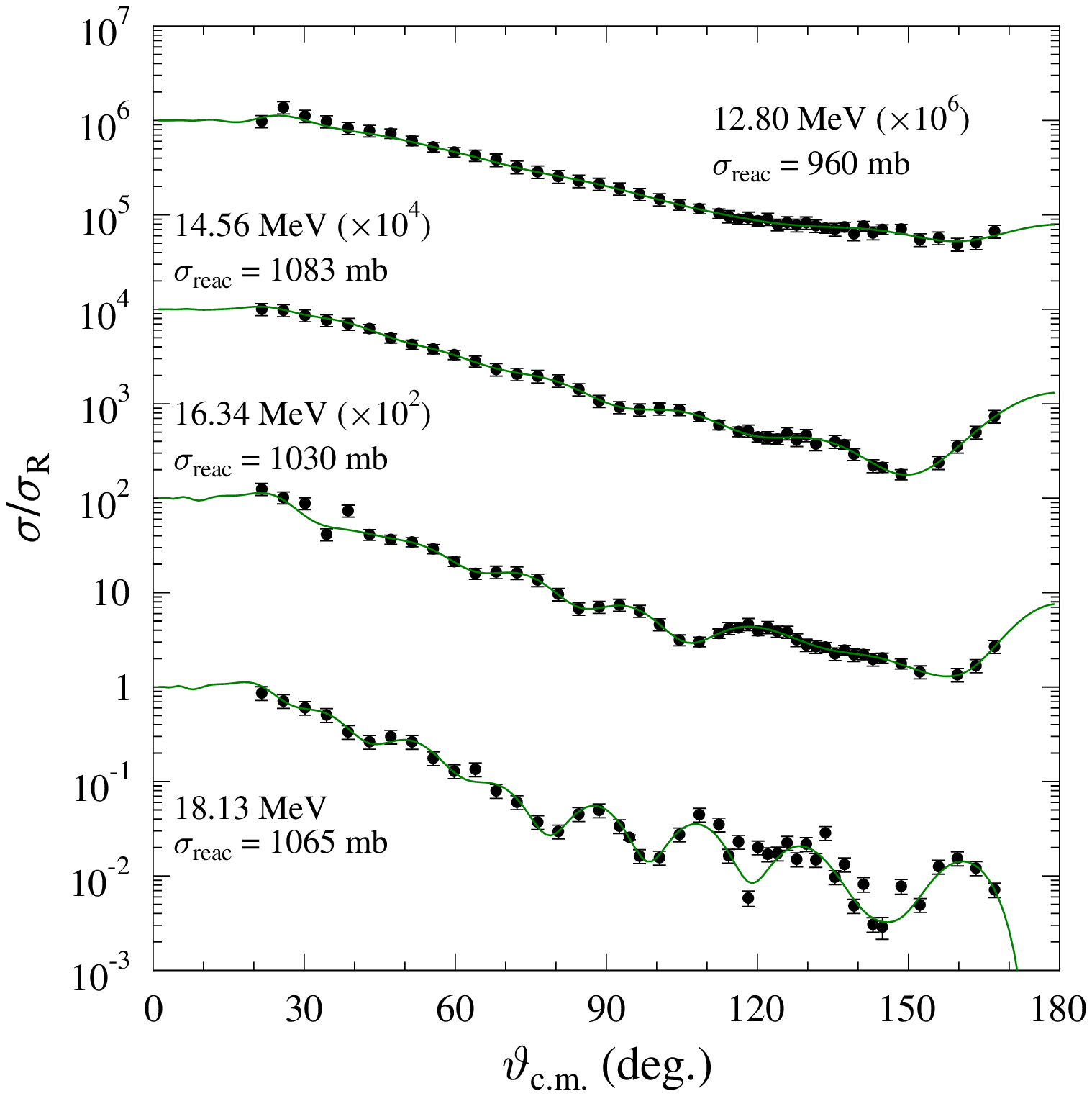}
\caption{
$^{50}$Cr\raa $^{50}$Cr elastic scattering: experimental data \cite{Bredbecka94}
  and a phase shift fit to determine the total reaction cross section
  \sreac\ using the method from \cite{Chi96}.
}
\label{fig:cr50scat}
\end{figure}

The experimental \sred\ data for $^{34}$S and $^{50}$Cr have been derived from
the experimental angular distributions of Bredbecka {\it et
  al.}\ \cite{Bredbecka94} by phase shift fits according to \cite{Chi96} (see
Figs.~\ref{fig:s34scat} and \ref{fig:cr50scat}). The
data point for $^{44}$Ti at \Ered\ $\approx 2.52$\,MeV has been taken from the
analysis of elastic $^{44}$Ti\raa $^{44}$Ti scattering in \cite{Rehm02}; it is
hardly visible because it overlaps with a 
data point for $^{34}$S. The low-energy data points for $^{44}$Ti will be
explained later (Sec.~\ref{sec:ti44}). As can be seen from
Fig.~\ref{fig:sred_heavy}, the \sred\ values for lighter targets seem to be
close above the general systematics with increasing differences towards lower
energies and lower masses. 

Interestingly, the relation between the reduced energy \Ered\ and the most
effective energy for the determination of astrophysical reaction rates (the
so-called Gamow window) is practically independent of the target mass $A_T$
and target charge number $Z_T$ for \al -induced reactions. In the mass range
under study the Gamow window is found at \Ered\ $\approx 0.29$\,MeV for $T_9 =
1$ (where $T_9$ is the temperature in Giga-Kelvin); the variation between
$^{21}$Ne (\Ered\ $ = 0.293$\,MeV) and $^{51}$V (\Ered\ $ = 0.279$\,MeV) is
practically negligible (and even for heavy nuclei like $^{208}$Pb a close value
of \Ered\ $= 0.264$\,MeV is found for $T_9 = 1$). $T_9 = 2$ corresponds to
\Ered\ $\approx 0.45$\,MeV, and $T_9 = 3$ corresponds to \Ered\ $\approx
0.59$\,MeV for all nuclei under study in this work. Further details on the
relation between the Gamow window and the corresponding reduced energy
\Ered\ will be given in \cite{Mohr15}.

\subsection{\al -nucleus potential}
\label{sec:pot}
For heavy nuclei it has been found that total reaction cross sections
\sreac\ and thus reduced cross sections \sred\ can be reproduced by almost any
reasonable \al -nucleus potential at energies above the Coulomb barrier. The
reason for this universal behavior is discussed in detail in
\cite{Mohr11,Mohr13b}. As an example the reduced cross sections for $^{140}$Ce
are calculated from three global \al -nucleus potentials. The dashed green
line in Fig.~\ref{fig:sred_heavy} shows the result from the ATOMKI-V1
potential which was derived from elastic scattering data in the mass range $A
\approx 90-150$ \cite{Mohr13}. The red dash-dotted line is calculated from the
many-parameter potential by Avrigeanu {\it et al.}\ in the version of
\cite{Avr10} which was derived from elastic scattering and reaction data in a
similar mass range, and the blue dotted line corresponds to the old and very
simple 4-parameter potential by McFadden and Satchler \cite{McF66}. It is
obvious that the results are very similar above \Ered\ $\approx 0.8$\,MeV, but
at lower energies significant discrepancies appear. As a typical result for
heavy nuclei it has been found that the McFadden/Satchler potential strongly
overestimates experimental cross sections at low energies. This may be a
consequence of the missing energy dependence of the imaginary part of the
McFadden/Satchler potential, and noticeable improvements have been achieved by
adding such an energy dependence (e.g., \cite{Som98,Sau11}). The ATOMKI-V1
potential has also a slight trend to overestimate experimental cross sections
at low energies \cite{Mohr13,Mohr13b}, and the Avrigeanu potential
\cite{Avr10,Avr14} typically slightly underestimates experimental data at very
low energies. 

As the ATOMKI-V1 potential and the Avrigeanu potential have not been optimized
for the mass range under study in this work, reduced cross sections \sred\ for
the nuclei $^{51}$V, $^{36}$Ar, and $^{21}$Ne have been calculated from the
McFadden/Satchler potential (full black lines in
Fig.~\ref{fig:sred_heavy}). It will be shown later that the McFadden/Satchler
potential gives excellent predictions in the whole mass range $A \approx 20 -
50$. The shown calculations for $^{51}$V, $^{36}$Ar, and $^{21}$Ne indicate a
slightly increased reduced cross section \sred\ at higher energies above
\Ered\ $\approx 1.5$\,MeV, as compared to the global systematics of heavy-ion
induced reactions. At energies below \Ered\ $\approx 1.5$\,MeV the
\sred\ for $A \approx 20 - 50$ are significantly increased, and this increase
becomes more pronounced for lighter targets. In the detailed study of the
available experimental data in the $A \approx 20 - 50$ mass range this trend
will be confirmed for most nuclei under study (see Sec.~\ref{sec:res}).

\subsection{Statistical model}
\label{sec:statmod}
Reaction cross sections of \al -induced reactions for heavy nuclei can be
calculated using the StM \cite{Haus52}.  In particular, this model has been
widely applied for the calculation of reaction cross sections and stellar
reaction rates in nuclear astrophysics \cite{Rau11} using the TALYS
\cite{TALYS} and NON-SMOKER \cite{NONSMOKER} codes. The applicability of the
StM to \al -induced reactions in the mass range $A \approx 20 - 50$ may be
limited because the level density in the compound nucleus may be not
sufficiently high. In such cases the cross section will be dominated by
individual resonances instead of many overlapping resonances. Consequently,
the StM model will not be able to predict the detailed shape of the excitation
function, but nevertheless the StM should be able to reproduce the general
trend of the energy dependence of the cross section. E.g., such a behavior has
been found for the $^{18}$Ne\rap $^{21}$Na reaction, see \cite{Mohr14b} and
Sec.~\ref{sec:ne18}. Further details on the applicability of the StM are given
in \cite{Rau97} (see Fig.~8 of \cite{Rau97} for \al -induced reactions).

In a schematic notation the reaction cross section in the StM is proportional
to 
\begin{equation}
\sigma(\alpha,X) \sim \frac{T_{\alpha,0} T_X}{\sum_i T_i} = T_{\alpha,0}
\times b_X
\label{eq:StM}
\end{equation}
with the transmission coefficients $T_i$ into the $i$-th open channel and the
branching ratio $b_X = T_X / \sum_i T_i$ for the decay into the channel $X$. The
$T_i$ are calculated from global optical potentials (particle channels) and
from the gamma-ray strength function for the photon channel. For
details of the definition of $T_i$, see \cite{Rau11}. $T_{\alpha,0}$ refers to
the entrance channel where the target nucleus is in its ground state under
laboratory conditions. The calculation of stellar reaction rates
\Nsv\ requires further modification of Eq.~(\ref{eq:StM}) \cite{Rau11}.

It is typical for \al -induced reactions on heavy nuclei that $T_{\alpha}$
(and thus $T_{\alpha,0}$) is
much smaller than the other $T_i$. A simple qualitative explanation is the
high Coulomb barrier in the \al\ channel. In the neutron channel a
Coulomb barrier is completely missing, and in the proton channel the barrier
is much lower. As a consequence, the cross section in the StM in
Eq.~(\ref{eq:StM}) factorizes into a production cross section of the compound
nucleus which is proportional to $T_{\alpha,0}$, and a decay branching ratio
$b_X = T_X/\sum_i T_i$ practically independent of $T_\alpha$ because
$T_\alpha$ only marginally contributes to the sum $\sum_i T_i$ in the above
nominator of $b_X$. The production cross section is thus entirely defined 
by the underlying \al -nucleus potential whereas the branching ratio $b_X$
does practically not depend on the chosen \al -potential but on all the other
ingredients of the StM (optical potentials for the other channels, gamma-ray
strength functions, level densities). Consequently, the cross sections of \al
-induced reactions are sensitive to the \al -nucleus potential, but in
addition each individual \rap , \ran , or \rag\ reaction has further and
sometimes complicated sensitivities to the other ingredients. A quantitative
estimate whether a calculated reaction cross is sensitive to a particular
ingredient, is the so-called sensitivity (as defined e.g.\ in \cite{Rau12}).
As the \al -nucleus potential affects directly the production cross section,
the sensitivity on the \al -nucleus potential is typically close to 1 for all
(\al ,$X$) reactions at energies around or below the Coulomb barrier.

The present study focuses on the deviation of the \sred\ values for \al
-induced reactions for $A \approx 20 - 50$ from the universal trend of
heavy-ion induced reactions. This deviation mainly appears for reduced
energies below 
\Ered\ $\approx 1$\,MeV. At these energies typically at least one particle
channel (proton or neutron channel) is open, and this open channel dominates
the sum $\sum_i T_i$ because $T_\alpha$ is suppressed by the Coulomb barrier
and $T_\gamma$ is usually much smaller than $T_X$ into particle channels. For
particle channels this means that Eq.~(\ref{eq:StM}) simplifies to
\begin{equation}
\sigma(\alpha,X) \sim \frac{T_{\alpha,0} T_X}{\sum_i T_i} \approx T_{\alpha,0}
\times \frac{T_X}{T_p+T_n}
\label{eq:StM2}
\end{equation}
Furthermore, because of the different $Q$-values of the \rap\ and
\ran\ reactions, often one channel is strongly dominating. In such cases $b_X
= T_X/(T_p+T_n) \approx 1$ for the dominating channel, and the \raX\ cross
section is almost identical to 
the total reaction cross section \sreac . Consequently, it is sufficient to
measure the total reaction cross section of the dominating particle channel
for these particular nuclei to determine the total reaction cross section
\sreac\ and the reduced cross section \sred .

\section{Experimental techniques}
\label{sec:exp}
Total reaction cross sections \sreac\ of \al -induced reactions at higher
energies far above the Coulomb barrier have been determined from transmission
data, see 
e.g.\ \cite{Ing00}. At lower energies \sreac\ can be derived from the analysis
of elastic scattering angular distributions, and it has been shown recently
that the result from elastic scattering is in agreement with the sum of the
cross sections of all open channels \cite{Gyu12}. As pointed out above, in
many cases one particular open channel is dominating, and then it is
sufficient to measure the total \rap\ or total \ran\ cross section to obtain
an excellent estimate of the total reaction cross section \sreac .

\subsection{Activation}
\label{sec:act}
Activation is a reliable and widely used technique for the measurement of
total \rap\ or \ran\ cross sections. A large fraction of the \rap\ and
\ran\ data for targets with $A \approx 20 - 50$ has been measured by
activation. However, activation experiments are obviously limited to reactions
with unstable residual nuclei. In a usual activation experiment many targets
are irradiated at different energies, and the excitation function is derived
from the activation yields. In several cases the so-called ``stacked-foil''
technique was applied which allows the determination of excitation functions
within a very limited beamtime because many target foils are stacked behind
each other and irradiated simultaneously. As will be shown below, this
technique provides good results at energies close to the projectile energy
before the stack of foils. However, the results for the lowest energies are
often not reliable, probably because of uncertainties in foil thickness and
resulting energy loss and straggling of the projectiles.

Various techniques can be used to determine the number of produced radioactive
nuclei. The chosen technique depends on the half-life and the decay properties
of the respective nucleus. In most cases $\gamma$-rays following the
$\beta^+$- or $\beta^-$-decay of the mother nuclide are detected using
high-resolution germanium detectors. In some cases without detectable
$\gamma$-ray branch, $X$-rays can be measured (following e.g.\ electron
capture from the $K$-shell). Also a direct detection of electrons from
$\beta^-$-decay or positrons from $\beta^+$-decay is possible. Finally, for
long half-lives the accelerator mass spectrometry technique allows to count
few nuclei with otherwise unrivaled sensitivity.

\subsection{Direct neutron measurements}
\label{sec:neutron}
\subsubsection{Neutron counting}
\label{sec:count}
As an alternative to activation, the total cross section of \ran\ reactions
can be measured by neutron thermalization and counting. In practice, this
technique is widely used, but experiments have to be done very carefully
because minor build-up of carbon on the target may lead to a significant
neutron yield from the $^{13}$C\ran $^{16}$O background reaction. Highly
enriched (and thus expensive) targets are required to avoid neutron yields
from other (more neutron-rich) isotopes.

\subsubsection{Time-of flight measurements}
\label{sec:tof}
A direct neutron detection using the time-of-flight (TOF) technique is also
possible. However, these experiments determine the differential cross section
$d\sigma/d\Omega$ of a particular $n_i$ channel at the detector angle
$\vartheta$ (where $n_0$ stands for the ground state of the residual nucleus,
$n_1$ for the first excited state, etc.). There are two basic problems to
determine the total \ran\ cross section from such data. First, the integration
of the differential cross section requires the knowledge of the full angular
distribution (i.e., measurements at many angles), and second, all exit
channels $n_i$ have to be summed up properly. Here weak channels may be
overlooked.

\subsection{Direct proton measurements}
\label{sec:proton}
Similar problems appear in direct measurements of \rap\ cross sections. Again,
the angular distribution has to integrated correctly, and all channels $p_i$
have to be summed up. This summation is even more critical for \rap\ reactions
because relatively high-lying final states lead to small proton energies which
may be difficult to detect (of course depending on the target thickness).

\subsection{Target thickness}
\label{sec:thickness}
The target thickness is a very important experimental parameter. The
experimental yield is given by the average cross section over the energy
interval $[E_\alpha,E_\alpha - \Delta E]$ where $\Delta E$ is the energy loss of
the \al\ projectile in the target. A precise determination of the cross
section $\sigma(E)$ clearly asks for a thin target, i.e.\ a small energy loss
$\Delta E$. However, for the application of the StM a sufficient number of
resonances must lie within the corresponding interval of excitation energies
in the compound nucleus. There is no problem for heavy target nuclei which
have high level densities, and any realistic target is sufficiently thick for
the applicability of the StM. However, at the lower end of the mass range $A
\approx 20 - 50$ the level density may be too low, in particular for very thin
targets. As a consequence, the experimental data for a thin target will be
governed by individual resonances, and the StM will not be able to reproduce
the details of the excitation function. Experimental data for a thick target
will average over the individual resonances, leading to a smooth energy
dependence of the excitation function. Such a behavior will be nicely
illustrated e.g.\ for the target $^{27}$Al (see Sec.~\ref{sec:al27}).

\subsection{(Infinitely) Thick-target yields}
\label{sec:thick}
In principle, it is possible to derive reaction cross sections from
thick-target yield curves by differentiation. However, in practice this leads
to significant uncertainties. If the thick-target yield curve is measured with
large energy steps (e.g., $E_2 \gg E_1$), the resulting cross section is
averaged over a broad energy interval $E_2 - E_1$. Smaller energy steps reduce
this uncertainty; but at the same time the yields $Y(E_2)$ and $Y(E_1)$ become
more and more similar, and the cross section has to be derived from the
difference $Y(E_2) - Y(E_1)$ which is a small number. The uncertainty in the
difference of two quite similar numbers is further amplified if yield curves
are not available numerically and have to be re-digitized from
figures. Therefore, thick-target yield curves are not considered in this work.

One exception is made for the data by Roughton {\it et al.}\ \cite{Roughton83}
because these data allow to include the doubly-magic nucleus $^{48}$Ca in this
study. Roughton {\it et al.}\ have measured thick-target yield curves for 36
nuclear reactions in relatively small energy steps. The data are available
numerically from Table II in \cite{Roughton83}, and the conversion from the
given thick-target yield to cross sections is precisely defined in Eq.~(3) of
an earlier study of proton-induced reactions \cite{Roughton79}; it is based on
the energy-loss formulae given in \cite{Zaidins74}. The resulting cross
sections are in reasonable agreement with other available data as long as the
energy difference between two subsequent yields is sufficiently large. As
expected, for very small energy differences the uncertainty of the derived
average cross section increases dramatically. Surprisingly it turns out
that practically all of these cross sections from small energy differences are
much lower than other experimental data. Despite of these obvious
inconsistencies, this allows at least to 
determine a good estimate of the $^{48}$Ca\ran $^{51}$Ti cross section from
Roughton {\it et al.}\ \cite{Roughton83}. However, because of the above
inconsistencies in the Roughton {\it et al.}\ data, in most cases the data are
only shown in 
the cross section plots without further discussion in the text (with the same
symbol ``star'' and same color ``olive-green'' in all figures), and the data
are omitted in the plots of reduced cross sections (see Sec.~\ref{sec:pres}).

\subsection{Elastic scattering}
\label{sec:elastic}
The determination of total reaction cross sections \sreac\ from elastic
scattering can be done under two prerequisites. First, the deviation of the
elastic scattering cross section from the Rutherford cross section of
point-like charges has to exceed the experimental uncertainty. Thus, at
energies below the Coulomb barrier experimental data with very small
uncertainties are required. Second, full angular distributions must be
available. Often total reaction cross sections are determined by fits of an
optical potential; but it should be kept in mind that the choice of the
parametrization of the optical potential already restricts the model space,
and thus the derived \sreac\ may become model-dependent. Such a model
dependence should always be checked by a model-independent phase shift
analysis. In the present work the formalism of Chiste {\it et
  al.}\ \cite{Chi96} was applied for this purpose. 

For completeness it should also be noted that the parameters of optical
potentials are not very well constrained from elastic scattering below the
Coulomb barrier. So-called continuous and discrete ambiguities are often found
(see e.g.\ \cite{Mohr97}). However, although the parameters of the optical
potential may remain uncertain, the resulting angular distributions are more
or less similar. It remains then possible to determine total cross sections
\sreac\ from elastic scattering even at low energies where it is impossible to
determine the parameters of the optical potential \cite{Mohr13b}.

\subsection{Availability of experimental data}
\label{sec:avail}
Fortunately, nowadays many experimental data are provided by the EXFOR database
\cite{EXFOR} which is a great facilitation for a literature overview. However,
it has to be kept in mind that the quality of the data in EXFOR depends
sensitively on the data source. Newer data are often provided by the authors
of the experimental paper. For earlier papers the original data are only
available if the data are listed in a table in the paper (or in an underlying
thesis or laboratory report; however, the latter are often not easily
accessible). If original data are not available, the EXFOR editors have often
re-digitized experimental data from figures. In such cases significant
uncertainties arise from the digitization procedure which may exceed the
experimental uncertainties of the original data. This holds in particular for
small figures in logarithmic scale. 

The determination of total reaction cross sections \sreac\ from elastic
scattering is particularly sensitive to the available data quality because the
experimental \raa\ angular distribution must be fitted in an optical model
calculation or phase shift analysis. In many cases re-digitized data in EXFOR
are listed without experimental error bars; then assumptions on the
uncertainties have to be made for the fitting procedure. Fortunately, for the
phase shift fits shown above in Figs.~\ref{fig:s34scat} and \ref{fig:cr50scat}
the original data of Bredbecka {\it et al.}\ \cite{Bredbecka94} could be
restored, and these original data were sent to EXFOR to replace the previously
available re-digitized data.

\section{Results}
\label{sec:res}
\subsection{General remarks on the presentation of results}
\label{sec:pres}
The results for \al -induced reactions for nuclei in the $A \approx 20 - 50$
mass range will 
be presented in the following way. For each nucleus the available data at
EXFOR will be briefly described. A comparison is made between the experimental
\rap\ and \ran\ cross sections and predictions from the StM. Here I use the
TALYS code \cite{TALYS} with standard parameters except the \al -nucleus
potential where the potential by McFadden/Satchler is selected. As the \al
-nucleus potential is the essential ingredient in most cases (see discussion
above), NON-SMOKER \cite{NONSMOKER} calculations lead to very similar results
because the McFadden/Satchler potential is the default option in NON-SMOKER.

In practically all cases the \rag\ cross section is much smaller than the
\rap\ or \ran\ cross section. Thus, \rag\ cross sections will be shown only in
few cases; an explanation will be given in the respective sections for these
special cases.

For some nuclei under study only very few or even no \rap\ or \ran\ data are
available in literature. For these nuclei it was attempted to obtain further
information on the total reaction cross section \sreac\ from the analysis of
elastic scattering angular distributions. In practice, this analysis can only
be performed at energies around or above the Coulomb barrier whereas at very
low energies below the Coulomb barrier the elastic scattering cross section
approaches the Rutherford cross section of point-like charges.

For each nucleus under study the available data for the \rap\ and
\ran\ reactions (and also the \rag\ data and total reaction cross sections
\sreac\ from elastic scattering) will be shown in a first figure. In this
figure also a comparison with a calculation in the StM will be made. The
energy in all these figures is $E_\alpha$ in the laboratory system; for
experiments in inverse kinematics, the corresponding energy $E_\alpha$ is
calculated from the given energy of the heavy projectile.

In addition to the comparison of reaction cross sections, reduced cross
sections \sred\ are shown versus the reduced energy \Ered\ in a second figure
for each nucleus under study. In these \sred\ figures the experimental data
are shown together with the three calculations of \sred\ from the total
reaction cross section \sreac\ for $^{21}$Ne, $^{36}$Ar, and
$^{51}$V (these calculations have already been shown as full lines in
Fig.~\ref{fig:sred_heavy}). The purpose of this presentation is to show how
the experimental data for nuclei with $A \approx 20 - 50$ move smoothly with
decreasing $A$ from the calculation for $^{51}$V to the calculation for
$^{21}$Ne with the few noticeable exceptions of $^{40}$Ar, $^{36}$Ar,
$^{33}$S, and $^{23}$Na (as already mentioned in the introduction). For this
purpose the scale of all graphs with \sred\ versus \Ered\ is exactly the same
to guide the eye, and the systematics for \al -induced reaction cross sections
for heavy targets (see Fig.~\ref{fig:sred_heavy}) is repeated in each graph
with small blue crosses.

In general, experimental data for \ran\ reactions will be shown as open
symbols, and \rap\ reactions will be shown as full symbols. Exceptions will be
indicated in the figure captions.

For several nuclei under study, data in the book of Levkovskij \cite{Lev94}
are referenced in EXFOR. Often significant deviations to other available data
sets are found for the Levkovskij data. As the book \cite{Lev94} is not
available to the author of this study, it is not possible to trace back to the
origin of these discrepancies. The data of \cite{Lev94} are omitted in the
following graphs.

\subsection{$^{50}$Cr}
\label{sec:cr50}
The $^{50}$Cr\ran $^{53}$Fe and $^{50}$Cr\rap $^{53}$Mn reactions can both be
measured by activation. However, the half-life of $^{53}$Mn is extremely long
($T_{1/2} = 3.74 \times 10^6$\,years), and no activation data are available
for the proton channel. Because of the strongly negative $Q$-value of the
\ran\ reaction and $Q \approx -0.44$\,MeV for the \rap\ reaction, the
\rap\ channel is dominating at low energies, but at energies above about
$6-7$\,MeV the \ran\ cross section also contributes significantly to the total
reaction cross section \sreac .

\begin{figure}[bht]
  \includegraphics{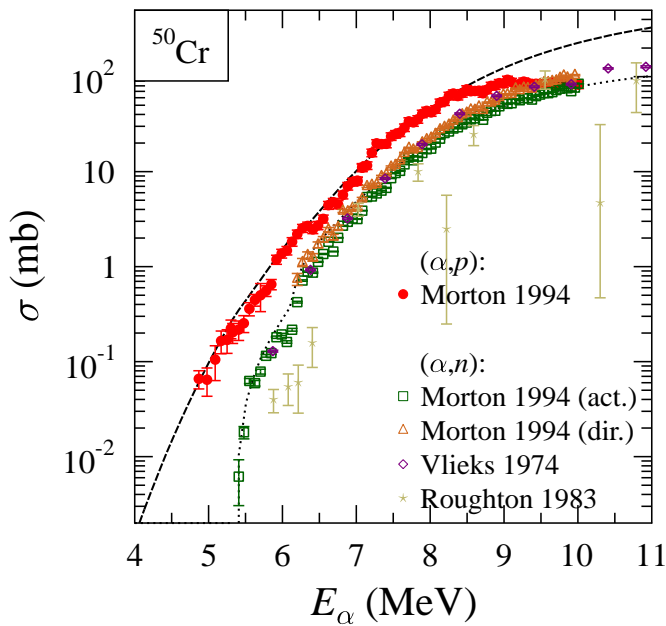}
\caption{
Cross sections of the $^{50}$Cr\ran $^{53}$Fe and $^{50}$Cr\rap$^{53}$Mn
reactions. The \ran\ data are shown with open symbols and dotted line, the
\rap\ data are shown with full symbols and dashed line. (The same style will
be used in all following figures except explicitly noted.) The experimental
data have been taken from \cite{Morton94,Vlieks74,Roughton83}.
Further discussion see text.
}
\label{fig:sig_50cr}
\end{figure}

A detailed study of both \al -induced reactions on $^{50}$Cr has been done by
Morton {\it el al.}\ \cite{Morton94}. The excitation function of the
$^{50}$Cr\rap $^{53}$Mn reaction has been measured using a silicon detector at
the angle of $\vartheta = 125^\circ$. The measured protons have been grouped
into $p_0$, $p_1$, $p_{2-4}$, $p_{5-7}$, $p_{8-15}$, and
$p_{16-28}$. Corrections for the angular distributions have been taken from
StM calculations; however, these corrections remain relatively small because
of the chosen angle. The total \rap\ cross section is determined by the sum
over the above proton groups. The result is shown in Fig.~\ref{fig:sig_50cr}.

The experimental $^{50}$Cr\rap $^{53}$Mn cross section is compared to a
StM calculation using the code TALYS with default parameters and the \al
-nucleus potential by McFadden and Satchler \cite{McF66} (dashed line in
Fig.~\ref{fig:sig_50cr}). It can be seen that the agreement is excellent at
low energies whereas above about 8\,MeV the calculation is higher than the
experimental data. As the StM calculations are typically very stable and
reliable at higher energies, this deviation may indicate that the
summation over the proton groups ($p_{0-28}$) is insufficient at the highest
energies. Alternatively, the uncertainty of the correction on the angular
distribution of the proton groups increases with energy \cite{Morton94}; this
uncertainty may be larger than estimated by the authors. 

\begin{figure}[b]
  \includegraphics{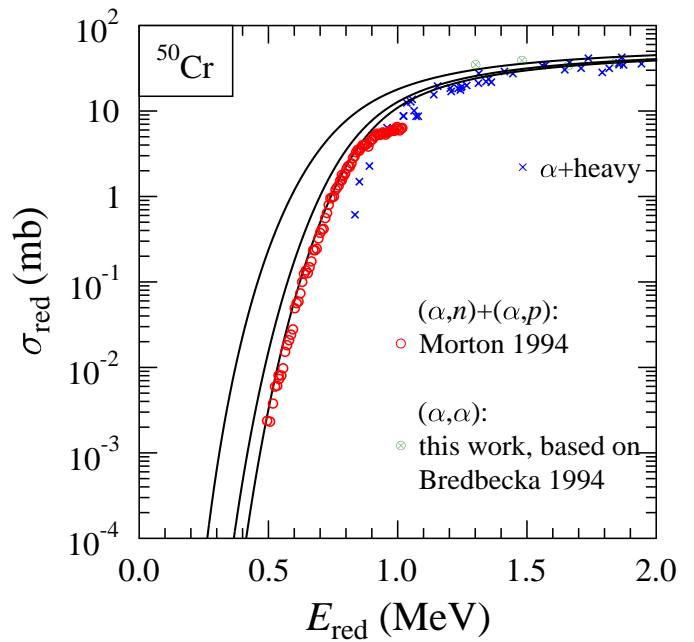}
\caption{
Reduced cross section \sred\ versus reduced energy \Ered\ for \al -induced
reactions on $^{50}$Cr. The experimental data have been taken from
\cite{Morton94}. In addition, the result of a reanalysis of the elastic
scattering data of \cite{Bredbecka94} is shown. Data points from elastic
scattering on heavy ($A > 60$) target nuclei are repeated from
Fig.~\ref{fig:sred_heavy} (blue crosses). As a guide to the eye, the
calculations for $^{21}$Ne, $^{36}$Ar, and $^{51}$V are also repeated from
Fig.~\ref{fig:sred_heavy} (full black lines).
Further discussion see text.
}
\label{fig:sred_50cr}
\end{figure}
The $^{50}$Cr\ran $^{53}$Fe reaction has been measured by Morton {\it et
  al.}\ using two independent techniques. Neutrons were counted directly with
$^3$He-filled proportional counters, and activation was observed with a
germanium detector by detection of the 378\,keV $\gamma$-ray in the decay of
$^{53}$Fe $\rightarrow$ $^{53}$Mn. Both data sets show the same energy
dependence, but unfortunately the two data sets deviate by about 20\,\% in
their absolute scale. The reason for this discrepancy remains unclear. An
earlier experiment by Vlieks {\it el al.}\ \cite{Vlieks74} has measured the
induced activity by positron counting. These earlier data are in better
agreement with the direct neutron data by Morton {\it et al.}
\cite{Morton94}. As a similar experiment has been done by the same authors on
$^{51}$V\ran $^{54}$Mn (see next Sec.~\ref{sec:v51}) where direct neutron
counting and activation are in excellent agreement, the discrepancy observed
for the $^{50}$Cr\ran $^{53}$Fe reaction is probably the consequence of an
incorrect decay branching of the analyzed  378\,keV $\gamma$-ray. Morton 
{\it et al.}\ have used $42 \pm 8$\,\%; the present ENSDF evaluation
recommends $42 \pm 3$\,\% \cite{ENSDF}, and good agreement between direct
neutron counting and activation would be obtained for a branching of about
$34$\,\%. 

It is obvious from Fig.~\ref{fig:sig_50cr} that the StM calculations reproduce
the experimental data of both reactions very well. In particular at low
energies the agreement is excellent, and a possible explanation for the
deviation at higher energies in the $^{50}$Cr\rap $^{53}$Mn channel has
already been given above. 

For the determination of the total reaction cross section \sreac\ and the
reduced cross section \sred , the \rap\ and \ran\ data of Morton {\it et
  al.}\ \cite{Morton94} were summed. The result is shown in
Fig.~\ref{fig:sred_50cr}. In addition, the data from elastic $^{50}$Cr\raa
$^{50}$Cr scattering (see Fig.~\ref{fig:cr50scat}) are shown. The \sred\ data
are slightly higher than the general trend which was derived from elastic
\raa\ scattering of heavy target nuclei. 
Similar to most nuclei under study in this work, the \sred\ data for $^{50}$Cr
do not show a peculiar behavior.

\subsection{$^{51}$V}
\label{sec:v51}
Many experimental data are available at EXFOR for the $^{51}$V\ran $^{54}$Mn
reaction 
which can be measured by activation. Contrary, only one data set is available
for the $^{51}$V\rap $^{54}$Cr reaction which was measured using the same
technique as for $^{50}$Cr\rap $^{53}$Mn (see previous Sec.~\ref{sec:cr50}).

\begin{figure}[ht]
  \includegraphics{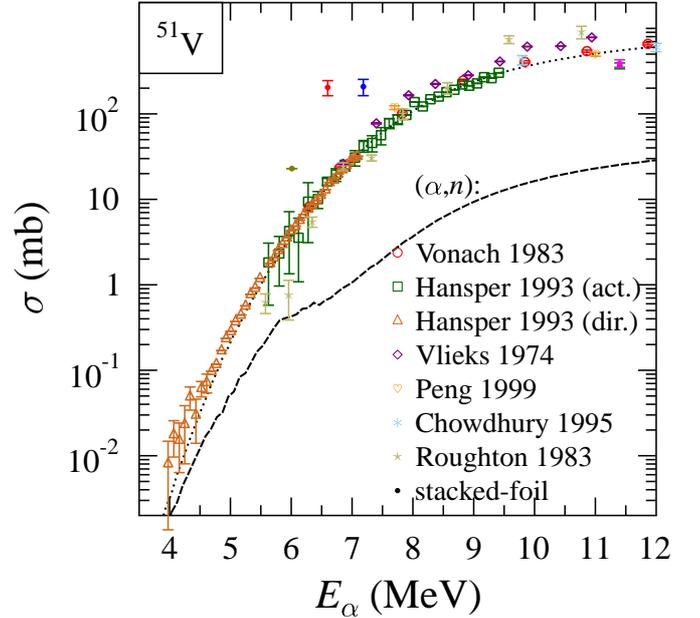}
\caption{
Cross sections of the $^{51}$V\ran $^{54}$Mn and $^{51}$V\rap$^{54}$Cr
reactions. 
The experimental data have been taken from
\cite{Vonach83,Hansper93,Vlieks74,Peng99,Chow95,Bowman69,Iguchi60,Ramarao87,Singh93,Singh95,Sonzogni93,Roughton83}. The
lowest data points of several stack-foil experiments
\cite{Bowman69,Iguchi60,Ramarao87,Singh93,Singh95,Sonzogni93} (shown as small
dots in different colors) are in disagreement with the other data.
Further discussion see text.
}
\label{fig:sig_51v}
\end{figure}

The $^{51}$V\ran $^{54}$Mn data are shown in Fig.~\ref{fig:sig_51v}. The
precision data by Vonach {\it et al.}\ \cite{Vonach83} were measured by
activation and are in excellent agreement with the activation data and the
direct neutron counting data by Hansper {\it et
  al.}\ \cite{Hansper93}. Earlier data by Vlieks {\it et al.}\ \cite{Vlieks74}
are slightly higher especially at energies above 9\,MeV. Recent stacked-foil
data by Peng {\it et al.}\ \cite{Peng99} and Chowdhury {\it et
  al.}\ \cite{Chow95} are in good agreement with the other data whereas
earlier stacked-foil data
\cite{Bowman69,Iguchi60,Ramarao87,Singh93,Singh95,Sonzogni93} deviate
significantly at the lowest energies. The agreement of the experimental data
with the StM calculation is excellent over the full shown energy range.
\begin{figure}[hb!]
  \includegraphics{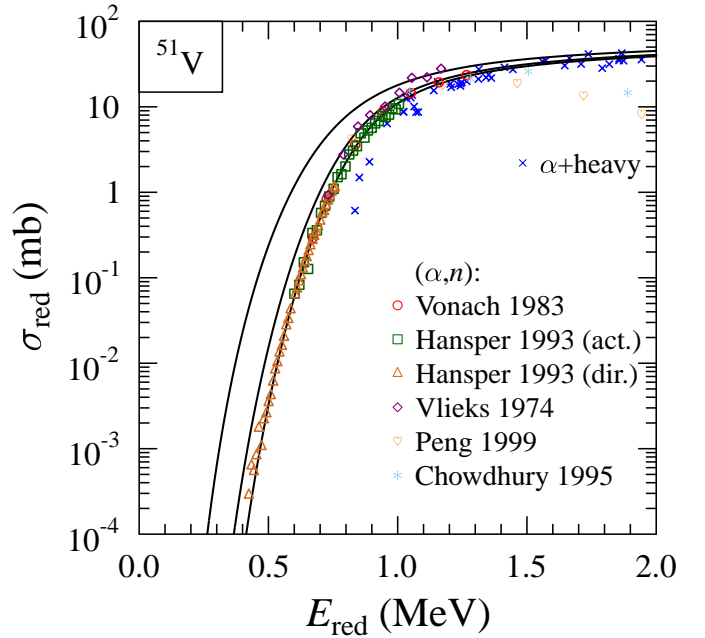}
\caption{
Same as Fig.~\ref{fig:sred_50cr}, but for \al -induced
reactions on $^{51}$V. The experimental data have been taken from
\cite{Vlieks74,Vonach83,Hansper93,Peng99,Chow95}. 
Further discussion see text.
}
\label{fig:sred_51v}
\end{figure}
The $^{51}$V\rap $^{54}$Cr data by Hansper {\it et al.}\ \cite{Hansper93} are
not shown in Fig.~\ref{fig:sig_51v} for several reasons. It is impossible to
derive the total \rap\ cross section from the information in the paper. Proton
groups ($p_0$, $p_1$, $p_2$, $p_{3-4}$, and $p_{5-7}$) are shown in the
spectrum (Fig.~3 of \cite{Hansper93}) but in the following cross section plots
the strong group $p_{5-7}$ is missing. In addition, the data at EXFOR are
re-digitized from the figures in \cite{Hansper93} which makes a point-by-point
addition of the proton groups practically impossible. Fortunately, the
$^{51}$V\rap $^{54}$Cr reaction contributes only very minor to the total
reaction cross section \sreac\ of $^{51}$V which is dominated by the
$^{51}$V\ran $^{54}$Mn reaction (see the shown calculations in
Fig.~\ref{fig:sig_51v}). Only at the lowest energies below about 5\,MeV the
\rap\ cross section becomes comparable to the \ran\ cross section whereas at
higher energies the \rap\ reaction contributes by less than 20\,\%
\cite{Vonach83}. The data points at the lowest energies
for the \rap\ reaction in \cite{Hansper93} are about $10^{-2}$\,mb
at $\approx 5$\,MeV with huge error bars, i.e.\ close to the theoretically
expected values (dashed line in Fig.~\ref{fig:sig_51v}).

Fig.~\ref{fig:sred_51v} shows the reduced cross section \sred\ for
$^{51}$V. For better readability only the \ran\ data of
\cite{Vlieks74,Vonach83,Hansper93,Peng99,Chow95} are shown. The agreement
between the experimental \ran\ data and the calculated total reaction cross
section \sreac\ and reduced cross section \sred\ is excellent up to
\Ered\ $\approx 1.5$\,MeV. At these energies other reaction channels open, and
thus the \ran\ cross section does not represent the total cross section
\sreac\ anymore. At even higher energies (above the range shown in
Fig.~\ref{fig:sred_51v}) the 
analysis of elastic $^{51}$V\raa $^{51}$V scattering leads to data points of
\sred\ $= 55 \pm 2$\,mb at \Ered\ $= 2.90$\,MeV \cite{Bilaniuk81} and
\sred\ $= 46 \pm 4$\,mb at \Ered\ $= 2.49$\,MeV \cite{Lep14}, again in
reasonable agreement with the theoretical expectations of 46.4\,mb and
43.8\,mb. Finally, it is interesting to note that the reduced cross sections
\sred\ for the semi-magic ($N=28$) nucleus $^{51}$V are very similar to the
non-magic neighboring nuclei $^{50}$Cr (see previous Sec.~\ref{sec:cr50}) and
$^{48}$Ti (see following Sec.~\ref{sec:ti48}).
Similar to most nuclei under study in this work, the \sred\ data for $^{51}$V
do not show a peculiar behavior.

\subsection{$^{50}$V}
\label{sec:v50}
The odd-odd ($Z = 23$, $N = 27$) nucleus $^{50}$V has a very low natural
abundance. Unfortunately, no data for \al -induced reactions are available
below 10\,MeV. The only available data set by Peng {\it et al.}\ \cite{Peng99}
covers the $^{50}$V($\alpha$,$2n$)$^{52}$Mn reaction from close above
threshold around $\approx 13$\,MeV to about 26\,MeV. The data have been
measured using the stacked-foil activation technique. The experimental data
are well reproduced by the StM calculation. However, the data do not restrict
the total reaction cross section \sreac\ of \al -induced reactions on
$^{50}$V, and thus no figure for cross sections or reduced cross sections
\sred\ is shown here. Nevertheless, from the nice agreement between the
experimental $^{50}$V($\alpha$,$2n$)$^{52}$Mn data and the StM calculation it
can be concluded that there is at least no evidence for a peculiar behavior of
the odd-odd nucleus $^{50}$V.

\subsection{$^{48}$Ti}
\label{sec:ti48}
\begin{figure}[htb]
  \includegraphics{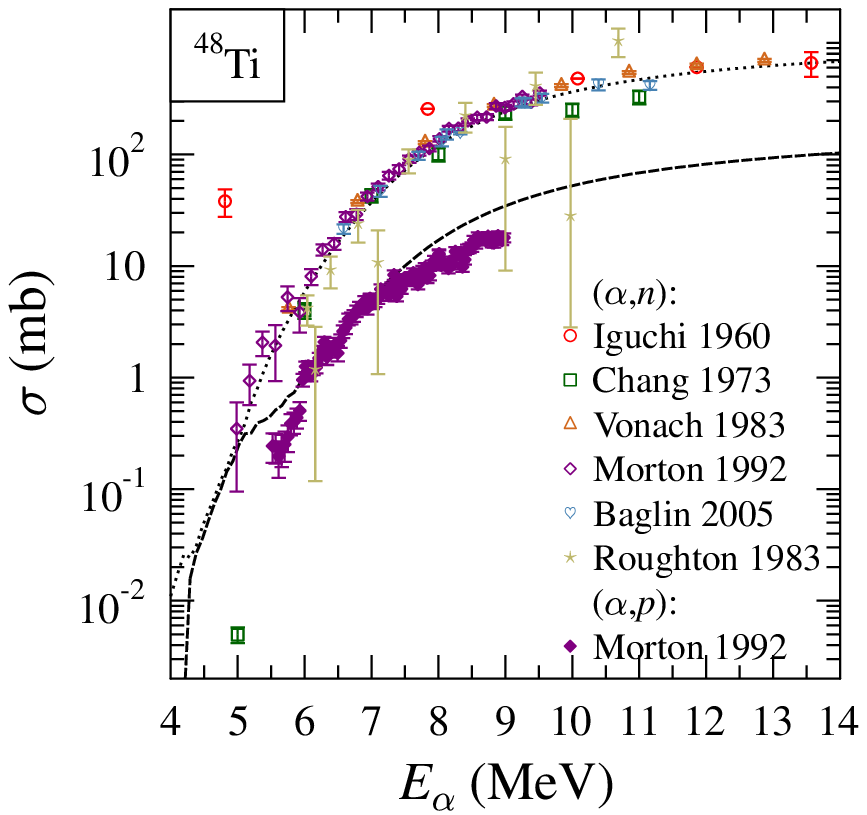}
\caption{
Cross sections of the $^{48}$Ti\ran $^{51}$Cr and $^{48}$Ti\rap$^{51}$V
reactions. 
The experimental data have been taken from
\cite{Iguchi60,Chang73,Vonach83,Morton92,Baglin05,Roughton83}.
Further discussion see text.
}
\label{fig:sig_48ti}
\end{figure}

Five data sets for the $^{48}$Ti\ran $^{51}$Cr reaction are available from
EXFOR. All experiments have applied activation techniques. The
precision data for the $^{48}$Ti\ran $^{51}$Cr reaction by Vonach {\it et
  al.}\ \cite{Vonach83} have been obtained by measuring the induced activity
by $\gamma$-spectroscopy. The same technique was used by Morton {\it et
  al.}\ \cite{Morton92} and Baglin {\it et al.}\ \cite{Baglin05} whereas Chang
{\it el al.}\ \cite{Chang73} used $X$-ray spectroscopy of the 4.95\,keV
$X$-ray which is emitted in $^{51}$V after the electron capture decay of
$^{51}$Cr. Iguchi {\it et al.}\ \cite{Iguchi60} used the stacked-foils
technique and $\gamma$-spectroscopy. It can be seen from
Fig.~\ref{fig:sig_48ti} that in general the data are in good
agreement. Exceptions are the lowest energy points of the stacked-foil
experiment \cite{Iguchi60} and also the lowest data point of the $X$-ray
experiment \cite{Chang73}; here the analysis at the lowest energy may be
hampered by the relatively thick target (485\,$\mu$g/cm$^2$). The agreement
with the StM calculation is excellent over the full energy range.
\begin{figure}[b]
  \includegraphics{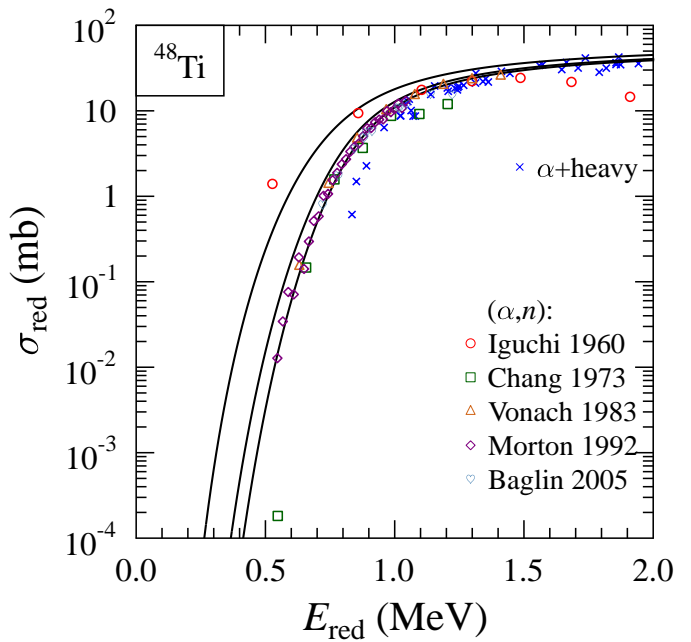}
\caption{
Same as Fig.~\ref{fig:sred_50cr}, but for \al -induced
reactions on $^{48}$Ti. The experimental data have been taken from
\cite{Iguchi60,Chang73,Vonach83,Morton92,Baglin05}.
Further discussion see text.
}
\label{fig:sred_48ti}
\end{figure}
The $^{48}$Ti\rap $^{51}$V reaction has been measured by Morton {\it et
  al.}\ \cite{Morton92}. Cross sections for the proton groups $p_0$ to $p_5$
are shown in Fig.~4 of \cite{Morton92}; the sum of these cross sections is
shown as \rap\ cross section in Fig.~\ref{fig:sig_48ti}. The proton groups
$p_{6-8}$ could not be resolved from background reactions, but should have
only a minor contribution to the total \rap\ cross section (see the spectrum
in Fig.~3 of \cite{Morton92}). The agreement with the StM calculation is good
for the $^{48}$Ti\rap $^{51}$V reaction
although the experimental results are slightly overestimated at the upper and
lower end of the measured energy interval. It is interesting to note that the
NON-SMOKER calculation for the $^{48}$Ti\rap $^{51}$Cr reaction deviates from
the TALYS calculation, leading to better agreement at higher and lower
energies, but underestimation in the middle.

Because the \ran\ cross section for $^{48}$Ti is much larger than the
\rap\ cross section, the total reaction cross section \sreac\ and the reduced
cross section \sred\ are taken from the \ran\ data. A contribution of less
than 20\,\% was estimated in \cite{Vonach83} for the \rap\ reaction. At higher
energies the low \sred\ values from \cite{Iguchi60} can be explained by
additional open channels. At even higher energies the analysis of elastic
scattering angular distributions in \cite{Bilaniuk81} leads to \sred\ $=
54.6$\,mb at \Ered\ $= 2.98$\,MeV (not shown in Fig.~\ref{fig:sred_48ti}).
Similar to most nuclei under study in this work, the \sred\ data for $^{48}$Ti
do not show a peculiar behavior.

\subsection{$^{46}$Ti}
\label{sec:ti46}
\begin{figure}[h]
  \includegraphics{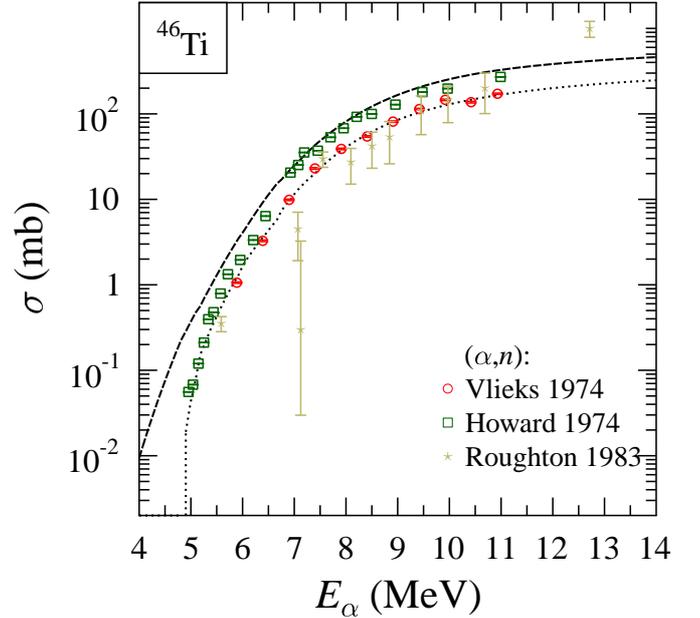}
\caption{
Cross section of the $^{46}$Ti\ran $^{49}$Cr reaction. 
The experimental data have been taken from
\cite{Vlieks74,Howard74,Roughton83}.
Further discussion see text.
}
\label{fig:sig_46ti}
\end{figure}

Only few data sets are available for $^{46}$Ti. The cross section of the
$^{46}$Ti\ran $^{49}$Cr reaction has been measured by activation and
annihilation spectroscopy by Vlieks {\it et al.}\ \cite{Vlieks74} and by
Howard {\it et al.}\ \cite{Howard74}. The data by Vlieks {\it et al.}\ are
significantly lower than the data by Howard {\it et al.}\ (see
Fig.~\ref{fig:sig_46ti}). The Vlieks {\it et al.}\ data are confirmed by the
thick-target yield measured by Roughton {\it et al.}\ \cite{Roughton83}. As
the experimental data from Vlieks {\it et al.}\ \cite{Vlieks74} for 
$^{50}$Cr (see Sec.~\ref{sec:cr50}),
$^{51}$V (see Sec.~\ref{sec:v51}), and
$^{45}$Sc (see Sec.~\ref{sec:sc45})
agree with other experimental data, the data of Vlieks {\it et al.}\ should be
adopted. In addition, the data by Vlieks {\it et al.}\ are available from a
table in \cite{Vlieks74} whereas the data by Howard {\it et al.}\ had to be
re-digitized from Fig.~3 of \cite{Howard74} (see also comment in
Sec.~\ref{sec:avail}). Furthermore, also for $^{40}$Ca and $^{35}$Cl the data
by Howard {\it et al.}\ \cite{Howard74} are slightly higher than other
available data (see Sec.~\ref{sec:ca40} and \ref{sec:cl35}).

\begin{figure}[b!]
  \includegraphics{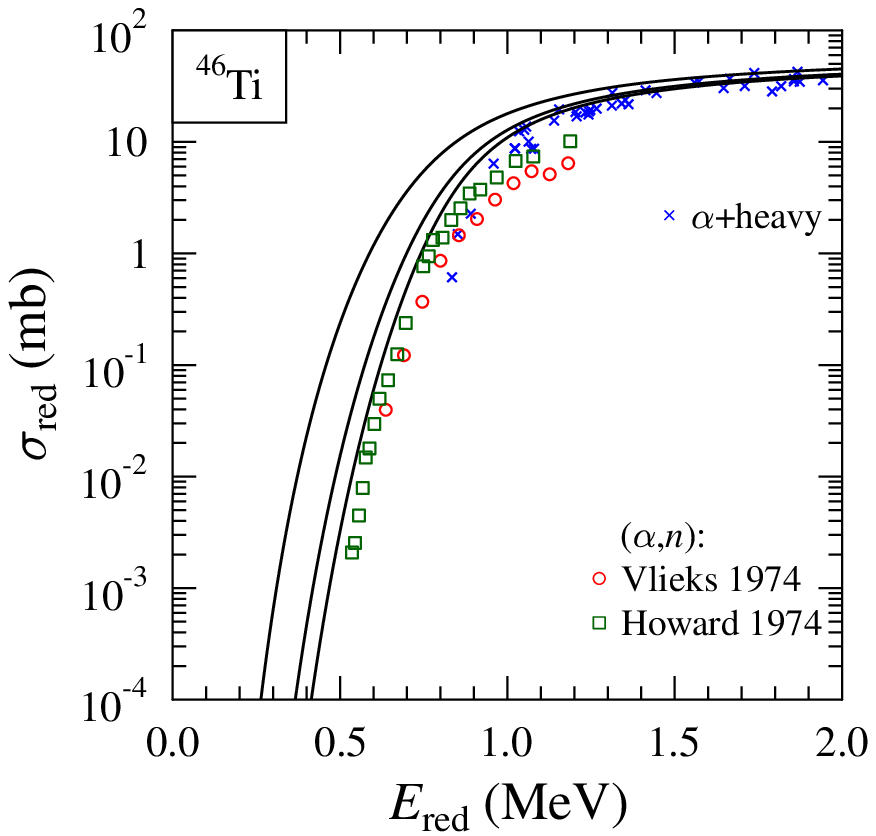}
\caption{
Same as Fig.~\ref{fig:sred_50cr}, but for \al -induced
reactions on $^{46}$Ti. The experimental data have been taken from
\cite{Howard74,Vlieks74}. 
According to StM calculations, the shown $^{46}$Ti\ran $^{49}$Cr cross section
is about a factor of two smaller than the dominating $^{46}$Ti\rap $^{49}$V
cross section. Consequently, the total reaction cross section \sreac\ should
be about a factor of three larger than the shown \ran\ cross sections.
Further discussion see text.
}
\label{fig:sred_46ti}
\end{figure}

The agreement between the data of Vlieks {\it et al.}\ \cite{Vlieks74} and the
StM calculation (dotted line in Fig.~\ref{fig:sig_46ti}) is again
excellent. However, the total reaction cross section \sreac\ and the reduced
cross section \sred\ are dominated by the $^{46}$Ti\rap $^{49}$V reaction
(dashed line in Fig.~\ref{fig:sig_46ti}) where no data are available from
EXFOR. As a consequence, the \sred\ data from 
the $^{46}$Ti\ran $^{49}$Cr reaction are significantly lower than the
expectation for the total reaction cross section (see
Fig.~\ref{fig:sred_46ti}).

\subsection{$^{44}$Ti}
\label{sec:ti44}
It is not surprising that only very few data are available for the radioactive
nucleus $^{44}$Ti. Nevertheless, a determination of the total reaction cross
section at low energies is possible from the dominating $^{44}$Ti\rap $^{47}$V
reaction. The $^{44}$Ti\ran $^{47}$Cr reaction has a strongly negative
$Q$-value and does not contribute to \sreac\ at low energies.
\begin{figure}[htb]
  \includegraphics{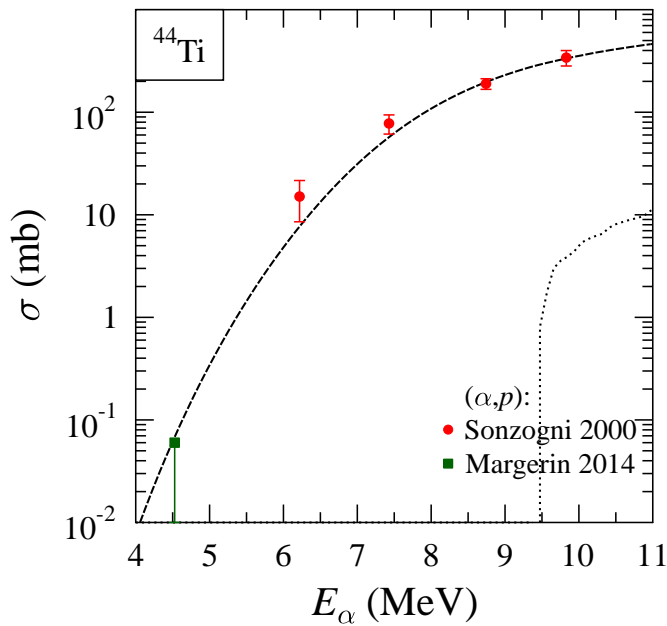}
\caption{
Cross section of the $^{44}$Ti\rap $^{47}$V reaction. 
The experimental data have been taken from
\cite{Sonzogni00,Margerin14}.
Further discussion see text.
}
\label{fig:sig_44ti}
\end{figure}

The cross section of the $^{44}$Ti\rap $^{47}$V reaction has been determined
by Sonzogni {\it et al.}\ \cite{Sonzogni00} using a radioactive $^{44}$Ti beam
in combination with a $^4$He gas cell and the Argonne fragment mass
analyzer. Very recently, at lower energies an upper limit was obtained by
Margerin {\it et al.}\ \cite{Margerin14} at CERN using also a $^{44}$Ti
beam. Further suggestions for experiments have been made in \cite{Abdullah14}
very recently. The two data sets \cite{Sonzogni00,Margerin14} are shown in
Fig.~\ref{fig:sig_44ti} and are compared to a StM calculation. Again very good
agreement between experiment and theory is found.

\begin{figure}[htb]
  \includegraphics{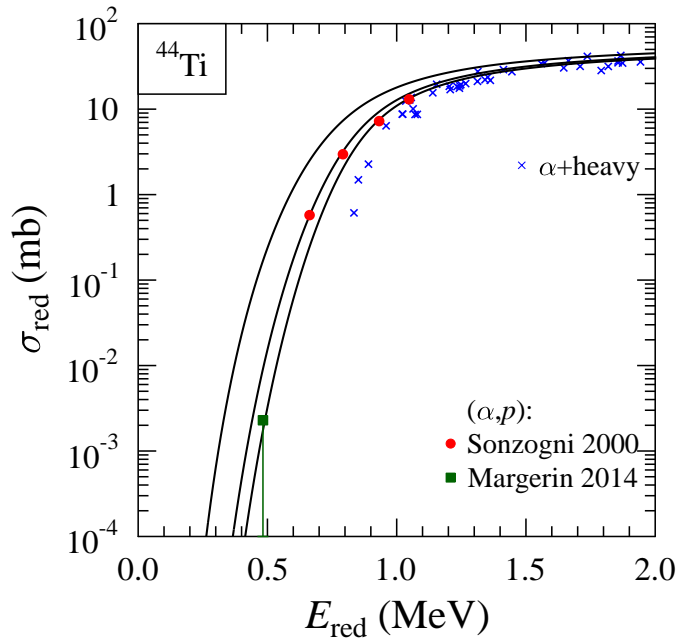}
\caption{
Same as Fig.~\ref{fig:sred_50cr}, but for \al -induced
reactions on $^{44}$Ti. The experimental data have been taken from
\cite{Sonzogni00,Margerin14}.
Further discussion see text.
}
\label{fig:sred_44ti}
\end{figure}
As the total reaction cross section \sreac\ is essentially defined by the
$^{44}$Ti\rap $^{47}$V cross section, reduced cross sections \sred\ can be
determined from the available \rap\ data \cite{Sonzogni00,Margerin14}. The
result is shown in Fig.~\ref{fig:sred_44ti}. An additional data point from
$^{44}$Ti\raa $^{44}$Ti elastic scattering (\sred\ $= 55.8$\, mb at \Ered\ $=
2.53$\,MeV; not shown in Fig.~\ref{fig:sred_44ti}) has already been presented
above in Sec.~\ref{sec:red}. 
Similar to most nuclei under study in this work, the \sred\ data for $^{44}$Ti
do not show a peculiar behavior.

\subsection{$^{45}$Sc}
\label{sec:sc45}
\begin{figure}[b]
  \includegraphics{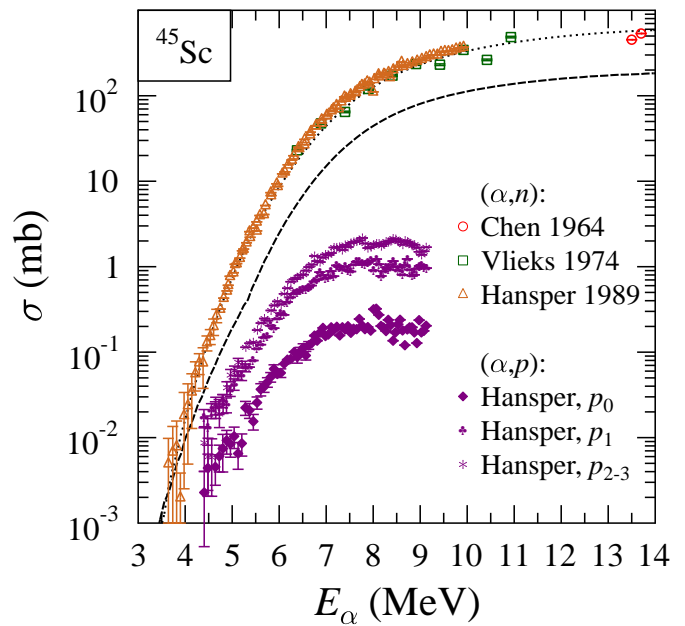}
\caption{
Cross sections of the $^{45}$Sc\ran $^{48}$V and $^{45}$Sc\rap $^{48}$Ti
reactions. The experimental data have been taken from
\cite{Chen64,Vlieks74,Hansper89}.
Further discussion see text.
}
\label{fig:sig_45sc}
\end{figure}

\al -induced reactions on $^{45}$Sc have been studied by Chen {\it et
  al.}\ \cite{Chen64}, Vlieks {\it et al.}\ \cite{Vlieks74}, and Hansper {\it
  et al.}\ \cite{Hansper89}. Chen {\it et al.}\ have used the stacked-foil
activation technique in combination with $\beta$-proportional counters, Vlieks
{\it et al.}\ used activation and annihilation spectroscopy, and Hansper {\it
  et al.}\ applied both direct neutron counting and activation in combination
with $\gamma$-ray spectroscopy. The different techniques provide results
which are in excellent agreement with each other (see
Fig.~\ref{fig:sig_45sc}). The StM calculation reproduces the experimental
cross sections of the $^{45}$Sc\ran $^{48}$V reaction very nicely.

The $^{45}$Sc\rap $^{48}$Ti reaction cannot be measured by activation because
the residual $^{48}$Ti is stable. Only one data set is available by Hansper
{\it et al.}\ \cite{Hansper89}. However, the shown data cover only the $p_0$,
$p_1$, and $p_{2-3}$ groups whereas the spectrum in Fig.~2 of \cite{Hansper89}
shows additional proton groups at higher excitation energies which cannot be
fully resolved from background. A rough estimate from that Fig.~2 shows that
about twice the strength of the shown $p_{0-3}$ groups is found in
higher-lying proton groups up to $p_{48}$ at $E_\alpha = 6.8$\,MeV. As
expected, the StM calculation for the total \rap\ cross section is far above
the experimental partial $p_0$, $p_1$, and $p_{2-3}$ cross sections. This is
consistent with the finding from the spectrum shown in Fig.~2 of
\cite{Hansper89}. The determination of the total \rap\ cross section is
further hampered by the fact that only re-digitized data are available for the
\rap\ cross sections.

\begin{figure}[t]
  \includegraphics{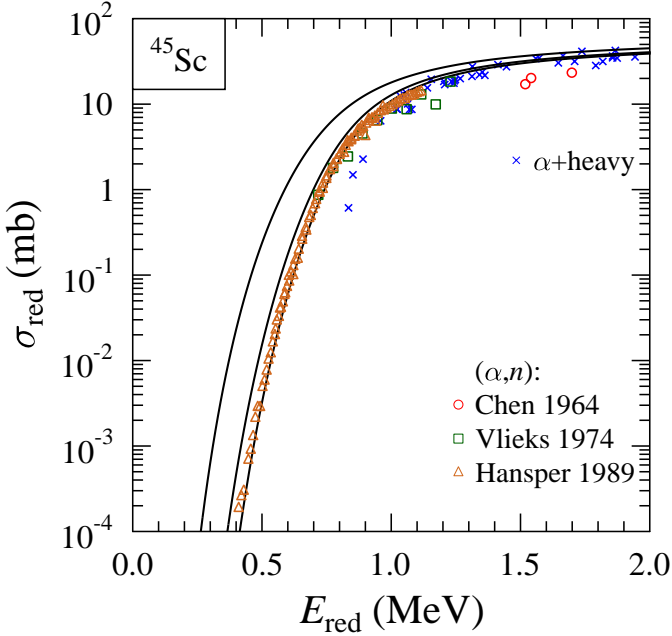}
\caption{
Same as Fig.~\ref{fig:sred_50cr}, but for \al -induced
reactions on $^{45}$Sc. The experimental data have been taken from
\cite{Chen64,Vlieks74,Hansper89}.
Further discussion see text.
}
\label{fig:sred_45sc}
\end{figure}
The calculated total \ran\ and \rap\ cross sections for $^{45}$Sc in
Fig.~\ref{fig:sig_45sc} show that the total reaction cross section \sreac\ and
the reduced cross section \sred\ are dominated by the \ran\ contribution at
energies above $\approx 7$\,MeV. However, at the lowest energies the \ran\ and
\rap\ cross sections are of comparable strength. The reduced cross sections
\sred\ in Fig.~\ref{fig:sred_45sc} are thus well-defined around
\Ered\ $\approx 1$\,MeV by the \ran\ cross section whereas at the lowest
energies in Fig.~\ref{fig:sred_45sc} around \Ered\ $\approx 0.5$\,MeV
\sred\ will be underestimated by about a factor of two. At higher energies one
further data point is obtained from elastic $^{45}$Sc\raa $^{45}$Sc
scattering: \sred\ $ = 52.1$\,mb at \Ered\ $= 3.06$\,MeV
\cite{Bilaniuk81}. 
Similar to most nuclei under study in this work, the
\sred\ data for $^{45}$Sc do not show a peculiar behavior.

\subsection{$^{48}$Ca}
\label{sec:ca48}
\begin{figure}[thb]
  \includegraphics{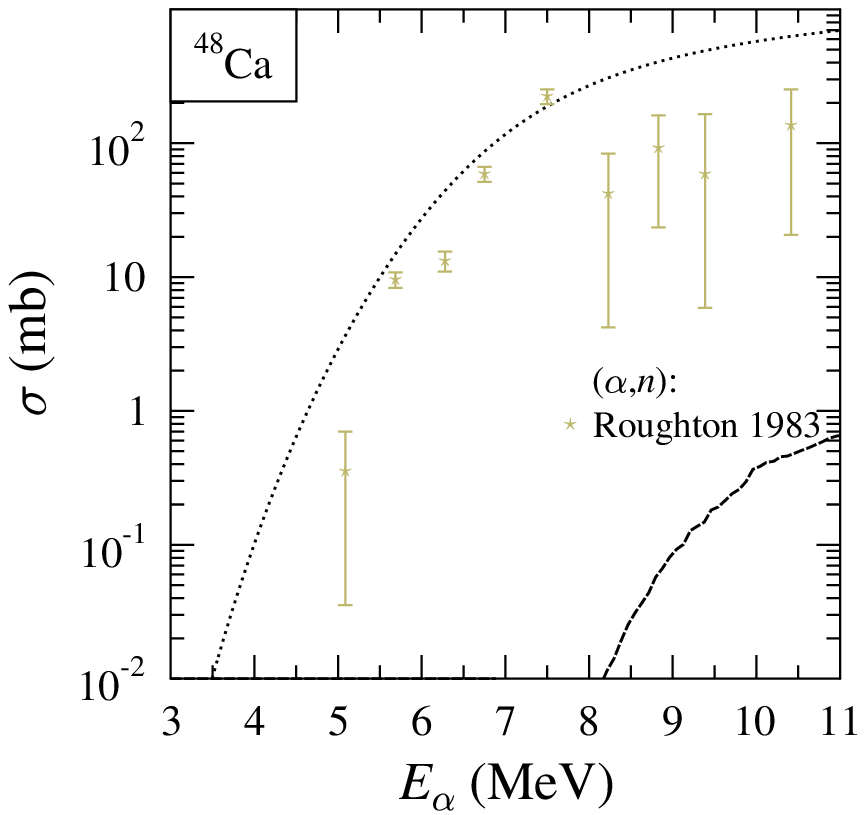}
\caption{
Cross sections of the $^{48}$Ca\ran $^{51}$Ti and $^{48}$Ca\rap $^{51}$Sc
reactions. 
The experimental data have been taken from \cite{Roughton83}.
Further discussion see text.
}
\label{fig:sig_48ca}
\end{figure}

$^{48}$Ca is a doubly-magic ($Z = 20$, $N = 28$) nucleus with a significant
neutron excess ($N/Z = 1.4$). As a consequence, the $^{48}$Ca\rap 
$^{51}$Sc reaction is strongly suppressed, and the total reaction cross section
\sreac\ at low energies is well-defined by its dominant $^{48}$Ca\ran
$^{51}$Ti contribution. Unfortunately, no data for this reaction are available
a EXFOR. The thick-target yield curve of Roughton {\it et
  al.}\ \cite{Roughton83} has been differentiated to extract the $^{48}$Ca\ran
$^{51}$Ti cross section (see Sect.~\ref{sec:thickness}). The result is shown
in Fig.~\ref{fig:sig_48ca}.
\begin{figure}[b!]
  \includegraphics{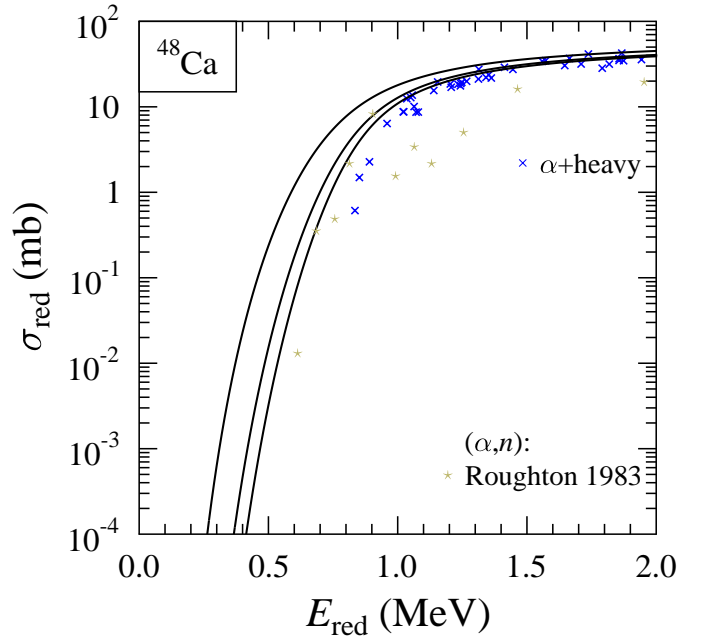}
\caption{
Same as Fig.~\ref{fig:sred_50cr}, but for \al -induced
reactions on $^{48}$Ca. The experimental data have been taken from
\cite{Roughton83}. 
Further discussion see text.
}
\label{fig:sred_48ca}
\end{figure}
Because of the few available reaction data for $^{48}$Ca, in addition the
elastic $^{48}$Ca\raa $^{48}$Ca scattering data of Gaul {\it et
  al.}\ \cite{Gaul69} at 18 to 29\,MeV were analyzed. Total cross sections
\sreac\ between 1365 and 1771\,mb were obtained for the four angular
distributions at 18.0, 22.0, 24.1, and 29.0\,MeV, corresponding to reduced
cross sections \sred\ of 50.0, 58.5, 64.9, and 59.3\,mb at \Ered\ $ = 2.17,
2.65, 2.90$, and 3.49\,MeV (above the shown range in
Fig.~\ref{fig:sred_48ca}). Although the experimental data are quite limited,
it can be concluded that there is no evidence for a peculiar behavior
of the reduced cross sections \sred\ for the doubly-magic nucleus $^{48}$Ca.

\subsection{$^{42}$Ca}
\label{sec:ca42}
The semi-magic $^{42}$Ca nucleus is characterized by relatively large negative
$Q$-values for the \ran\ ($-5.18$\,MeV) and \rap\ ($-2.34$\,MeV)
reactions. Therefore, at very low energies the $^{42}$Ca\rag $^{46}$Ti
reaction plays also an important role in the determination of the total
reaction cross section.

\begin{figure}[thb]
  \includegraphics{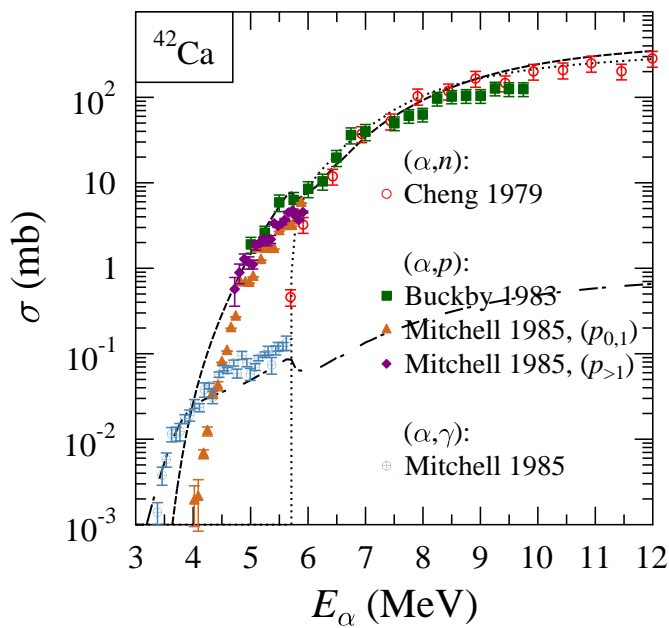}
\caption{
Cross sections of the $^{42}$Ca\ran $^{45}$Ti, $^{42}$Ca\rap $^{45}$Sc, and
$^{42}$Ca\rag $^{46}$Ti reactions. The experimental data have been taken from
\cite{Cheng79,Buckby83,Mitchell85}. The additional dash-dotted line shows the
StM calculation for the \rag\ reaction.
Further discussion see text.
}
\label{fig:sig_42ca}
\end{figure}
\begin{figure}[htb]
  \includegraphics{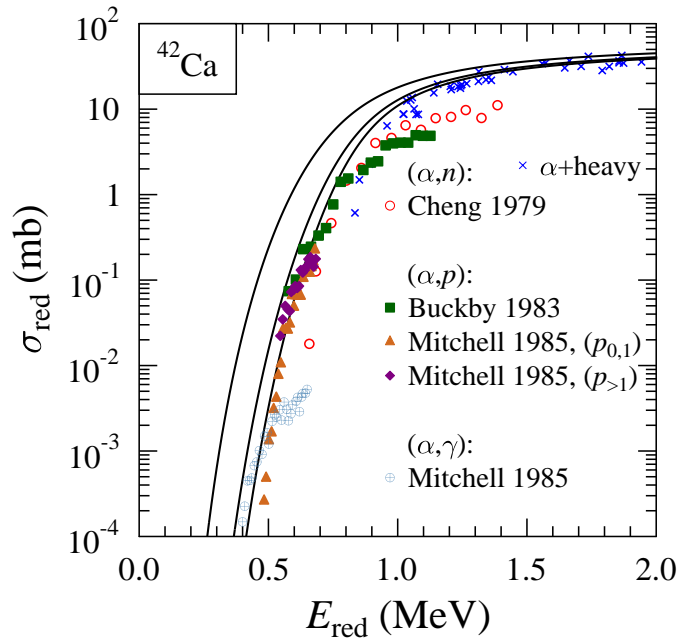}
\caption{
Same as Fig.~\ref{fig:sred_50cr}, but for \al -induced
reactions on $^{42}$Ca. The experimental data have been taken from
\cite{Cheng79,Buckby83,Mitchell85}. 
At energies above the \ran\ threshold, the cross sections of the $^{42}$Ca\ran
$^{45}$Ti and $^{42}$Ca\rap $^{45}$Sc reactions are very similar. Below the
\rap\ threshold, the total cross section \sreac\ is dominated by the
$^{42}$Ca\rag $^{46}$Ti reaction.
Further discussion see text.
}
\label{fig:sred_42ca}
\end{figure}
The $^{42}$Ca\ran $^{45}$Ti cross section has been determined by Cheng {\it et
  al.}\ \cite{Cheng79} using activation in combination with annihilation
spectroscopy. The energy range starts close above the \ran\ threshold. A
comparison with a StM calculation shows excellent agreement (see
Fig.~\ref{fig:sig_42ca}). 

Buckby {\it et al.}\ \cite{Buckby83} have measured excitation functions for
the $^{42}$Ca\rap $^{45}$Sc reaction at five different angles. Unfortunately, no
spectrum is shown in \cite{Buckby83}, but total cross sections for the
\rap\ reaction are reported, and it is stated that ``sufficient counts were
obtained for the smallest proton peaks in the spectrum'' and ``The
contribution of any remaining missed proton groups was then estimated by
reference to their percentage contribution to the total yield at higher
energies.'' At energies above the \ran\ threshold the \rap\ and \ran\ cross
sections are of comparable strength whereas at energies below the
\ran\ threshold (and obviously above the \rap\ threshold) the total reaction
cross section is dominated by the \rap\ cross section.

The \rap\ data by Buckby {\it et al.}\ are extended towards lower energy by
Mitchell {\it et al.}\ \cite{Mitchell85}. Here two techniques were
applied. The $^{42}$Ca($\alpha$,$p_{0,1}$)$^{45}$Sc cross section was measured
by a proton detector at $\vartheta = 145^\circ$, and isotropy of the angular
distributions was checked at few energies with a five-detector array. The
$^{42}$Ca($\alpha$,$p_{>1}$)$^{45}$Sc reaction was measured by the
$\gamma$-ray yield of de-exciting $^{45}$Sc residual nuclei. The total
$^{42}$Ca\rap $^{45}$Sc cross section is then obtained from the sum of the
measured ($\alpha$,$p_{0,1}$) and ($\alpha$,$p_{>1}$) cross
sections. Figs.~\ref{fig:sig_42ca} and \ref{fig:sred_42ca} show the individual
($\alpha$,$p_{0,1}$) and ($\alpha$,$p_{>1}$) cross sections. As both
contributions are almost equal, the total \rap\ cross section should be about
a factor of 2 higher than the two individual cross sections. However, this sum
is slightly higher than the result of Buckby {\it et al.}\ \cite{Buckby83}.

At energies below about $E_\alpha \approx 4$\,MeV, the \rap\ cross section
approaches its threshold ($Q = -2.34$\,MeV), and thus the cross section is
further suppressed by the Coulomb barrier in the exit channel. As a
consequence, the \rag\ cross section exceeds the \rap\ cross
section. Experimental data for the $^{42}$Ca\rag $^{46}$Ti cross section have
also been measured by Mitchell {\it et al.}\ \cite{Mitchell85} by
summing the intensities of the $\gamma$-rays to the ground state and the
$\gamma$-ray from the first excited $2^+$ state to the $0^+$ ground state in
$^{46}$Ti.

The total reaction cross section \sreac\ and the reduced cross section
\sred\ for $^{42}$Ca are essentially given by the \rag\ cross section at very
low energies, by the \rap\ cross section between about 5 and 6.5\,MeV, and the
sum of \rap\ and \ran\ cross sections at energies above the neutron
threshold. The individual \ran , \rap , and \rag\ cross sections are shown as
reduced cross sections \sred\ in Fig.~\ref{fig:sred_42ca}.
Similar to most nuclei under study in this work, the
\sred\ data for $^{42}$Ca do not show a peculiar behavior.

\subsection{$^{40}$Ca}
\label{sec:ca40}
\begin{figure}[ht]
  \includegraphics{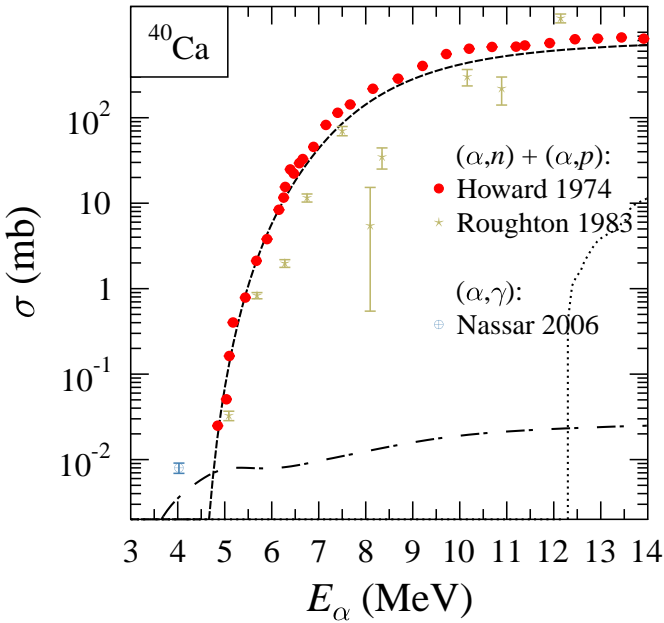}
\caption{
Cross sections of the $^{40}$Ca\ran $^{43}$Ti, $^{40}$Ca\rap $^{43}$Sc, and
$^{40}$Ca\rag $^{44}$Ti reactions. The experimental data have been taken from
\cite{Howard74,Nassar06,Roughton83}. The additional dash-dotted line shows the
StM calculation for the \rag\ reaction.
Further discussion see text.
}
\label{fig:sig_40ca}
\end{figure}

Similar to the semi-magic $^{42}$Ca, the doubly-magic $^{40}$Ca nucleus is
also characterized by strongly negative $Q$-values for the
\rap\ ($-3.52$\,MeV) and the \ran\ ($-11.17$\,MeV) reactions. As both residual
nuclei of the \rap\ and \ran\ reaction are unstable, and $^{43}$Ti has a very
short half-life of less than 1 second, the activation technique can be applied
to measure the sum of the \rap\ and \ran\ cross sections by detection of the
induced $^{43}$Sc activity. Annihilation spectroscopy was used by Howard {\it
  et al.}\ \cite{Howard74} for this purpose. The result is shown in
Fig.~\ref{fig:sig_40ca}. The agreement with the StM calculation is excellent
for lower energies. At higher energies the experimental data are
slightly underestimated. As the StM calculation shows, the \ran\ cross section
is practically negligible for $^{40}$Ca.
\begin{figure}[h]
  \includegraphics{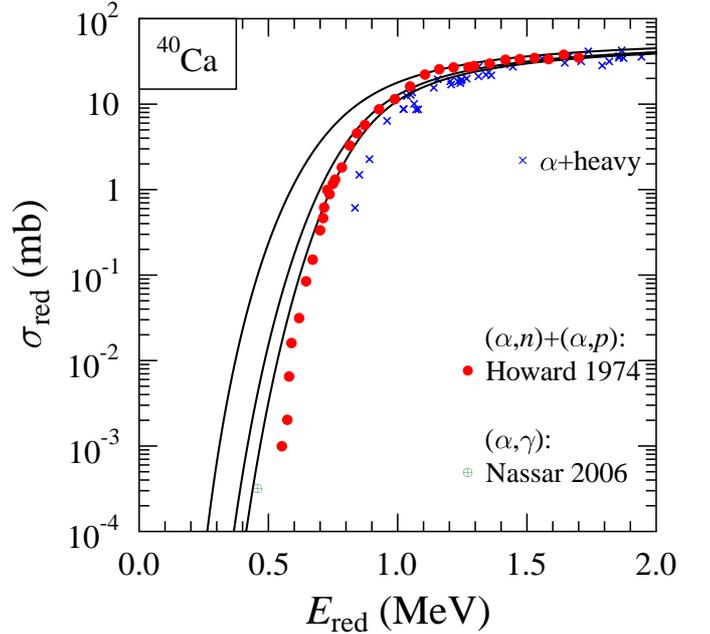}
\caption{
Same as Fig.~\ref{fig:sred_50cr}, but for \al -induced
reactions on $^{40}$Ca. The experimental data have been taken from
\cite{Howard74,Nassar06}. 
Further discussion see text.
}
\label{fig:sred_40ca}
\end{figure}
Again similar to $^{42}$Ca, at very low energies close above the
\rap\ threshold, the \rap\ reaction is further suppressed by the Coulomb
barrier in the exit channel, and consequently the total reaction cross section
is significantly affected by the \rag\ cross section. However, average cross
sections for the $^{40}$Ca\rag $^{44}$Ti reaction are very rare in literature,
and the focus of recent \rag\ experiments was the astrophysically very
important resonance triplet around 4.5\,MeV and the properties of resonances
\cite{Simpson71,Dixon80,Nassar06,Vock07,Hof10,Rob12,Schmidt13}. The
thick-target data point of Nassar {\it et al.}\ \cite{Nassar06} leads to an
average cross section of $8.0 \pm 1.1$\,$\mu$b for the broad energy range from
about 2.1 to 4.2\,MeV in the center-of-mass system. Using the calculated
average energy dependence of the \rag\ cross section, the effective energy is
$E_{\rm{c.m.}} \approx 3.45$\,MeV. This data point is shown in
Figs.~\ref{fig:sig_40ca} and \ref{fig:sred_40ca}.

As pointed out above, the total reaction cross section \sreac\ and the reduced
cross section \sred\ are well-defined by the \rap\ cross section over a wide
energy range. The reduced cross section for $^{40}$Ca is shown in
Fig.~\ref{fig:sred_40ca}. At higher energies above \Ered\ $\approx 1$\,MeV,
the obtained \sred\ is close to the other nuclei in the $A \approx 20 - 50$
mass range. However, at lower energies below \Ered\ $\approx 1$\,MeV, the
reduced cross section \sred\ for $^{40}$Ca is slightly lower than for
neighboring nuclei. This reflects the doubly-magic nature of $^{40}$Ca.

\subsection{$^{41}$K}
\label{sec:k41}
Whereas the previously studied nuclei are located completely in the
$fp$-shell, $^{41}$K with $Z = 19$ and $N = 22$ enters the transition region
between the $sd$-shell and the $fp$-shell. Several experimental data sets are
available for the $^{41}$K\ran $^{44}$Sc reaction. Scott {\it et
  al.}\ \cite{Scott91} have measured the \ran\ cross section by neutron
counting and by activation in combination with $\gamma$-ray detection of the
1157\,keV $\gamma$-ray in the decay $^{44}$Sc $\rightarrow$ $^{44}$Ca. Both
experimental techniques provide consistent results in the energy range under
study by Scott {\it et al.}; a small correction of a few per cent was applied
to the activation data to take into account a long-living $J^\pi = 6^+$ isomer
with $T_{1/2} = 2.44$\,days. 
\begin{figure}[htb]
  \includegraphics{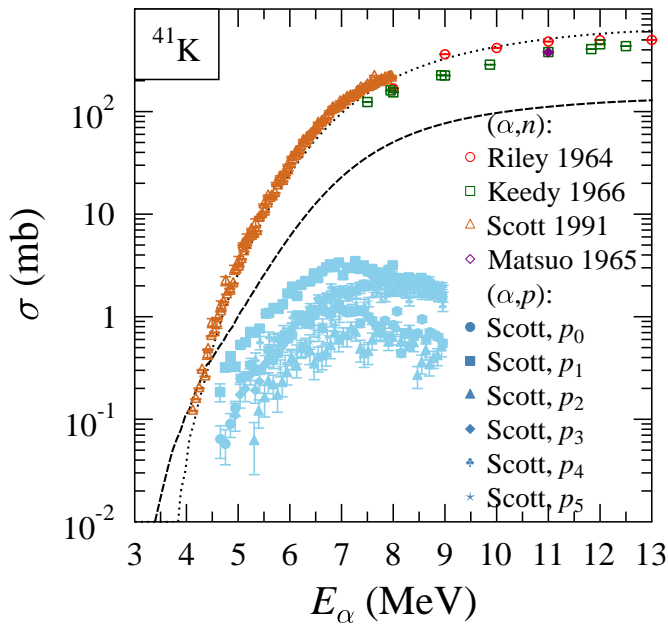}
\caption{
Cross sections of the $^{41}$K\ran $^{44}$Sc and $^{41}$K\rap $^{44}$Ca
reactions. 
The experimental data have been taken from
\cite{Riley64,Keedy66,Scott91,Matsuo65} and are the sum of the ground state
and isomer cross sections in the \ran\ channel.
Further discussion see text.
}
\label{fig:sig_41k}
\end{figure}

At higher energies yields for the ground state
and isomeric state have been measured separately. Keedy {\it et
  al.}\ \cite{Keedy66} used activation in combination with annihilation
spectroscopy whereas Riley {\it et al.}\ \cite{Riley64} and Matsuo {\it et
  al.}\ \cite{Matsuo65} used also activation,
but in combination with $\gamma$-ray spectroscopy. All experimental data sets
are in reasonable agreement within the experimental uncertainties, with about
$10-25$\,\% lower cross sections in \cite{Keedy66} and \cite{Matsuo65}. The
results are shown in Fig.~\ref{fig:sig_41k}; the ground state cross section
and the isomeric cross section have been added to provide the total
\ran\ cross section. The agreement between the experimental data and a StM
calculation is again excellent.

\begin{figure}[htb]
  \includegraphics{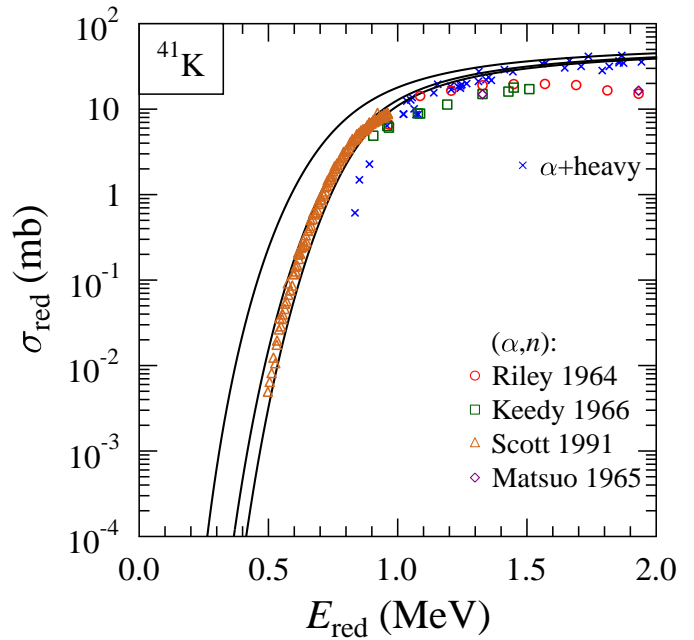}
\caption{
Same as Fig.~\ref{fig:sred_50cr}, but for \al -induced
reactions on $^{41}$K. The experimental data have been taken from
\cite{Riley64,Keedy66,Scott91,Matsuo65} and include the ground state and
isomer contributions of the $^{41}$K\ran $^{44}$Sc reaction. 
Further discussion see text.
}
\label{fig:sred_41k}
\end{figure}
Scott {\it et al.}\ \cite{Scott91} also provide cross sections for the
$^{41}$K\rap $^{44}$Ca reaction from an excitation function measurement at one
angle $\vartheta = 125^\circ$. Fig.~4 of \cite{Scott91} shows only the cross
sections of the lowest $p_0$, $p_1$, $p_2$, $p_3$, $p_4$, and $p_5$ proton
groups. However, the spectrum in Fig.~3 of \cite{Scott91} shows a dominating
peak for $p_{6-9}$, another strong peak for $p_{10-14}$, and further peaks for
$p_{15-35}$. The results for the low-lying proton groups are shown in
Fig.~\ref{fig:sig_41k}; as expected, they are significantly lower than the StM
calculation for the total \rap\ cross section. A point-by-point summing of the
$p_i$ channels is again hampered by the fact that the EXFOR data had to be
re-digitized from Fig.~4 of \cite{Scott91}.

The total reaction cross section \sreac\ and the reduced cross section
\sred\ are dominated by the $^{41}$K\ran $^{44}$Sc reaction for energies above
5\,MeV, i.e., over almost the entire energy range under study. The results for
\sred\ from the \ran\ channel are shown in Fig.~\ref{fig:sred_41k}. Only at
the lowest energies a significant contribution of the \rap\ channel is found.
Similar to most nuclei under study in this work, the \sred\ data for $^{41}$K
do not show a peculiar behavior.

\subsection{$^{40}$K}
\label{sec:k40}
The odd-odd ($Z = 19$, $N = 21$) nucleus $^{40}$K has a very low natural
abundance, and in addition it is unstable. The 1461\,keV $\gamma$-ray of the
$^{40}$K $\rightarrow$ $^{40}$Ar decay is a prominent background line in almost
any $\gamma$-ray spectrum. Therefore, only very few experimental data are
available for $^{40}$K. Elastic $^{40}$K\raa $^{40}$K scattering has been
measured by Oeschler {\it et al.}\ \cite{Oeschler72} at
24\,MeV. Unfortunately, the experimental data are only presented as a line in
Fig.~1 of \cite{Oeschler72} which had to be re-digitized for EXFOR. As this is
the only data set for $^{40}$K, a phase shift fit was made to the angular
distribution (as provided by EXFOR). The result of \sred\ $\ 67.5$\,mb at
\Ered\ $= 2.88$\,MeV is within the expected range. Because of the lack of
experimental data, no figure is shown for \sred\ of $^{40}$K. But from the
only available data set it can be concluded that there is at least no evidence
for a peculiar behavior of \sred\ for the odd-odd $^{40}$K.

\subsection{$^{39}$K}
\label{sec:k39}
\begin{figure}[htb]
  \includegraphics{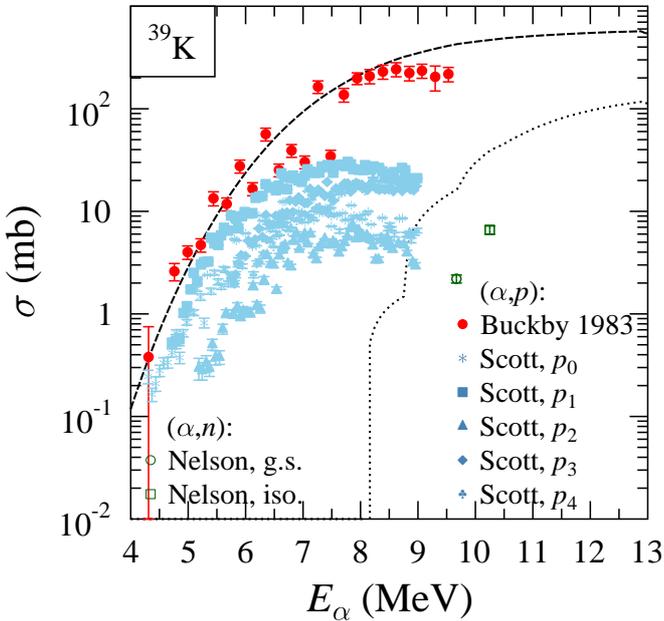}
\caption{
Cross sections of the $^{39}$K\ran $^{42}$Sc and $^{39}$K\rap $^{42}$Ca
reactions. 
The experimental data have been taken from
\cite{Buckby83,Scott87,Nelson65}; the \ran\ of \cite{Nelson65} data are
separated for the ground state and the isomer in $^{42}$Sc.
Further discussion see text.
}
\label{fig:sig_39k}
\end{figure}

Contrary to $^{41}$K with the dominating \ran\ cross section, the semi-magic
$(N=20$) $^{39}$K has a much larger \rap\ cross section. The $^{39}$K\ran
$^{42}$Sc reaction has a strongly negative $Q$-value ($Q = -7.33$\,MeV). Only
two data points are available by Nelson {\it et al.}\ \cite{Nelson65}, one for
the ground state and one for the $(7)^+$ isomer in $^{42}$Sc which decays to a
$6^+$ state in $^{42}$Ca. The two data points are shown in
Fig.~\ref{fig:sig_39k}; both experimental \ran\ data are far below the
theoretical expectation from the StM model. As the \ran\ cross section is by
far more than one order of magnitude below the \rap\ cross section, this
deviation does fortunately not affect the determination of the total
reaction cross section \sreac\ which is close to the \rap\ cross section.
\begin{figure}[htb]
  \includegraphics{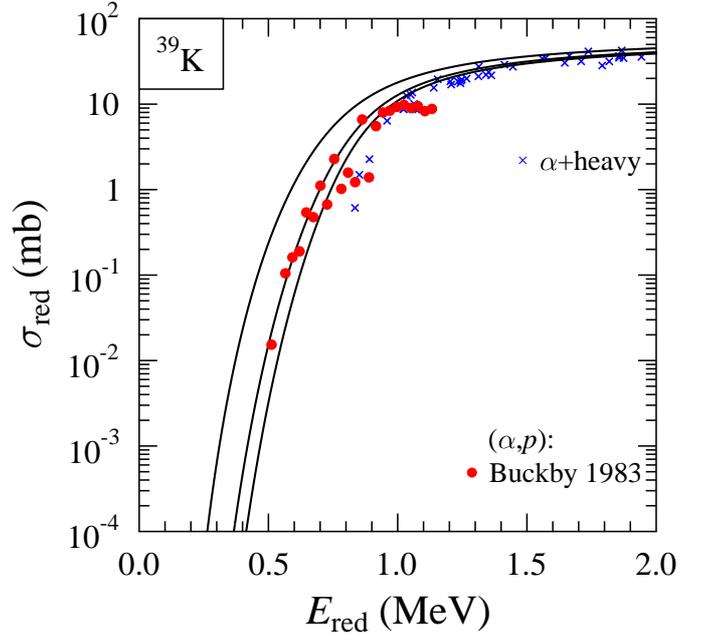}
\caption{
Same as Fig.~\ref{fig:sred_50cr}, but for \al -induced
reactions on $^{39}$K. The experimental data have been taken from
\cite{Buckby83}.
Further discussion see text.
}
\label{fig:sred_39k}
\end{figure}
The $^{39}$K\rap $^{42}$Ca reaction has been studied by Buckby {\it et
  al.}\ \cite{Buckby83} and Scott {\it et al.}\ \cite{Scott87}. Buckby {\it et
  al.}\ have measured five-point angular distributions (for a discussion of
the experimental procedure, see Sec.~\ref{sec:ca42}). Scott {\it et al.}\ 
have measured excitation functions at one angle $\vartheta = 125^\circ$, and
results are reported for the proton groups $p_0$, $p_1$, $p_2$, $p_3$, and
$p_4$. However, it can be seen from the spectrum in Fig.~1 of \cite{Scott87},
that significant contributions to the total \rap\ cross section come also from
the $p_{5-9}$ and $p_{10}$ proton groups, and it is stated in the text that
additional proton groups $p_{11-18}$, $p_{19-28}$, and $p_{29-32}$ have been
observed at higher energies. In addition, the EXFOR data of Scott {\it et
  al.}\ had to be re-digitized. Thus, it is impossible to determine the total
\rap\ cross section from these data. But it can be seen in
Fig.~\ref{fig:sig_39k} that the Scott {\it et al.}\ data for each proton group
are -- as expected -- below the total \rap\ cross section reported by Buckby
{\it et al.} in their Table 1 of \cite{Buckby83}.

For the determination of the total reaction cross section \sreac\ and the
reduced cross section \sred\ only the data by Buckby {\it et
  al.}\ \cite{Buckby83} are used. The results are shown in
Fig.~\ref{fig:sred_39k}. 
Similar to most nuclei under study in this work, the
\sred\ data for $^{39}$K do not show a peculiar behavior.

\subsection{$^{40}$Ar}
\label{sec:ar40}
\begin{figure}[htb]
  \includegraphics{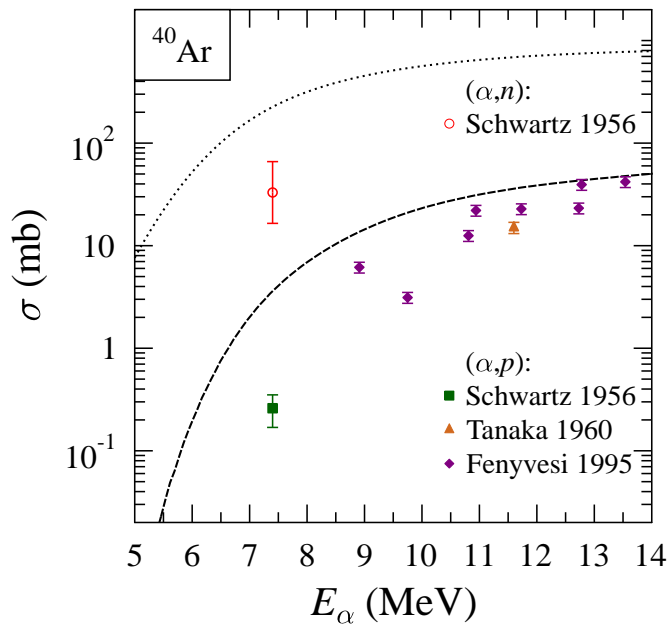}
\caption{
Cross sections of the $^{40}$Ar\ran $^{43}$Ca and $^{40}$Ar\rap $^{43}$K
reactions. 
The experimental data have been taken from
\cite{Schwartz56,Tanaka60,Feny95}.
The energy of the data points by Schwartz
{\it et al.}\ \cite{Schwartz56} at $E_\alpha = 7.4$\,MeV should be corrected
by about 500\,keV because of the energy loss of the beam in the entrance
window of the gas target and in the gas cell.
Further discussion see text.
}
\label{fig:sig_40ar}
\end{figure}
For the neutron-rich $^{40}$Ar it is obvious that the \ran\ cross section
dominates whereas the \rap\ cross section is much smaller. Unfortunately, only
few experimental data for $^{40}$Ar are available at EXFOR. The $^{40}$Ar\ran
$^{43}$Ca cross section was determined by Schwartz {\it et
  al.}\ \cite{Schwartz56} at $E_\alpha = 7.4$\,MeV using a argon-filled gas
target. The emitted particles were detected by nuclear track counting at one
angle ($\vartheta = 90^\circ$), and isotropy was assumed. The result of 33\,mb
has an uncertainty of a factor of two and is much lower than the prediction
from the StM (see Fig.~\ref{fig:sig_40ar}). It should be noted that the energy
loss of the beam in the entrance window of the gas target and in the gas cell
was not taken into account in \cite{Schwartz56}. This energy loss should be of
the order of 500\,keV for the entrance window. The target thickness in
\cite{Schwartz56} is given with 130\,keV leading to an effective energy which
is about 65\,keV lower. These corrections of more than 500\,keV bring the data
point closer to the StM prediction.

In addition to the Schwartz {\it et al.}\ \cite{Schwartz56} experiment at
7.4\,MeV, the $^{40}$Ar\rap $^{43}$K reaction was also measured at higher
energies by Tanaka {\it et al.}\ \cite{Tanaka60} and Fenyvesi {\it et
  al.}\ \cite{Feny95}. Both experiments used a stacked-target technique and
$4\pi$-$\beta$-counting in \cite{Tanaka60} and $\gamma$-spectroscopy in
\cite{Feny95} for the detection of the decay of the residual $^{43}$K
nucleus. The results of the experiments of \cite{Tanaka60,Feny95} are in
good agreement, and also the 7.4\,MeV data point of Schwartz {\it et
  al.}\ \cite{Schwartz56} seems roughly to follow the expected energy
dependence (in particular, if the energy of this data point is corrected by
about 500\,keV as discussed above). Similar to the $^{40}$Ar\ran $^{43}$Ca
cross section, also the $^{40}$Ar\rap $^{43}$K cross section is overestimated
by the StM calculation at low energies. For completeness it should be noted
that the StM calculations using either TALYS or NON-SMOKER are almost identical
for $^{40}$Ar.

\begin{figure}[htb]
  \includegraphics{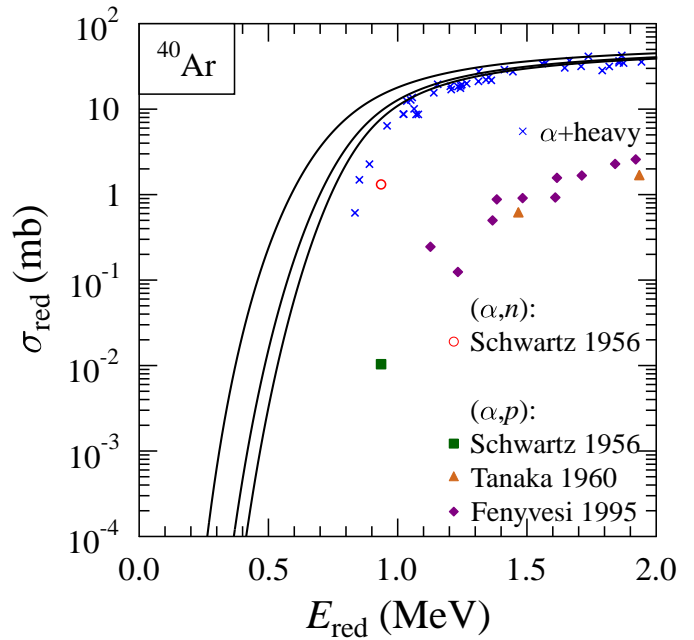}
\caption{
Same as Fig.~\ref{fig:sred_50cr}, but for \al -induced
reactions on $^{40}$Ar. The experimental data have been taken from
\cite{Schwartz56,Tanaka60,Feny95}. The energy of the data points by Schwartz
{\it et al.}\ \cite{Schwartz56} at $E_\alpha = 7.4$\,MeV should be corrected
by about 500\,keV because of the energy loss of the beam in the entrance
window of the gas target and in the gas cell. Only one data point is available
for the dominating $^{40}$Ar\ran $^{43}$Ca reaction. The $^{40}$Ar\rap $^{43}$K
reaction contributes only minor to the total reaction cross section \sreac .
Further discussion see text.
}
\label{fig:sred_40ar}
\end{figure}
The total reaction cross section \sreac\ and the reduced cross section
\sred\ can be derived from the $^{40}$Ar\ran $^{43}$Ca cross section. However,
there is only one data point with large uncertainties \cite{Schwartz56} which
may need a correction of the energy. The results for \sred\ are shown in
Fig.~\ref{fig:sred_40ar}. Based on the one data point by Schwartz {\it et
  al.}\ \cite{Schwartz56} with its large error bars, the reduced cross section
of $^{40}$Ar seems to be smaller than for most neighboring nuclei in the $A
\approx 20-50$ mass region. Improved data for $^{40}$Ar are highly desirable.

Because of the noticeable behavior of \sred\ of $^{40}$Ar at low energies,
\sred\ of $^{40}$Ar was additionally studied at higher energies using
$^{40}$Ar\raa $^{40}$Ar elastic scattering. Data at relatively low energies
are available in \cite{Bucurescu70,Seidlitz58}, and data above 20\,MeV have
also been measured in \cite{Gaul69,Oeschler72}. Only the 18\,MeV data by
Seidlitz {\it et al.}\ \cite{Seidlitz58} are available numerically from Table
I of \cite{Seidlitz58}; the other data had to be re-digitized from small figures
with logarithmic scale in \cite{Bucurescu70,Gaul69,Oeschler72}, and no error
bars are available in the EXFOR data. Therefore the following study is
restricted to the data by Seidlitz {\it et al.}\ \cite{Seidlitz58} and the
data at the lowest energies by Bucurescu {\it et al.}\ \cite{Bucurescu70}.
It turns out that the limited angular range of the angular distributions by
Bucurescu {\it et al.}\ \cite{Bucurescu70} is not sufficient to derive the
total reaction cross section \sreac\ from these data. The angular distribution
of Seidlitz {\it et al.}\ \cite{Seidlitz58} covers the full angular range;
however, at forward angles the measured cross section is almost twice the
Rutherford cross section of pointlike charges. Therefore, Gaul {\it et
  al.}\ \cite{Gaul69} have suggested to scale the angular distribution by
Seidlitz {\it et al.}\ by a factor of 0.55; a similar scaling factor of 0.57 is
suggested from the best fit obtained in this work. The total reaction cross
section is \sreac\ $= 1468$\, mb at $E_\alpha = 17.98$\,MeV, corresponding to
a reduced cross section \sred\ $= 58.5$\,mb at \Ered\ $= 2.27$\,MeV. This
value fits nicely into the general systematics of reduced cross sections at
higher energies around \Ered\ $\approx 2$\,MeV (see
Fig.~\ref{fig:sred_heavy}). Thus, the behavior of \sred\ for $^{40}$Ar is
extraordinary only at low energies.

\subsection{$^{36}$Ar}
\label{sec:ar36}
\begin{figure}[htb]
  \includegraphics{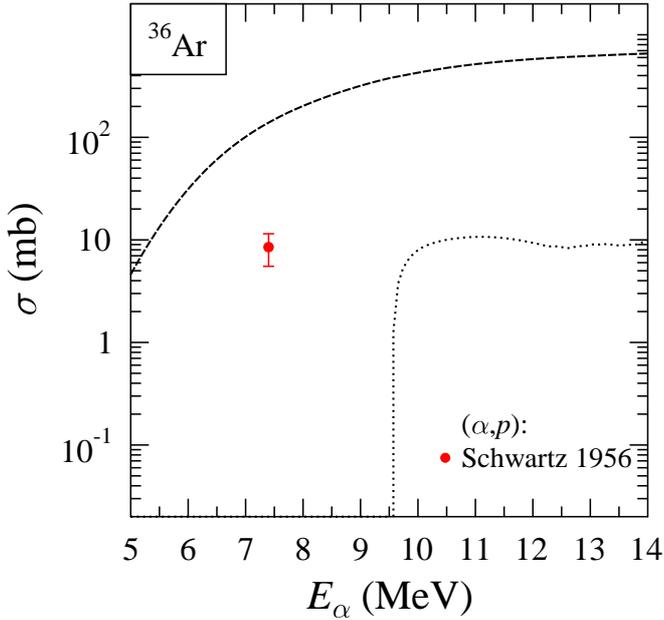}
\caption{
Cross sections of the $^{36}$Ar\ran $^{39}$Ca and $^{36}$Ar\rap $^{39}$K
reactions. 
The experimental data point has been taken from \cite{Schwartz56}.
The energy of the data point by Schwartz
{\it et al.}\ \cite{Schwartz56} at $E_\alpha = 7.4$\,MeV should be corrected
by about 500\,keV because of the energy loss of the beam in the entrance
window of the gas target and in the gas cell. 
Further discussion see text.
}
\label{fig:sig_36ar}
\end{figure}

Contrary to the neutron-rich nucleus $^{40}$Ar, the dominating channel for
$^{36}$Ar is the $^{36}$Ar\rap $^{39}$K reaction. The $^{36}$Ar\ran $^{39}$Ca
reaction has a strongly negative $Q$-value ($Q = -8.60$\,MeV) and thus cannot
contribute to the total cross section \sreac\ for $^{36}$Ar at low
energies. There is only one data point for the $^{36}$Ar\rap $^{39}$K reaction
at 7.4\,MeV by Schwartz {\it et al.}\ \cite{Schwartz56} with large
uncertainties. The energy of this data point should be corrected by about
500\,keV because of the energy loss of the beam in the entrance window of the
gas target and in the gas cell (see discussion in the previous
Sec.~\ref{sec:ar40}). Even with the correction of the energy, the experimental
data point of \cite{Schwartz56} is significantly below the StM calculation
(see Fig.~\ref{fig:sig_36ar}).

The determination of the total reaction cross section \sreac\ and reduced
cross section \sred\ for $^{36}$Ar is possible from the dominating
$^{36}$Ar\rap $^{39}$K cross section. However, this determination is obviously
hampered by the availability of experimental data, and
improved data for the $^{36}$Ar\rap $^{39}$K reaction are highly desirable.
\begin{figure}[htb]
  \includegraphics{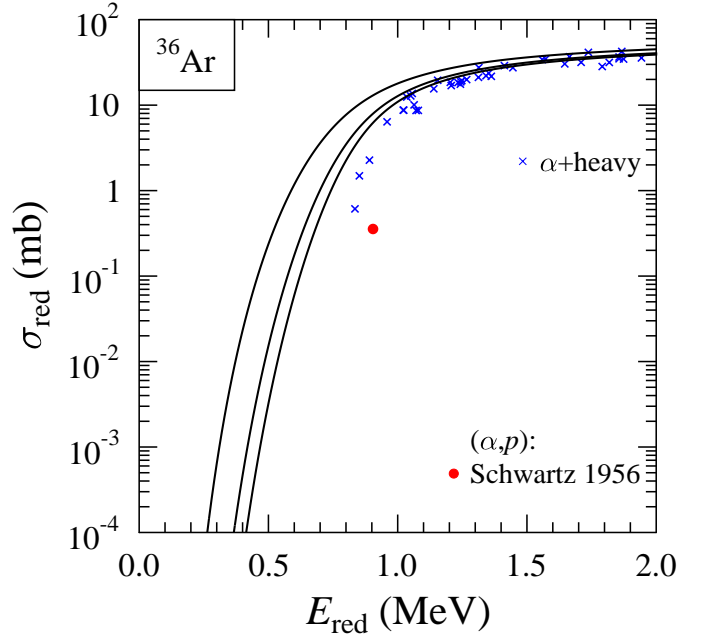}
\caption{
Same as Fig.~\ref{fig:sred_50cr}, but for \al -induced
reactions on $^{36}$Ar. The experimental data have been taken from
\cite{Schwartz56}. 
The energy of the data point by Schwartz
{\it et al.}\ \cite{Schwartz56} at $E_\alpha = 7.4$\,MeV should be corrected
by about 500\,keV because of the energy loss of the beam in the entrance
window of the gas target and in the gas cell. 
Further discussion see text.
}
\label{fig:sred_36ar}
\end{figure}

Similar to $^{40}$Ar, the noticeable behavior of \sred\ for $^{36}$Ar at low
energies requires further studies. Elastic scattering data are available by
Gaul {\it et al.}\ \cite{Gaul69}, Oeschler {\it et al.}\ \cite{Oeschler72},
and at higher energies by Kocher {\it et al.}\ \cite{Kocher92}. A reasonable
description of the 18\,MeV data by Gaul {\it et al.}\ \cite{Gaul69} was only
obtained in the later analysis by Kocher {\it et
  al.}\ \cite{Kocher92}. Therefore, the total reaction cross section
\sreac\ at 18\,MeV was obtained by repeating the optical model calculation in
\cite{Kocher92}. This leads to \sreac\ $= 1241$\,mb or \sred\ $= 51.9$\,mb at
\Ered\ $= 2.20$\,MeV. Also this value fits nicely into the general systematics
of reduced cross sections at higher energies around \Ered\ $\approx 2$\,MeV
(see Fig.~\ref{fig:sred_heavy}). Thus, also the behavior of \sred\ for
$^{36}$Ar is extraordinary only at low energies.

\subsection{$^{37}$Cl}
\label{sec:cl37}
Surprisingly, no experimental data are available in EXFOR for the
$^{37}$Cl\ran $^{40}$K and $^{37}$Cl\rap $^{40}$Ar reactions. This may be
related to the fact that both experiments cannot be done by activation because
$^{40}$K is quasi-stable with its half-life of more than 1 billion years and
$^{40}$Ar is stable. An excitation function for the $^{37}$Cl\rag $^{41}$K
reaction is available by Zyskind {\it et al.}\ \cite{Zyskind79}; therefore,
the presentation of results for $^{37}$Cl deviates from the usual restriction
of this work on \rap\ and \ran\ cross sections.
\begin{figure}[htb]
  \includegraphics{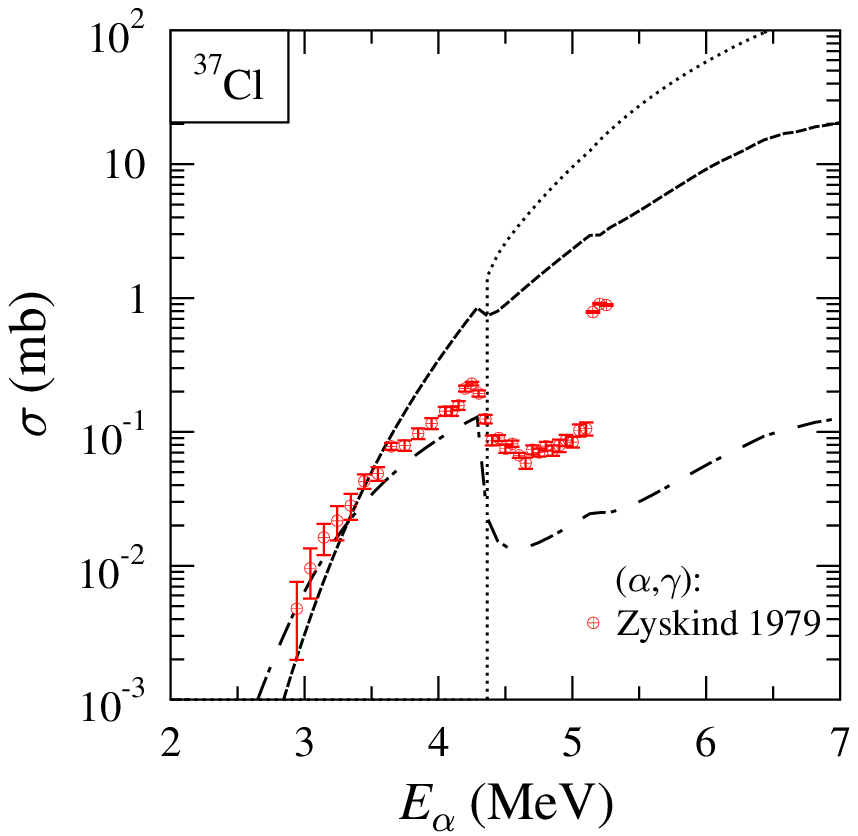}
\caption{
Cross sections of the $^{37}$Cl\ran $^{40}$K, $^{37}$Cl\rap $^{40}$Ar, and
$^{37}$Cl\rag $^{41}$K reactions. 
The experimental data for the $^{37}$Cl\rag $^{41}$K reaction have been taken
from \cite{Zyskind79}. No data are available in EXFOR for the $^{37}$Cl\ran
$^{40}$K and $^{37}$Cl\rap $^{40}$Ar reactions.
The additional dash-dotted line shows the
StM calculation for the \rag\ reaction.
Further discussion see text.
}
\label{fig:sig_37cl}
\end{figure}
\begin{figure}[h]
  \includegraphics{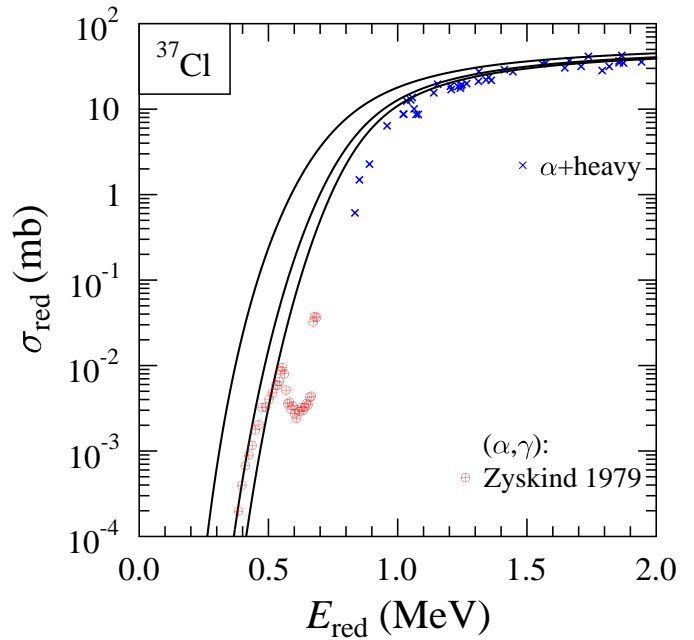}
\caption{
Same as Fig.~\ref{fig:sred_50cr}, but for \al -induced
reactions on $^{37}$Cl. The experimental data have been taken from
\cite{Zyskind79} for the $^{37}$Cl\rag $^{41}$K reaction. No data are
available at EXFOR for the $^{37}$Cl\ran $^{40}$K and $^{37}$Cl\rap $^{40}$Ar
reactions.
Further discussion see text.
}
\label{fig:sred_37cl}
\end{figure}
Zyskind {\it et al.}\ \cite{Zyskind79} used a Ge(Li) detector to measure
excitation functions for five strong $\gamma$-transitions at the angle of
$\vartheta = 55^\circ$, i.e.\ at a zero of the $P_2(\cos{\vartheta})$
Legendre polynomial. The total cross section of the $^{37}$Cl\rag $^{41}$K
reaction was derived from the sum of the five strong
transitions. Additionally, careful corrections were made for weak transitions
which were measured in special very long runs; these corrections were of the
order of about 25\,\%. The data are compared to a StM calculation in
Fig.~\ref{fig:sig_37cl}. At low energies below the \ran\ threshold, good
agreement is found. It has to be noted that the shown data by Zyskind {\it et
  al.}\ from the EXFOR database are taken from a table in the underlying
Ph.D.\ thesis; the three data points at the highest energies which deviate
from the expected energy dependence are not shown in
the paper \cite{Zyskind79}.

According to the StM calculations, at very low energies below about 3\,MeV the
\rag\ reaction is dominating. However, above 3\,MeV up to the \ran\ threshold,
the \rap\ cross section is comparable or even larger than the \rag\ cross
section, and above the \ran\ threshold the \ran\ reaction becomes
dominant. Due to the lack of other experimental data, the reduced cross
section \sred\ is taken from the $^{37}$Cl\rag $^{41}$K cross section (see
Fig.~\ref{fig:sred_37cl}). At the lowest energies the real \sred\ should be
only slightly larger than the shown data points from the
\rag\ reaction. Around \Ered\ $\approx 0.5$\,MeV one can see a weak kink in
the excitation function, indicating that a contribution of the \rap\ cross
section is missing here. At energies above \Ered\ $\approx 0.6$\,MeV there is
a strong cusp indicating the \ran\ threshold. Although the limited
availability of experimental data somewhat hampers the analysis for $^{37}$Cl,
it can nevertheless be stated that similar to most nuclei under study in this
work, the \sred\ data for $^{37}$Cl do not show a peculiar behavior.

\subsection{$^{35}$Cl}
\label{sec:cl35}
Three data sets are available for \al -induced reactions on $^{35}$Cl, but the
experimental data cover the $^{35}$Cl\ran $^{38}$K reaction only. Because of
the negative $Q$-value of the \ran\ reaction ($Q = -5.86$\,MeV), the
\ran\ data cannot restrict the total reaction cross section \sreac\ and the
reduced cross section \sred\ at low energies.

\begin{figure}[!b]
  \includegraphics{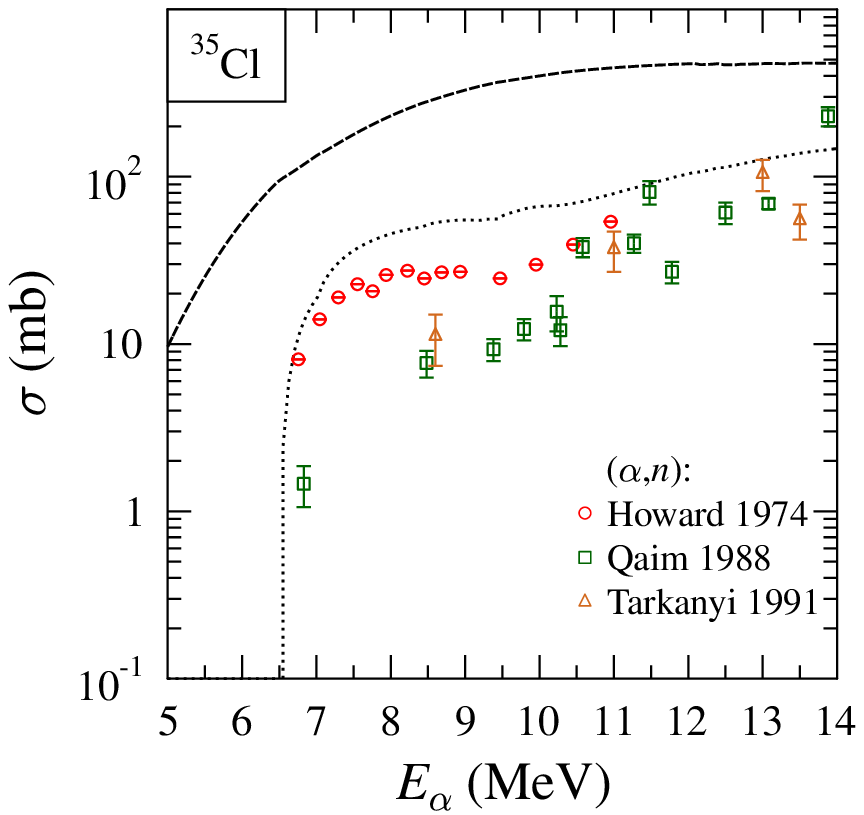}
\caption{
Cross sections of the $^{35}$Cl\ran $^{38}$K and $^{35}$Cl\rap $^{38}$Ar
reactions. 
The experimental data have been taken from \cite{Howard74,Qaim88,Tarkanyi91}.
Further discussion see text.
}
\label{fig:sig_35cl}
\end{figure}
Howard {\it et al.}\ \cite{Howard74} used the activation technique in
combination with annihilation spectroscopy to measure the $^{35}$Cl\ran
$^{38}$K cross section from about 7 to 11\,MeV. At higher energies the
stacked-foil activation technique has been used by Qaim {\it et
  al.}\ \cite{Qaim88} and by T{\'a}rk{\'a}nyi {\it et al.}\ \cite{Tarkanyi91};
the activity of the residual $^{38}$K nucleus was observed in both cases by
$\gamma$-spectroscopy. The agreement between the different data sets is not
very good; deviations are of the order of at least a factor of two (see
Fig.~\ref{fig:sig_35cl}).

As already stated above, a determination of the total reaction cross section
\sreac\ and reduced cross section \sred\ is not possible from the available
$^{35}$Cl\ran $^{38}$K data because the $^{35}$Cl\rap $^{38}$Ar reaction is
dominating. Taking into account the factor of about $5-7$ from the StM
calculation in Fig.~\ref{fig:sig_35cl}, the estimated reduced cross section
\sred\ is close to the expected values. The $^{35}$Cl\ran $^{38}$K data thus
do not show evidence for a peculiar behavior of \sred\ for $^{35}$Cl.
\begin{figure}[htb]
  \includegraphics{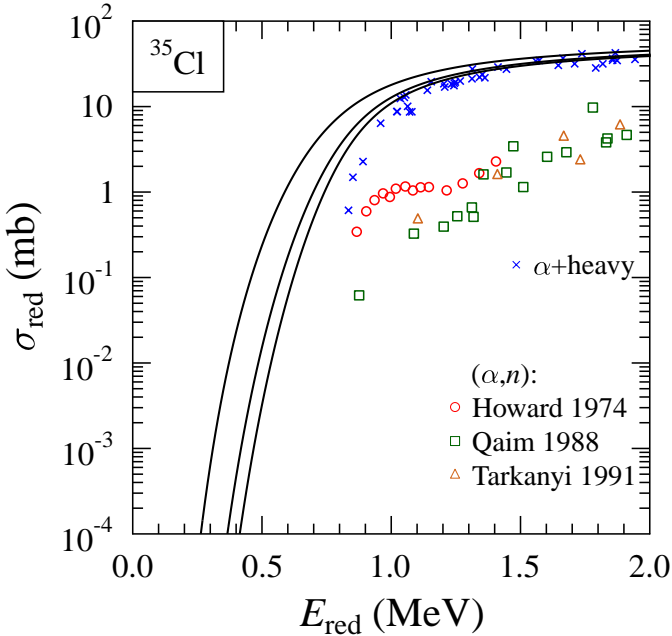}
\caption{
Same as Fig.~\ref{fig:sred_50cr}, but for \al -induced
reactions on $^{35}$Cl. The experimental data have been taken from
\cite{Howard74,Qaim88,Tarkanyi91} for the $^{35}$Cl\ran $^{38}$K reaction. The
$^{35}$Cl\ran $^{38}$K reaction is only a small ($\approx 10-15$\,\%)
contribution to the total reaction cross section \sreac\ which is dominated by
the $^{35}$Cl\rap $^{38}$Ar reaction. 
Further discussion see text.
}
\label{fig:sred_35cl}
\end{figure}

\subsection{$^{34}$S}
\label{sec:s34}
The $^{34}$S\ran $^{37}$Ar, $^{34}$S\rap $^{37}$Cl, and $^{34}$S\rag $^{38}$Ar
reactions have been measured simultaneously by Scott {\it
  et.}\ \cite{Scott93}. These data should allow to determine the total
reaction cross section \sreac\ and the reduced cross section \sred\ with small
uncertainties. 
\begin{figure}[htb]
  \includegraphics{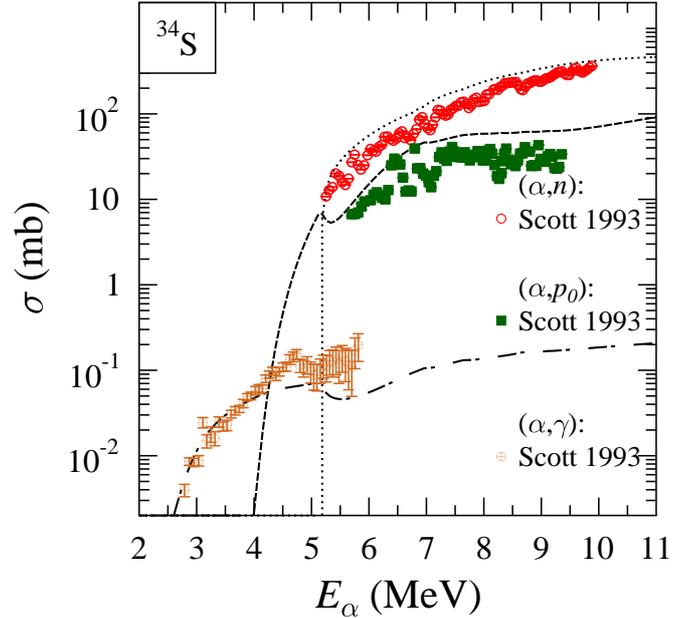}
\caption{
Cross sections of the $^{34}$S\ran $^{37}$Ar, $^{34}$S\rap $^{37}$Cl, and
$^{34}$S\rag $^{38}$Ar reactions.
The experimental data have been taken from \cite{Scott93}.
The additional dash-dotted line shows the
StM calculation for the \rag\ reaction.
Further discussion see text.
}
\label{fig:sig_34s}
\end{figure}
\begin{figure}[!b]
  \includegraphics{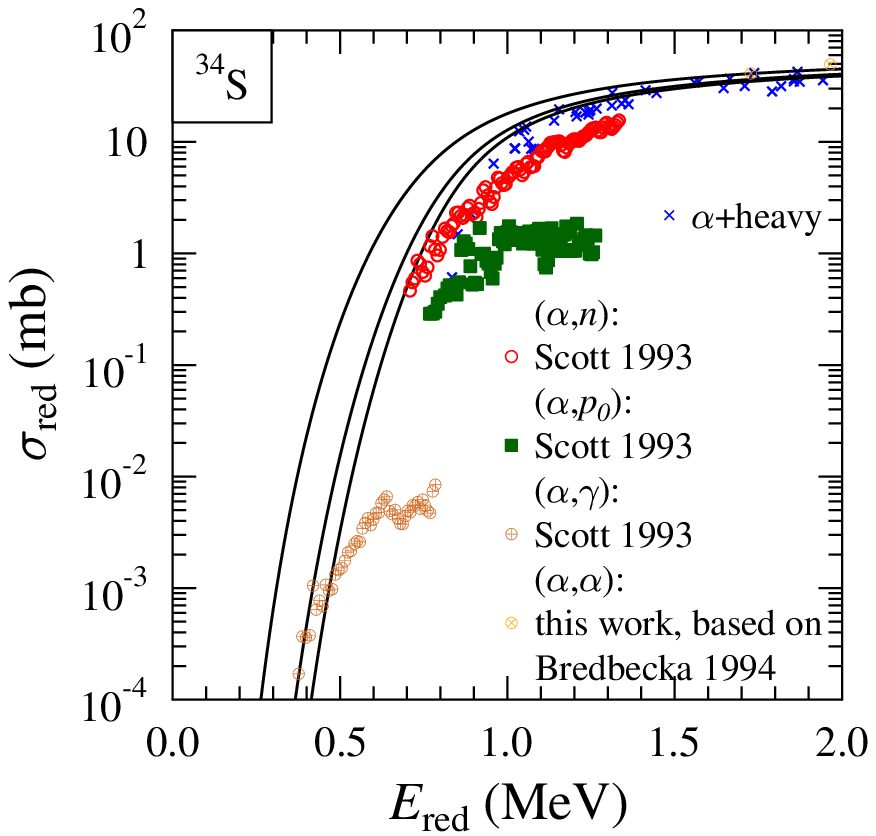}
\caption{
Same as Fig.~\ref{fig:sred_50cr}, but for \al -induced
reactions on $^{34}$S. The experimental data have been taken from
\cite{Scott93,Bredbecka94}. Above the \ran\ threshold, the total reaction
cross section is dominated by the $^{34}$S\ran $^{37}$Ar reaction. For
completeness, the $^{34}$S\rap $^{37}$Cl and $^{34}$S\rag $^{38}$Ar cross
sections and data points from $^{34}$S\raa $^{34}$S elastic scattering (see
also Figs.~\ref{fig:sred_heavy} and \ref{fig:s34scat}) are also shown. 
Further discussion see text.
}
\label{fig:sred_34s}
\end{figure}

The $^{34}$S\ran $^{37}$Ar reaction has been measured in \cite{Scott93} by
direct neutron counting from the \ran\ threshold ($Q = -4.63$\,MeV) up to
about 10\,MeV. Although minor problems with background from the $^{13}$C\ran
$^{16}$O reaction are reported in \cite{Scott93}, an overall uncertainty of
about 16\,\% is estimated in \cite{Scott93}. The data are shown in
Fig.~\ref{fig:sig_34s}. The comparison with the StM calculation shows that the
energy dependence is nicely reproduced; however, the absolute values of the
cross section are slightly overestimated by the StM. It is obvious that the
total reaction cross section \sreac\ is dominated by the \ran\ cross section
as soon as the energy is a few hundred keV above the threshold. It is
interesting to note that the scatter in the experimental data points is
probably related to the appearance of individual resonances. The target in the
experiment of Scott {\it et al.}\ \cite{Scott93} is not thick enough to
average over a sufficient number of resonances because of the relatively low
level density in the semi-magic ($N = 20$) $^{38}$Ar compound nucleus.

Below the \ran\ threshold, the $^{34}$S\rap $^{37}$Cl reaction dominates. The
total \rap\ cross section was derived from the excitation function of the
$^{34}$S($\alpha$,$p_0$)$^{37}$Cl$_{\rm{g.s.}}$ reaction which was measured at one
particular angle ($\vartheta = 125^\circ$). Corrections for the angular
distribution of the emitted protons and for proton groups $p_{i>0}$ were
estimated to be small in \cite{Scott93}. The shown data in
Fig.~\ref{fig:sig_34s} represent the $p_0$ channel only which contributes to
about 90\,\% to the total \rap\ cross section \cite{Scott93}. Similar to the
$^{34}$S\ran $^{37}$Ar cross section, also the $^{34}$S\rap $^{37}$Cl cross
section is slightly overestimated by the StM.

At energies below about 4\,MeV the $^{34}$S\rap $^{37}$Cl cross section
approaches its threshold ($Q = -3.03$\,MeV), and consequently the total
reaction cross section \sreac\ is essentially given by the $^{34}$S\rag
$^{38}$Ar reaction. The experimental data in \cite{Scott93} are restricted
to the analysis of the 2168\,keV $\gamma$-ray from the decay of the first
excited state in $^{38}$Ar to the ground state. Corrections for capture events
which bypass the first $2^+$ in $^{38}$Ar were
estimated to be of the order of about 20\,\%. Excellent agreement with the StM
calculation is found for the low-energy region below the \ran\ and
\rap\ thresholds.

The reduced cross section of $^{34}$S has been extracted from the available data
of Scott {\it et al.}\ \cite{Scott93}. Additional data points have been
obtained from the analysis of the $^{34}$S\raa $^{34}$S elastic scattering
data (see
also Figs.~\ref{fig:sred_heavy} and \ref{fig:s34scat}). Whereas the elastic
scattering data are in the expected range, the data of Scott {\it et
  al.}\ \cite{Scott93} are somewhat lower than expected. This holds in
particular at the highest energies of the Scott {\it et al.}\ experiment where
unobserved contributions of higher proton groups $p_{>0}$ in the $^{34}$S\rap
$^{37}$Ar reaction may be relevant.
Similar to most nuclei under study in this work, the
\sred\ data for $^{34}$S do not show a peculiar behavior.

\subsection{$^{33}$S}
\label{sec:s33}
\begin{figure}[htb]
  \includegraphics{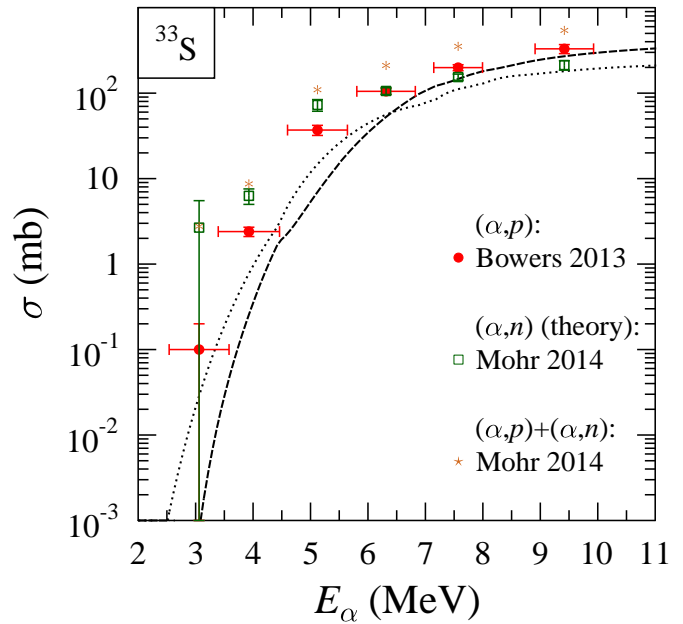}
\caption{
Cross section of the $^{33}$S\ran $^{36}$Ar and $^{33}$S\rap $^{36}$Cl reactions.
The experimental data for the \rap\ reaction have been taken from
\cite{Bowers13}; the \ran\ cross section was calculated from the experimental
\rap\ cross section and theoretical ratios between the \ran\ and \rap\ cross
section \cite{Mohr14}.
Further discussion see text.
}
\label{fig:sig_33s}
\end{figure}
Recently, the cross section of the $^{33}$S\rap $^{36}$Cl reaction was
measured by Bowers {\it et al.}\ \cite{Bowers13}. A $^4$He gas target was
irradiated with a $^{33}$S beam in inverse kinematics, and the residual
$^{36}$Cl nuclei were captured in an aluminum catcher foil. The number of
produced $^{36}$Cl nuclei was determined using accelerator mass
spectrometry. The result is shown in Fig.~\ref{fig:sig_33s}; here the cross
section is presented as a function of $E_\alpha$ (i.e., in forward
kinematics). The observed cross section shows a smooth energy dependence; no
individual resonances are visible. Compared to $^{34}$S in the previous
Sec.~\ref{sec:s34}, the level density in the odd-even compound nucleus
$^{37}$Ar is higher, and the target thickness is larger in the $^{33}$S
experiment. Thus, the experimental results are average cross sections,
averaged over a sufficient number of resonances within the energy spread of
the experimental data points, and the StM should provide excellent results.

The experimental data of \cite{Bowers13} are compared to a StM calculation in
Fig.~\ref{fig:sig_33s}. At the upper end of the energy range of
\cite{Bowers13} close to 10\,MeV, the StM calculation slightly underestimates
the experimental data. However, towards lower energies the discrepancy
increases significantly.

The $Q$-values of the \rap\ and \ran\ reactions on $^{33}$S are similar ($Q =
-1.93$\,MeV and $-2.00$\,MeV). Thus, at low energies very close above the
almost common threshold, the \ran\ cross section dominates. But at somewhat
higher energies around 6\,MeV the \rap\ cross section becomes dominant because
of the larger number of states in the exit channel towards the odd-odd nucleus
$^{36}$Cl. The $^{33}$S\ran $^{36}$Ar cross section was estimated in
\cite{Mohr14} from the experimental $^{33}$S\rap $^{36}$Cl cross section
\cite{Bowers13} and the theoretical ratio between the \ran\ and \rap\ cross
section (for details see \cite{Mohr14}). Finally, the total reaction cross
section \sreac\ is determined from the sum of \rap\ and \ran\ cross
sections. Because the ratio between the \ran\ and \rap\ cross sections is well
constrained by theory \cite{Mohr14}, the uncertainties for the \ran\ cross
section and the total reaction cross section \sreac\ remain acceptable.

\begin{figure}[htb]
  \includegraphics{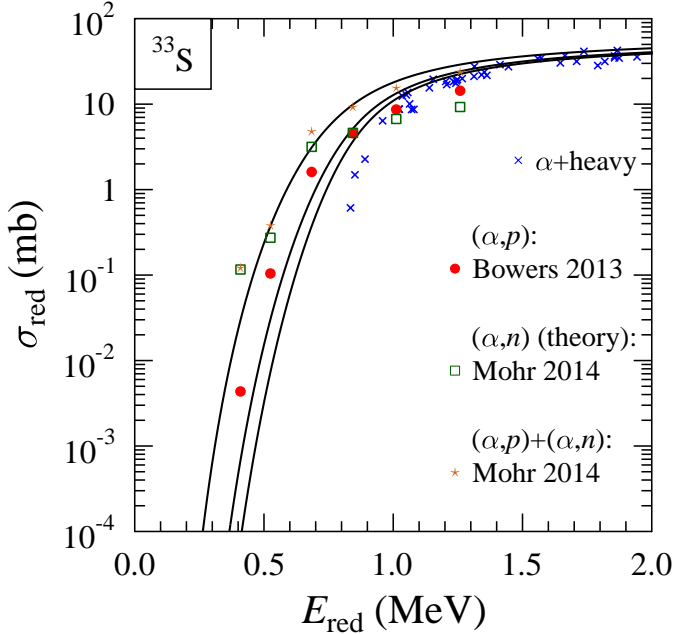}
\caption{
Same as Fig.~\ref{fig:sred_50cr}, but for \al -induced
reactions on $^{33}$S. The experimental data have been taken from
\cite{Bowers13} for the $^{33}$S\rap $^{36}$Cl reaction. In addition, the
$^{33}$S\ran $^{36}$Ar cross section has been estimated using a theoretical
branching ratio from the StM \cite{Mohr14}, and the total reaction cross
section is calculated from the sum of the \rap\ and \ran\ contributions.
Further discussion see text and \cite{Mohr14}.
}
\label{fig:sred_33s}
\end{figure}
The reduced cross section \sred\ was derived from the total reaction cross
section \sreac . The result is shown in Fig.~\ref{fig:sred_33s}. At the
highest energies the results for $^{33}$S lie within the typical range which
is indicated by the reduced cross sections for $^{21}$Ne and $^{51}$V in
Fig.~\ref{fig:sred_33s}. However, at lower energies the \sred\ values are
significantly above the typical range. This means that
the reduced cross section \sred\ of $^{33}$S behaves significantly
different compared to most other nuclei in the $A \approx 20 - 50$ mass range
under study, and also the energy dependence is unusually flat for $^{33}$S. A
detailed discussion of this unexpected behavior is given in \cite{Mohr14}.

\subsection{$^{32}$S}
\label{sec:s32}
\begin{figure}[b]
  \includegraphics{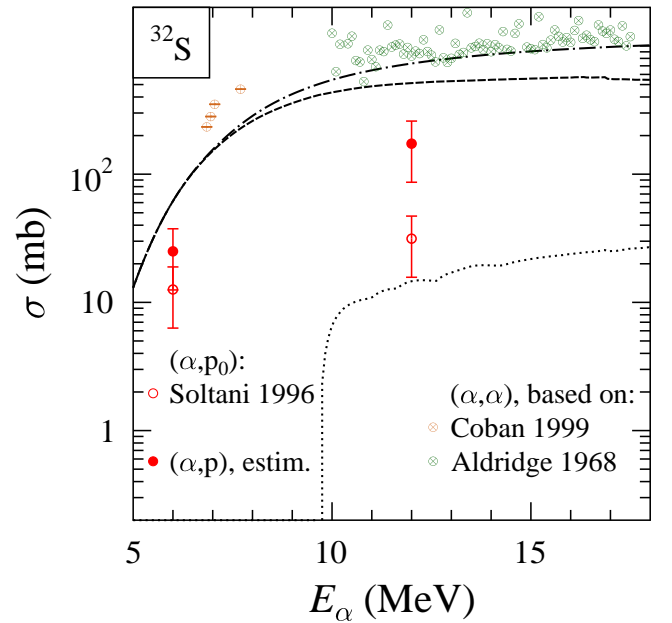}
\caption{
Cross section of the $^{32}$S\rap $^{35}$Cl reaction and total reaction cross
sections from $^{32}$S\raa $^{32}$S elastic scattering.
The experimental data have been taken from \cite{Solt96} for the
$^{32}$S($\alpha$,$p_0$)$^{35}$Cl reaction with the 
additional estimates for $^{32}$S\rap $^{35}$Cl as discussed in the text; the
upper data point should be considered as a lower limit only. Elastic
$^{32}$S\raa $^{32}$S data from \cite{Coban99,Aldridge68} have been
re-analyzed in this study to determine the total reaction cross section
\sred\ and should be compared to the corresponding calculation (dash-dotted
line) which is shown in addition to the \rap\ and \ran\ cross sections.
}
\label{fig:sig_32s}
\end{figure}
Individual resonances shape the energy dependence of \al -induced cross
sections for $^{32}$S at low energies. Because of the strongly negative
$Q$-value of the \ran\ reaction ($Q = -8.61$\,MeV), the total reaction cross
section \sreac\ is governed by the $^{32}$S\rap $^{35}$Cl reaction wich was
measured by Soltani-Farshi {\it et al.}\ \cite{Solt96}. The proton
groups $p_0$, $p_1$, $p_2$, $p_3$, $p_4$, $p_5$, and $p_6$ are nicely resolved
in the spectrum at $E_\alpha = 12$\,MeV (Fig.~2 of \cite{Solt96}), and
excitation functions have been measured from about 6 to 12\,MeV at six
angles. Unfortunately, only the $p_0$ cross section data are shown in
Fig.~3 of \cite{Solt96}. The EXFOR database provides this minor part by
re-digitization and states that it was impossible to obtain data tables from
the authors. 

Some estimates can nevertheless be made from available information of
Soltani-Farshi {\it et al.}\ \cite{Solt96}. The average differential cross
section at 12\,MeV is about $(d\sigma/d\Omega) \approx 2.5$\,mb/sr. This leads
to an angle-integrated cross section of 31.4\,mb with an uncertainty below a
factor of two. In the shown spectrum (Fig.~2 of \cite{Solt96}) the $p_0$ group
has a about the same intensity as the $p_3$, $p_4$, $p_5$, and $p_6$ groups;
the $p_1$ and $p_2$ groups are much weaker. This leads to a total \rap\ cross
section of about $5.5 \times 31.4\,{\rm{mb}} \approx 173$\,mb (assuming that
higher-lying proton groups do not contribute). In addition, at this energy the
\ran\ channel is already open; i.e., the value at 12\,MeV should be
considered as a lower limit. This value corresponds to a reduced cross
section \sred\ $> 7.6$\,mb at \Ered\ $= 1.59$\,MeV. At the lowest energy of
6\,MeV $(d\sigma/d\Omega) \approx 1$\,mb/sr, leading to $\sigma(p_0) \approx
12.6$\,mb. Higher-lying proton groups should be much weaker, leading to a
total \rap\ cross section of about a factor of two larger than the $p_0$ cross
section: $\sigma$\rap\ $\approx 25$\,mb. This corresponds to \sred\ $\approx
1.1$\,mb at \Ered\ $= 0.79$\,MeV. These roughly estimated data points are
shown in Fig.~\ref{fig:sig_32s}. The StM calculation is slightly above the
estimated experimental data. Although only roughly estimated, it can be
seen that the data are at least not far above the expectations.

Because of the few available reaction data for $^{32}$S, in addition 
$^{32}$S\raa $^{32}$S elastic scattering data were analyzed. Low-energy
angular distributions are available from Coban {\it et al.}\ \cite{Coban99}
and Aldridge {\it et al.}\ \cite{Aldridge68}. Unfortunately, both data had to
be re-digitized from figures in \cite{Coban99,Aldridge68}, and the EXFOR data
are provided without the original error bars. Because the angular
distributions have to be fitted for the determination of the total reaction
cross section \sreac , the resulting numbers have significant uncertainties
from the digitizing error and from the missing original uncertainties (for the
fitting procedure a fixed uncertainty of 5\,\% was used for all data points).

\begin{figure}[htb]
  \includegraphics{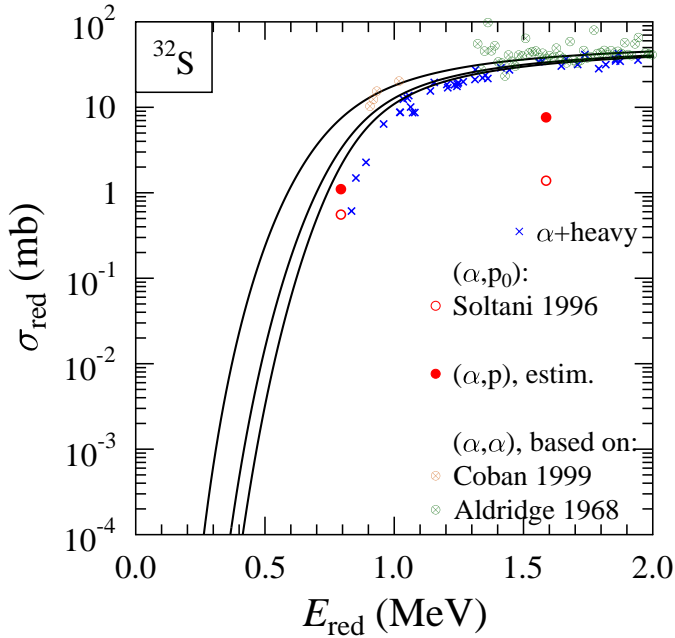}
\caption{
Same as Fig.~\ref{fig:sred_50cr}, but for \al -induced
reactions on $^{32}$S. 
The experimental data have been taken from \cite{Solt96,Coban99,Aldridge68}.
Further discussion see text.
}
\label{fig:sred_32s}
\end{figure}
Resonant structures in the excitation function of the $^{32}$S\raa $^{32}$S
elastic scattering were analyzed in the study of Coban {\it et
  al.}\ \cite{Coban99}. At resonance energies detailed angular distribution
measurements were carried out for a $J^\pi$ assignment. Therefore, the total
cross sections \sreac\ from these detailed angular distributions around 7\,MeV
in \cite{Coban99} (corresponding to reduced energies \Ered\ $\approx 0.9 -
1$\,MeV, shown as data points in Fig.~\ref{fig:sred_32s}) should be considered
as resonance-based upper limits. At the same energies the resonance properties
were confirmed by an enhanced yield in the ($\alpha$,$p_0$) cross section
which was observed in \cite{Coban99}.

Angular distributions in 100\,keV steps were measured by Aldridge {\it et
  al.}\ \cite{Aldridge68} at about 30\,angles. A phase shift fit of these
angular distributions is at the limits of numerical stability because the
number of adjustable parameters (two parameters per partial wave: reflexion
coefficient $\eta_L$ and phase shift $\delta_L$) is not much lower than the
number of experimental data points. Nevertheless, the general trend of the
reduced cross sections \sred\ is close to the expected values, and at least in
the overlap region with \cite{Solt96} the outliers are correlated with maxima
in the excitation 
function of the ($\alpha$,$p_0$) cross section of \cite{Solt96}. The resulting
\sred\ values for $^{32}$S are shown in Fig.~\ref{fig:sred_32s}.

Concluding the analysis of $^{32}$S, it can be said that the
\sred\ data for $^{32}$S do not show evidence for unusual behavior. Thus,
for the even-even sulfur isotopes $^{32,34}$S the \sred\ values behave
regularly or even low, whereas \sred\ for $^{33}$S is significantly enhanced
in particular at low energies.

\subsection{$^{31}$P}
\label{sec:p31}
The present study focuses on low-energy cross sections of \al -induced
energies around \Ered\ $\approx 0.5 - 1$\,MeV. The corresponding energies
$E_\alpha$ decrease towards lighter nuclei. Simultaneously, the level density
decreases towards lighter nuclei. Whereas for heavier nuclei a smooth energy
depencende of the \al -induced reaction cross sections was found, for lighter
nuclei individual resonances become more and more important. Calculations in
the StM model cannot reproduce these individual resonances; nevertheless, the
general trend of the data should be reproduced.
\begin{figure}[bht]
  \includegraphics{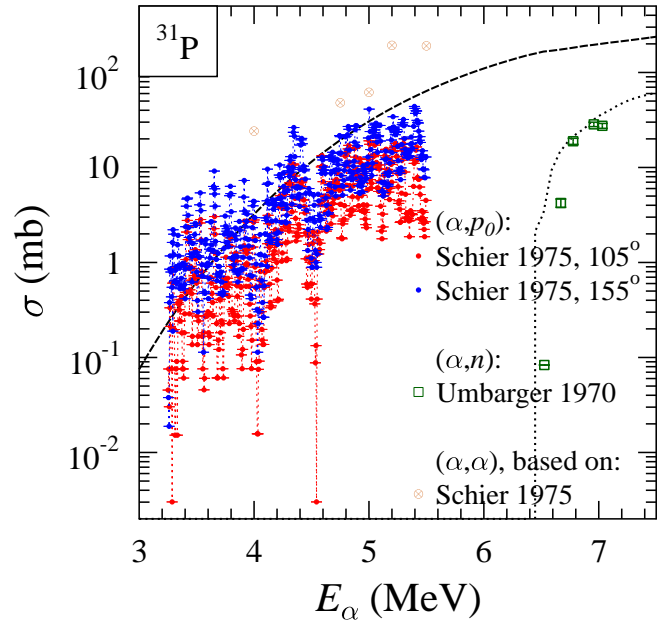}
\caption{
Cross section of the $^{31}$P\ran $^{34}$Cl and $^{31}$P\rap $^{34}$S reactions. 
The experimental data have been taken from the \ran\ data of \cite{Umbarger70}
and from the \rapo\ and \raa\ data of \cite{Schier75}. 
Further discussion see text.
}
\label{fig:sig_31p}
\end{figure}
\begin{figure}[htb]
  \includegraphics{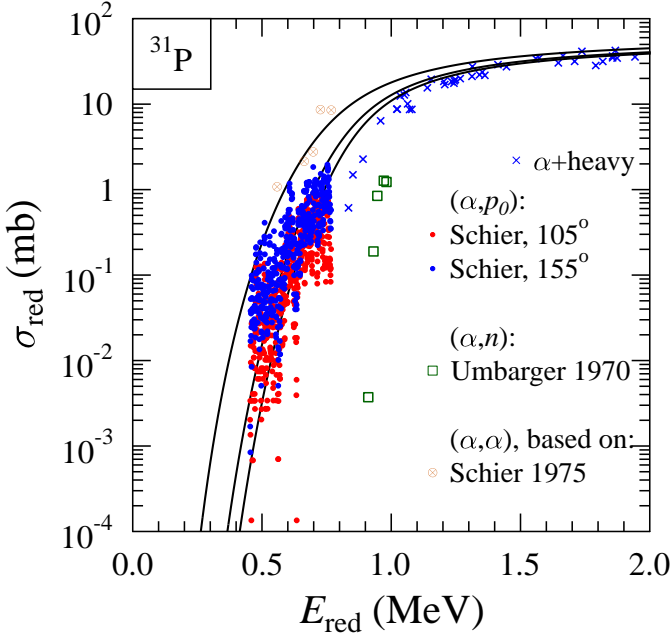}
\caption{
Same as Fig.~\ref{fig:sred_50cr}, but for \al -induced
reactions on $^{31}$P. The experimental data have been taken from
\cite{Schier75,Umbarger70}: the differential cross sections $(d\sigma/d\Omega)$
at $\vartheta = 105^\circ$ and $155^\circ$ of the dominating
$^{31}$P($\alpha$,$p_0$)$^{34}$S reaction has been converted to the total
cross section assuming isotropy. 
Further discussion see text.
}
\label{fig:sred_31p}
\end{figure}

Experimental data for the $^{31}$P\rapo $^{34}$S reaction have been measured
by Schier {\it el al.}\ \cite{Schier75}. Two excitation functions at
$\vartheta = 105^\circ$ and $155^\circ$ are shown in their Fig.~2. As no
spectrum is shown in \cite{Schier75}, it is not possible to determine the
contributions of higher-lying proton groups from this experiment. However, at
low energies close above the threshold the $p_0$ channel should be
dominating. The experimental data are shown in Fig.~\ref{fig:sig_31p} together
with a StM calculation. As expected, at higher energies the \rapo\ data of
\cite{Schier75} are below the theoretical estimate whereas at low energies
there is -- on average -- reasonable agreement between theory and
experiment. At lower energies resonance parameters of the $^{31}$P\rap
$^{34}$S reaction were determined by McMurray {\it et
  al.}\ \cite{McMurray71}.

Because of the relatively high \ran\ threshold ($Q = -5.65$\,MeV), the
$^{31}$P\ran $^{34}$Cl reaction does practically not contribute to the total
reaction cross section \sreac\ and the reduced cross section \sred . 
The \ran\ data by Umbarger {\it et al.}\ \cite{Umbarger70} are also shown in
Figs.~\ref{fig:sig_31p} and \ref{fig:sred_31p}. 
In addition, five data points from a re-analysis of
the elastic scattering data of Schier {\it et al.}\ \cite{Schier75} are shown
which have been measured simultaneously with the \rapo\ cross
section. However, this re-analysis is hampered by the limited number of data
points in the angular distributions which had to be re-digitized from Fig.~1
of \cite{Schier75}. As four of the five angular distributions in Fig.~1 of
\cite{Schier75} are measured in resonances (see the corresponding yield maxima
in the \rapo\ excitation functions in Fig.~2 of \cite{Schier75}), the
resulting \sreac\ and \sred\ should again be considered as resonance-based
upper limits. The off-resonance point at $E_\alpha = 4.75$\,MeV corresponds to
\Ered\ $= 0.70$\,MeV and \sred\ $= 2.77$\,mb. Although the eye may be mislead
by the many resonant data points of the \rapo\ cross section in
Fig.~\ref{fig:sred_31p}, the \sred\ data for $^{31}$P behave on average
similar to most nuclei under study in the $A \approx 20 - 50$ mass region.

\subsection{$^{30}$Si}
\label{sec:si30}
\begin{figure}[tbh]
  \includegraphics{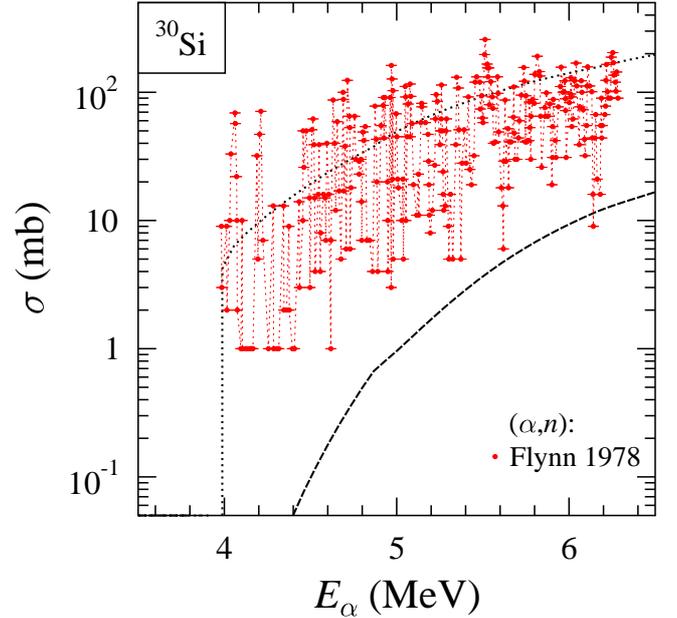}
\caption{
Cross section of the $^{30}$Si\ran $^{33}$S and $^{30}$Si\rap $^{33}$P reactions. 
The experimental data have been taken from \cite{Flynn78}.
Further discussion see text.
}
\label{fig:sig_30si}
\end{figure}
The $^{30}$Si\ran $^{33}$S cross section was measured by Flynn {\it et
  al.}\ \cite{Flynn78} by direct neutron counting. Because a very thin target
was used in this experiment, the measured cross section is governed by many
resonances. It has been shown already in \cite{Flynn78} that the average cross
section is well reproduced by a StM calculation, and a similar result is
obtained in this study. The experimental data of \cite{Flynn78} and the
present StM calculation are shown in Fig.~\ref{fig:sig_30si}.
\begin{figure}[b]
  \includegraphics{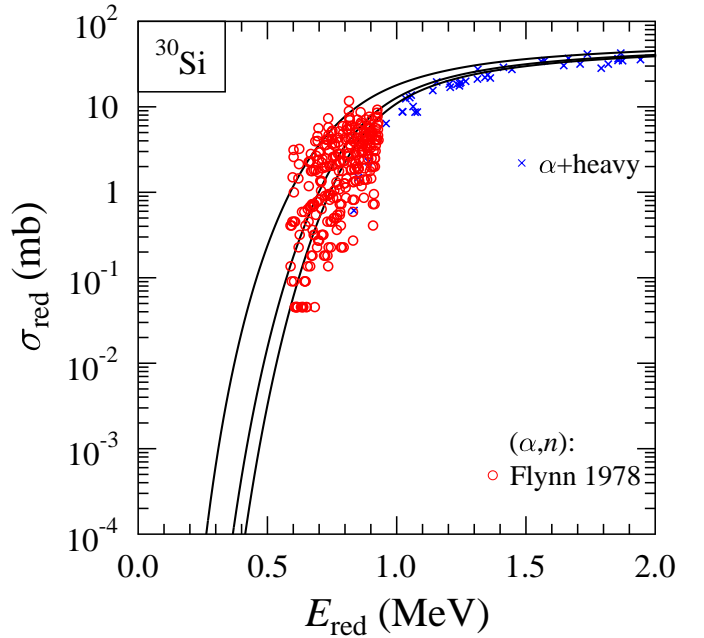}
\caption{
Same as Fig.~\ref{fig:sred_50cr}, but for \al -induced
reactions on $^{30}$Si. The experimental data have been taken from
\cite{Flynn78}.
Further discussion see text.
}
\label{fig:sred_30si}
\end{figure}
It is obvious from Fig.~\ref{fig:sig_30si} that the $^{30}$Si\ran $^{33}$S
reaction dominates the total reaction cross section \sreac\ and the reduced
cross section \sred . The $^{30}$Si\rap $^{33}$P reaction has a cross section
which is about one order of magnitude smaller over the entire measured energy
range of the experiment in \cite{Flynn78}. Hence, the reduced cross section
\sred\ of $^{30}$Si is well-defined by the experimental data of
\cite{Flynn78}. 

Similar to most nuclei under study in this work, the
\sred\ data for $^{30}$Si -- on average -- do not show a peculiar
behavior. This is also confirmed by the analysis of $^{30}$Si\raa $^{30}$Si
elastic scattering at 15.7\,MeV \cite{Wuehr74}. The analysis of the EXFOR data
leads to a relatively high \sred\ $= 68.7$\,mb at \Ered =
$2.32$\,MeV. However, the $\chi^2/F$ of the phase shift fit can be reduced by
about a factor of three if the experimental data of \cite{Wuehr74} are scaled
by a factor of $\approx 0.6$, leading to a lower value of \sred\ $= 61.2$\,mb.

\subsection{$^{29}$Si}
\label{sec:si29}
%
%
%
\begin{figure}[b]
  \includegraphics{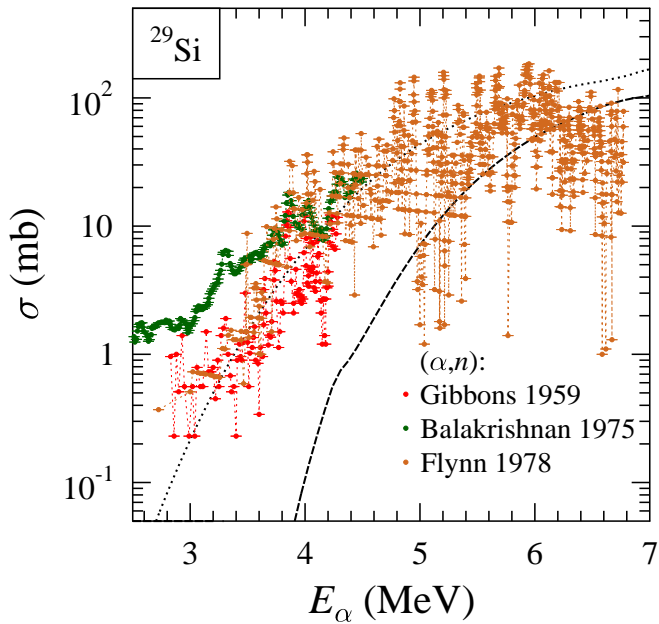}
\caption{
Cross section of the $^{29}$Si\ran $^{32}$S and $^{29}$Si\rap $^{32}$P reactions. 
The experimental data have been taken from \cite{Gibbons59,Bala75,Flynn78}.
Further discussion see text.
}
\label{fig:sig_29si}
\end{figure}
Three experimental data sets are available at EXFOR for the $^{29}$Si\ran
$^{32}$S reaction. The main focus of these experiments was the determination
of resonance properties from the measured neutron yield, and thus relatively
thin targets were used. Gibbons and Macklin \cite{Gibbons59} provide data from
about 2 to 4.5\,MeV, obtained with a 43\,$\mu$g/cm$^2$ target. Balakrishnan
{\it et al.}\ \cite{Bala75} identify 134 resonances for $E_\alpha = 2.15 -
5.25$\,MeV using a very thin target ($\Delta E \approx 5$\,keV, corresponding
to less than 5\,$\mu$g/cm$^2$), and Flynn {\it et al.}\ \cite{Flynn78} show
data for $E_\alpha \approx 2.75 - 7$\,MeV using a 9\,$\mu$g/cm$^2$ target. In
addition, for the measurements close above the threshold, a thicker target
with 113\,$\mu$g/cm$^2$ was used in \cite{Flynn78}. At energies around 4\,MeV
the newer data by Flynn {\it et al.}\ and Balakrishnan {\it et al.}\ are in
good agreement. However, at lower energies the data by Balakrishnan {\it et
  al.}\ are much higher. Flynn {\it et al.}\ \cite{Flynn78} state in
their discussion that the data by Balakrishnan {\it et al.}\ show structures
which are also visible in the $^{13}$C\ran $^{16}$O reaction, and thus the
Balakrishnan {\it et al.}\ data are contaminated by background. The early data
by Gibbons {\it et al.}\ \cite{Gibbons59} are about a factor of two below the
later Flynn {\it et al.}\ data around 4\,MeV where the cross section of the
$^{13}$C\ran $^{16}$O reaction is small. At lower energies the agreement
between the Flynn {\it et al.}\ data and the Gibbons and Macklin data may be
considered as accidental because also the Gibbons and Macklin experiment seems
to suffer from $^{13}$C background at lower energies (as discussed in
\cite{Flynn78}). In a further experiment McMurray {\it et
  al.}\ \cite{McMurray71} have determined resonance properties of the
$^{29}$Si\ran $^{32}$S reaction.

\begin{figure}[htb]
  \includegraphics{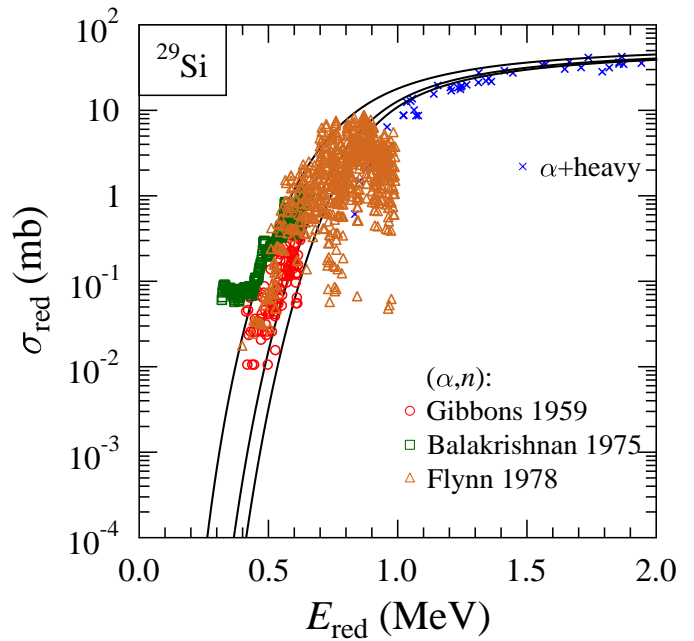}
\caption{
Same as Fig.~\ref{fig:sred_50cr}, but for \al -induced
reactions on $^{29}$Si. The experimental data have been taken from
\cite{Gibbons59,Bala75,Flynn78}.
Further discussion see text.
}
\label{fig:sred_29si}
\end{figure}
The experimental data of \cite{Gibbons59,Bala75,Flynn78} are shown in
Fig.~\ref{fig:sig_29si}. The data are -- on average -- in reasonable
agreement with the StM calculation. According to the StM calculation, the
cross section of the $^{29}$Si\rap $^{32}$P reaction is much lower in the
entire energy range under study. Unfortunately, no data for the $^{29}$Si\rap
$^{32}$P reaction are available at EXFOR.

The $^{29}$Si\ran $^{32}$S data are shown as reduced cross sections \sred\ in
Fig.~\ref{fig:sred_29si}. 
An additional data point can be taken from the analysis of $^{29}$Si\raa
$^{29}$Si elastic scattering. The fit to the angular distribution at $E_\alpha
= 26.6$\,MeV in \cite{Siemaszko94} leads to \sred\ $\approx 55$\,mb at the
relatively high reduced energy \Ered\ $\approx 3.89$\,MeV.
Similar to most nuclei under study in this work, the
\sred\ data for $^{29}$Si do not show a peculiar behavior.

\subsection{$^{28}$Si}
\label{sec:si28}
Data for the $^{28}$Si\rap $^{31}$P and $^{28}$Si\ran $^{31}$S reactions are
available at EXFOR. Because of the relatively high negative $Q$-value of the
\ran\ reaction ($Q = -9.10$\,MeV), at low energies the \rap\ reaction is
dominating. At very low energies also the \rap\ channel is closed, and the
only open reaction channel is $^{28}$Si\rag $^{32}$S. However, only resonance
strengths are available for the \rag\ reaction
\cite{Vernotte67,Toevs71,Rogers77,Babilon02}.
\begin{figure}[htb]
  \includegraphics{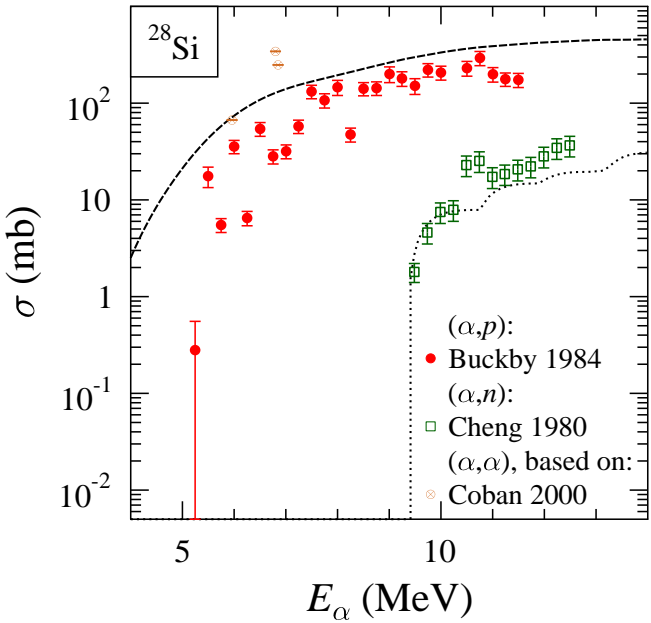}
\caption{
Cross section of the $^{28}$Si\ran $^{31}$S and $^{28}$Si\rap $^{31}$P
reactions. Three additional data points have been derived from a phase shift
analysis of $^{28}$Si\raa $^{28}$Si elastic scattering. 
The experimental data have been taken from \cite{Buckby84,Cheng80,Coban00}.
Further discussion see text.
}
\label{fig:sig_28si}
\end{figure}

The $^{28}$Si\rap $^{31}$P reaction was measured by Buckby {\it et
  al.}\ \cite{Buckby84}. The total \rap\ cross section was derived from
proton angular distributions (further discussion see also Sec.~\ref{sec:ca42},
\ref{sec:k39}, and \cite{Buckby83}). The results are shown in
Fig.~\ref{fig:sig_28si}. The StM calculation slightly overestimates the 
experimental data, in particular at the lowest energies. 

Contrary to the finding for the \rap\ reaction, the $^{28}$Si\ran $^{31}$S
data by Cheng {\it et al.}\ \cite{Cheng80} are nicely reproduced by the
StM. Here the \ran\ cross section was determined by activation in combination
with annihilation spectroscopy.

Because of the deviation between the experimental \rap\ data and the StM model
calculation, additionally $^{28}$Si\raa $^{28}$Si elastic scattering data are
studied. The angular distributions measured by Coban {\it et
  al.}\ \cite{Coban00} around $E_\alpha \approx 6 - 7$\,MeV can be nicely
fitted by a phase shift analysis. The result at 5.96\,MeV is in almost perfect
agreement with the StM calculation; at this energy the excitation functions in
\cite{Coban00} do not show strong resonances. At the higher energies (6.80 and
6.85\,MeV) a broad resonance (perhaps a dublett of resonances) can be seen
in the excitation functions; the existence of this resonance is confirmed by
the excitation function at backward angles measured by K\"allman {\it et
  al.}\ \cite{Kallman03}. Therefore it is not surprising that the total
reaction cross sections \sreac\ from elastic scattering at resonant energies
are significantly above the StM prediction.
\begin{figure}[htb]
  \includegraphics{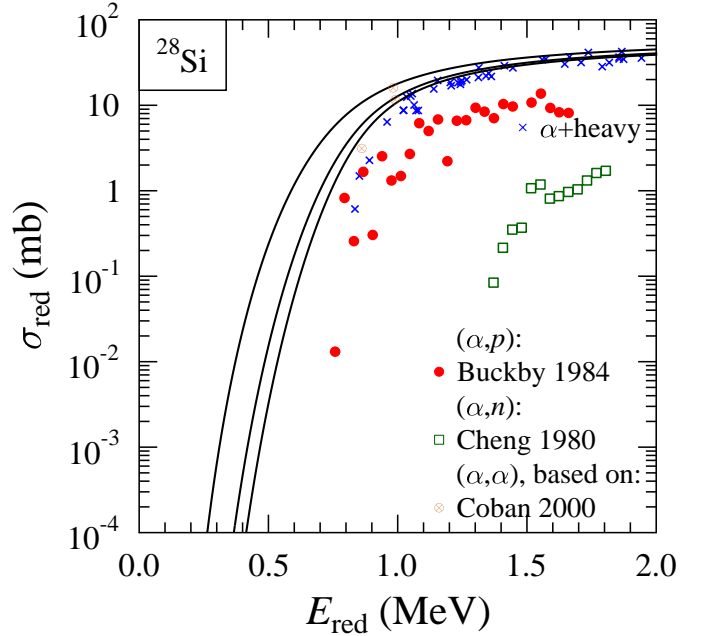}
\caption{
Same as Fig.~\ref{fig:sred_50cr}, but for \al -induced
reactions on $^{28}$Si. The experimental data have been taken from
\cite{Buckby84,Cheng80,Coban00}.
Further discussion see text.
}
\label{fig:sred_28si}
\end{figure}

Around 6\,MeV a discrepancy of about a factor of two is found between the
\rap\ cross section by Buckby {\it et al.}\ \cite{Buckby84} and the total
reaction cross section \sreac\ from the analysis of elastic scattering data by
Coban {\it et al.}\ \cite{Coban00}. Unfortunately, there is no simple
explanation for this discrepancy.

The available \al -induced cross sections are converted to reduced cross
sections. Fig.~\ref{fig:sred_28si} shows that the \sred\ values for $^{28}$Si
are smaller than for neighboring nuclei. Obviously, this effect is more
pronounced when one considers the lower \rap\ cross sections of Buckby {\it et
  al.}\ \cite{Buckby84} compared to the higher results from the analysis of
elastic scattering data by Coban {\it et al.}\ \cite{Coban00}.

\subsection{$^{27}$Al}
\label{sec:al27}
\begin{figure}[htb]
  \includegraphics{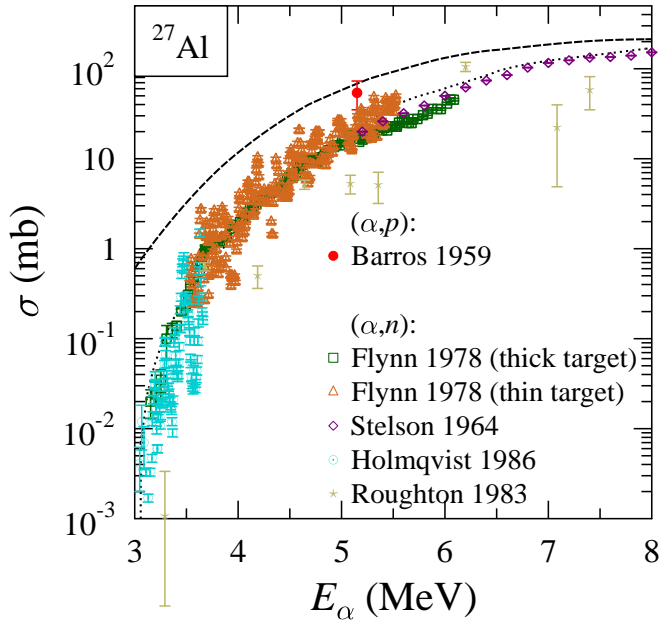}
\caption{
Cross section of the $^{27}$Al\ran $^{30}$P and $^{27}$Al\rap $^{30}$Si
reactions. 
The experimental data have been taken from
\cite{Barros59,Flynn78,Stelson64,Holmqvist86,Roughton83}.
Further discussion see text.
}
\label{fig:sig_27al}
\end{figure}

Although the EXFOR database contains a lot of data for \al -induced reactions
on $^{27}$Al, there are practically no data for the $^{27}$Al\rap $^{30}$Si
reaction which dominates the total reaction cross section \sreac\ at low
energies. Barros {\it et al.}\ \cite{Barros59} have measured angular
distributions for a relatively thick ($70\,\mu$g/cm$^2$) $^{27}$Al target
using nuclear track detection. From the measured angular distributions of the
resolved $p_0$, $p_1$, $p_2$, $p_3$, and $p_4$ proton groups a total
\rap\ cross section for $^{27}$Al can be roughly estimated to about
54\,mb. This result is very close to the StM calculation (see
Fig.~\ref{fig:sig_27al}). Unfortunately, only resonance parameters were
exracted from the experimental low-energy data by Kuperus \cite{Kuperus65}.

The $^{27}$Al\ran $^{30}$P reaction was already reviewed recently in the first
NACRE compilation of astrophysical reaction rates \cite{NACRE} (the later
NACRE-II compilation \cite{NACRE2} is restricted to lower masses up to $A <
16$). The following discussion is thus shortened and focuses on information
which is particularly relevant for this study.

The $^{27}$Al\ran $^{30}$P reaction was measured by Flynn {\it et
  al.}\ \cite{Flynn78} by direct neutron counting from about 3.5 to 6\,MeV
with two different targets. Whereas the thin-target (thickness
27\,$\mu$g/cm$^2$) measurement shows resonant structures, the thick-target
(442\,$\mu$g/cm$^2$) measurement averages over the resonances, and the
resulting excitation function shows a smooth energy dependence. The smooth
thick-target measurement is in good agreement with a StM calculation.

Earlier data by Stelson {\it et al.}\ \cite{Stelson64} extend the data by
Flynn {\it et al.}\ \cite{Flynn78} towards higher energies. Stelson {\it et
  al.}\ have used a similar experimental technique and also thick targets. In
the overlap region, the Flynn {\it et al.}\ data are slightly lower but
roughly compatible within the experimental uncertainties.

\begin{figure}[htb]
  \includegraphics{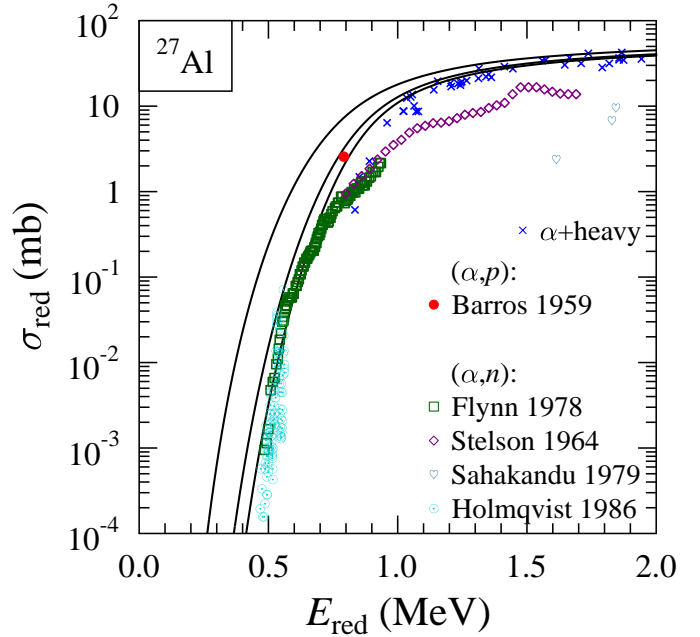}
\caption{
Same as Fig.~\ref{fig:sred_50cr}, but for \al -induced
reactions on $^{27}$Al. The experimental data have been taken from
\cite{Barros59} for the dominating $^{27}$Al\rap $^{30}$Si reaction and from
\cite{Flynn78,Stelson64,Saha79,Holmqvist86} for the $^{27}$Al\ran $^{30}$P
reaction.
Further discussion see text.
}
\label{fig:sred_27al}
\end{figure}
At even higher energies (only shown in Fig.~\ref{fig:sred_27al}) Sahakandu
{\it et al.}\ \cite{Saha79} have used activation and the stacked-foil
technique. As often, the lowest data points of the stacked-foil experiment
deviate significantly from the other available data whereas at higher energies
the agreement becomes much better.

A similar energy range as in Flynn {\it et al.}\ \cite{Flynn78} and in Stelson
{\it et al.}\ \cite{Stelson64} was investigated by Howard {\it et
  al.}\ \cite{Howard74}. The data are in good agreement with the other
experiments but show larger uncertainties. Following the NACRE
recommendations, these data are disregarded because of their larger
uncertainties \cite{NACRE}.

The above data are extended towards lower energies by Holmqvist and Ramstr\"om
\cite{Holmqvist86}. An infinitely thick $^{27}$Al target was irradiated in
small energy steps, and the observed neutron yield as a function of energy was
differentiated to obtain the cross section of the $^{27}$Al\ran $^{30}$P
reaction. The experimental data by Holmqvist and Ramstr\"om agree nicely with
the other available data sets.

%
%
An attempt was made to add further data points at low energies from the
analysis of $^{27}$Al\raa $^{27}$Al elastic scattering data. Unfortunately,
the available low-energy angular distribution by Dyachkov {\it et
  al.}\ \cite{Dyachkov13} does not provide a sufficient number of experimental
data points for a stable phase shift fit or optical model analysis. The
phase shift analysis of the angular distribution by Gailar {\it et
  al.}\ \cite{Gailar58} at $E_\alpha = 18.82$\,MeV leads to \sreac\ $=
1278$\,mb, corresponding to \Ered\ $= 2.89$\,MeV and \sred\ $= 60.7$\,mb,
i.e.\ a result in the expected range.

The reduced cross section \sred\ from the $^{27}$Al\ran $^{30}$P data (see
Fig.~\ref{fig:sred_27al}) is obviously below the typical range of
\sred\ values for $A \approx 20 - 50$ nuclei because the $^{27}$Al\rap
$^{30}$Si reaction is the dominating channel. But from the ratio of the
\ran\ and \rap\ cross sections in the StM calculation it can be concluded that
the \sred\ values for $^{27}$Al do not show an unusual behavior.

\subsection{$^{26}$Mg}
\label{sec:mg26}
\begin{figure}[hbt]
  \includegraphics{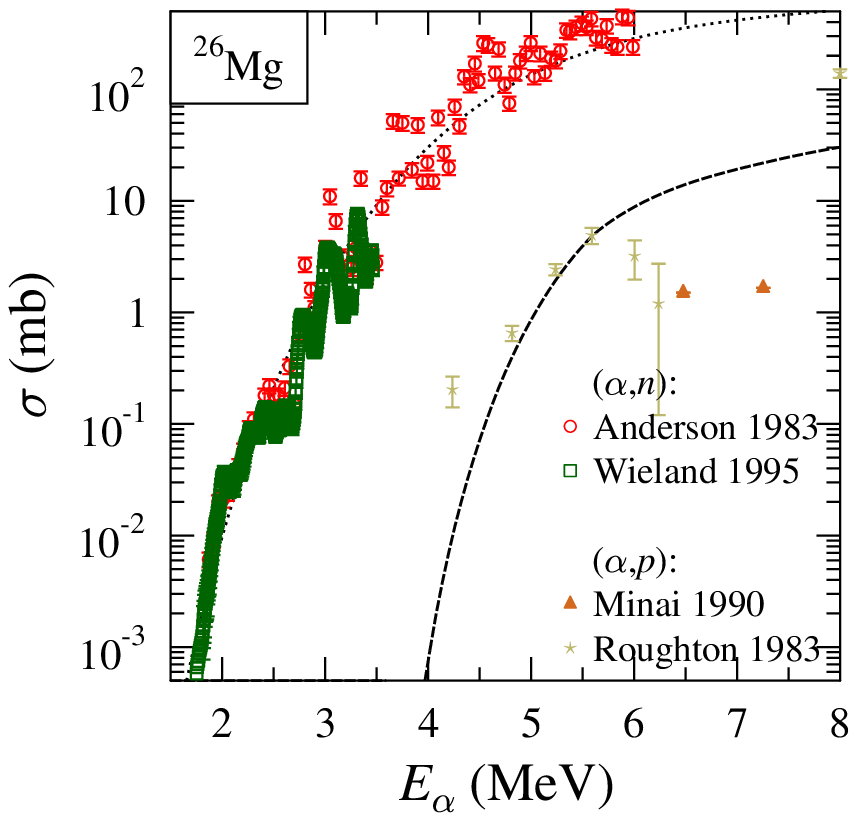}
\caption{
Cross section of the $^{26}$Mg\ran $^{29}$Si and $^{26}$Mg\rap $^{29}$Al
reactions. 
The experimental data have been taken from
\cite{Anderson83,Wieland95,Minai90,Roughton83}.
Further discussion see text.
}
\label{fig:sig_26mg}
\end{figure}

Because $^{26}$Mg is a relatively neutron-rich nucleus, the low-energy cross
section is dominated by the $^{26}$Mg\ran $^{29}$Si reaction. Data for the
$^{26}$Mg\rap $^{29}$Al reaction are available at higher energies. Minai {\it
  et al.}\ \cite{Minai90} cover the energy range between 6 and 34\,MeV, and
the data of Probst {\it et al.}\ \cite{Probst76} are restricted to energies
above 10\,MeV (i.e., above the range shown in Fig.~\ref{fig:sig_26mg}). Both
experiments used the stacked-foil activation technique in combination with
$\gamma$-ray spectroscopy for the detection of the residual $^{29}$Al
nuclei. Surprisingly, the experimental \rap\ data are almost one order of
magnitude lower than the StM calculation. This holds for both TALYS and
NON-SMOKER. However, the \rap\ cross sections from the thick-target yield of
Roughton {\it et al.}\ \cite{Roughton83} are in reasonable agreement with the
StM calculation. Fortunately, this discrepancy does not affect the conclusions
on the total reaction cross section \sreac\ which is dominated by the
\ran\ channel.

The $^{26}$Mg\ran $^{29}$Si reaction is included in the NACRE compilation
\cite{NACRE}. At low energies the data by Anderson {\it et
  al.}\ \cite{Anderson83} and the unpublished data by Wieland \cite{Wieland95}
are recommended. Both experiments used direct neutron counting. According to
NACRE, at very low energies the Wieland data should be preferred because special
care was taken to minimize background from the $^{13}$C\ran $^{16}$O
reaction. In the overlap region there is reasonable agreement between both
data sets. As the \ran\ cross section at low energies is dominated by
resonances, the StM calculation is only able to reproduce the average trend of
the data (see Fig.~\ref{fig:sig_26mg}). For completeness it has to be noted
that recently much lower data for the $^{26}$Mg\ran $^{29}$Si reaction have
been reported in an unpublished PhD thesis by Falahat
\cite{Falahat10}. 
\begin{figure}[htb]
  \includegraphics{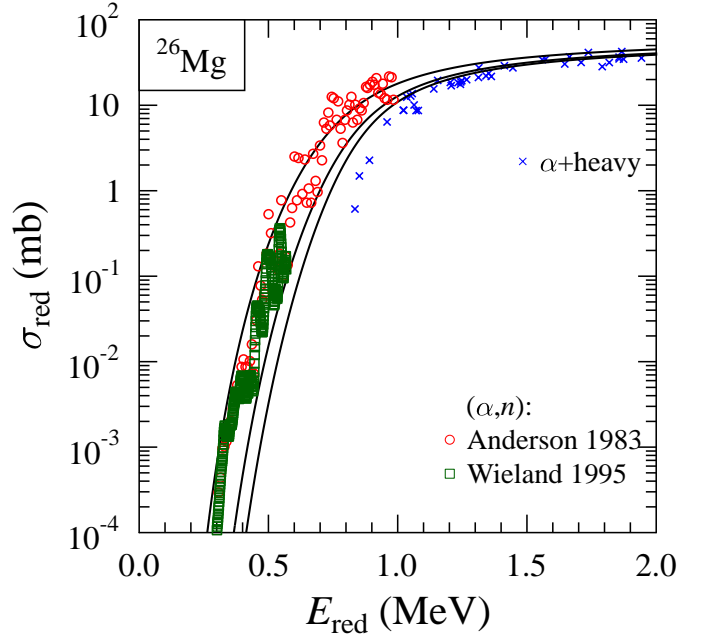}
\caption{
Same as Fig.~\ref{fig:sred_50cr}, but for \al -induced
reactions on $^{26}$Mg. The experimental data have been taken from
\cite{Anderson83,Wieland95}.
Further discussion see text.
}
\label{fig:sred_26mg}
\end{figure}

The total reaction cross section \sreac\ for $^{26}$Mg is well-defined from
the available $^{26}$Mg\ran $^{29}$Si data. The derived reduced cross sections
\sred\ are shown in Fig.~\ref{fig:sred_26mg}.
Similar to most nuclei under study in this work, the
\sred\ data for $^{26}$Mg do not show a peculiar behavior.

\subsection{$^{25}$Mg}
\label{sec:mg25}
\begin{figure}[htb]
  \includegraphics{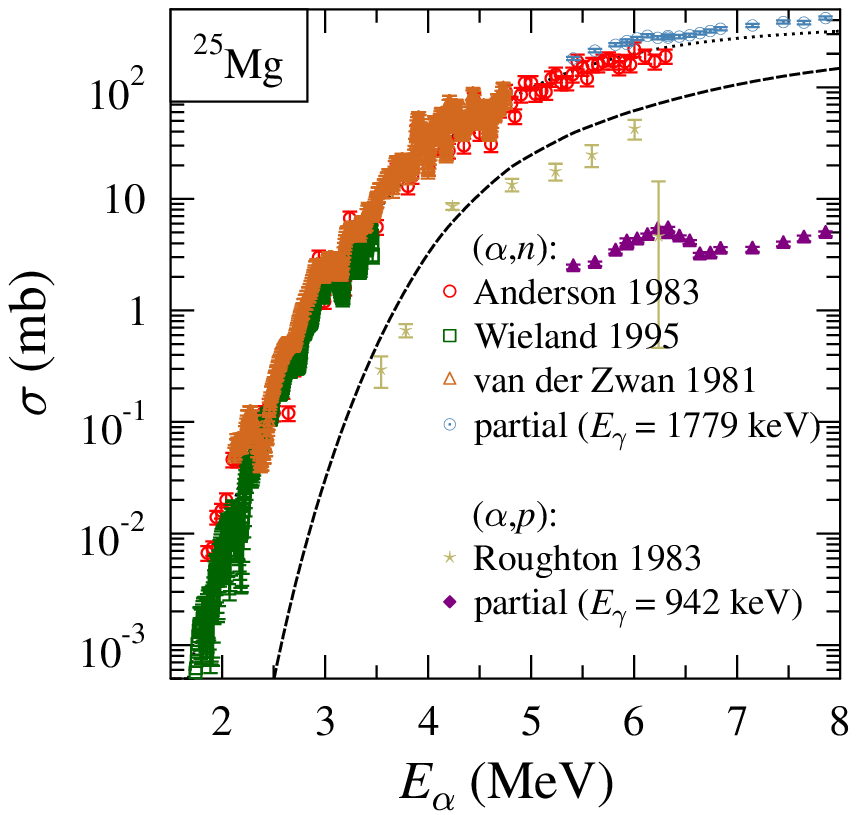}
\caption{
Cross section of the $^{25}$Mg\ran $^{28}$Si and $^{25}$Mg\rap $^{28}$Al
reactions. 
The experimental data have been taken from
\cite{Anderson83,Wieland95,Zwan81,Negret13,Roughton83}.
Further discussion see text.
}
\label{fig:sig_25mg}
\end{figure}

Similar to $^{26}$Mg in the previous section \ref{sec:mg26}, at low energies
the $^{25}$Mg\ran $^{28}$Si cross section is much larger than the
$^{25}$Mg\rap $^{28}$Al cross section. The NACRE compilation \cite{NACRE}
recommends the experimental data by Anderson {\it et al.}\ \cite{Anderson83},
Wieland \cite{Wieland95}, and an additional data set which is available from
van der Zwan {\it et al.}\ \cite{Zwan81}. The present status of the
$^{25}$Mg\ran $^{28}$Si reaction is also reviewed in \cite{Ili11}. It is
concluded in \cite{Ili11} that all data sets agree well, in particular if the
\rang\ data of \cite{Anderson83} are considered which are less sensitive to
background than the \ran\ data.
The average trend of the data is nicely reproduced by the StM calculation (see
Fig.~\ref{fig:sig_25mg}). Similar to $^{26}$Mg, the recent PhD thesis by
Falahat \cite{Falahat10} reports lower results, and because of these
discrepancies new experimental efforts have been started by Caciolli {\it et
  al.}\ \cite{Caciolli14}.

\begin{figure}[htb]
  \includegraphics{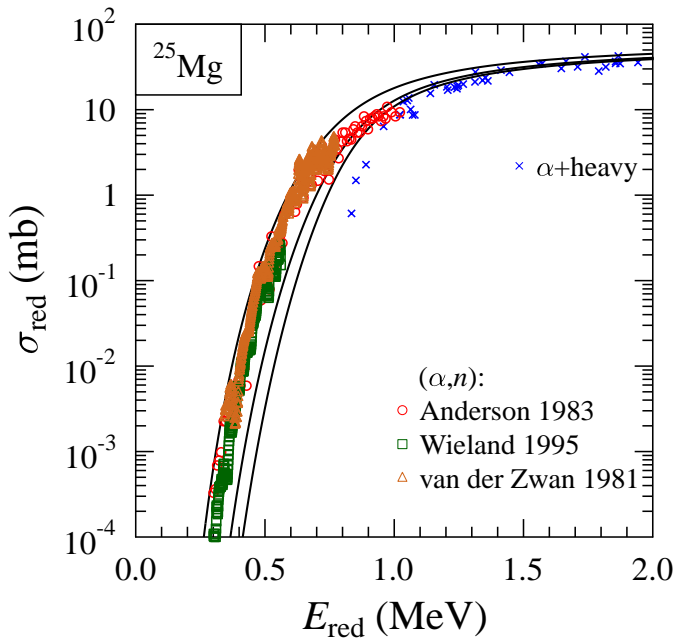}
\caption{
Same as Fig.~\ref{fig:sred_50cr}, but for \al -induced
reactions on $^{25}$Mg. The experimental data have been taken from
\cite{Anderson83,Wieland95,Zwan81}.
Further discussion see text.
}
\label{fig:sred_25mg}
\end{figure}
The determination of the total reaction cross section \sreac\ for $^{25}$Mg is
well defined by the above \ran\ data. But because of the huge deviation
between the StM calculation and experimental data for the $^{26}$Mg\rap
$^{29}$Al reaction in the previous section, the $^{25}$Mg\rap $^{28}$Al
reaction is also analyzed here. Unfortunately, data are available only for
a partial \rap\ cross section. Recently, Negret {\it et al.}\ \cite{Negret13}
have measured the $\gamma$-ray yields after bombardment of $^{25}$Mg by \al
-particles. The yield of the 1779\,keV $\gamma$-ray from the first excited
$2^+$ state in $^{28}$Si to the $0^+$ ground state corresponds to almost the
total $^{25}$Mg\ran $^{28}$Si cross section because practically all excited
states in $^{28}$Si decay through the first $2^+$
state. Fig.~\ref{fig:sig_25mg} shows that the partial \ran\ cross section from
the 1779\,keV $\gamma$-ray yield is even slightly above the \ran\ cross
sections from direct neutron counting. In a similar way, a partial \rap\ cross
section can be derived from the yield of the 942\,keV $\gamma$-ray in the
$^{25}$Mg\rap $^{28}$Al reaction. This $\gamma$-ray corresponds to the
transition from the second exited state ($J^\pi = 0^+$, $E^\ast = 972.4$\,keV) to
the first excited state ($J^\pi = 2^+$, $E^\ast = 30.6$\,keV); the ground
state of $^{28}$Al has $J^\pi = 3^+$. This $\gamma$-ray transition should also
represent a considerable amount of the total \rap\ cross section at energies
sufficiently far above the \rap\ threshold. Surprisingly, similar to the
$^{26}$Mg case, the experimental $^{25}$Mg\rap $^{28}$Al data are
overestimated by the StM by about one order of magnitude. Again, this
statement holds for TALYS and NON-SMOKER calculations. However, again similar
to $^{26}$Mg, the estimated \rap\ cross sections from the thick-target yield of
Roughton {\it et al.}\ \cite{Roughton83} show reasonable agreement with the
StM calculation. 

The total reaction cross section \sreac\ of $^{25}$Mg is well defined by the
dominating $^{25}$Mg\ran $^{28}$Si data, leading to the reduced cross sections
\sred\ shown in Fig.~\ref{fig:sred_25mg}.
Similar to most nuclei under study in this work, the
\sred\ data for $^{25}$Mg do not show a peculiar behavior.

\subsection{$^{24}$Mg}
\label{sec:mg24}
\begin{figure}[b]
  \includegraphics{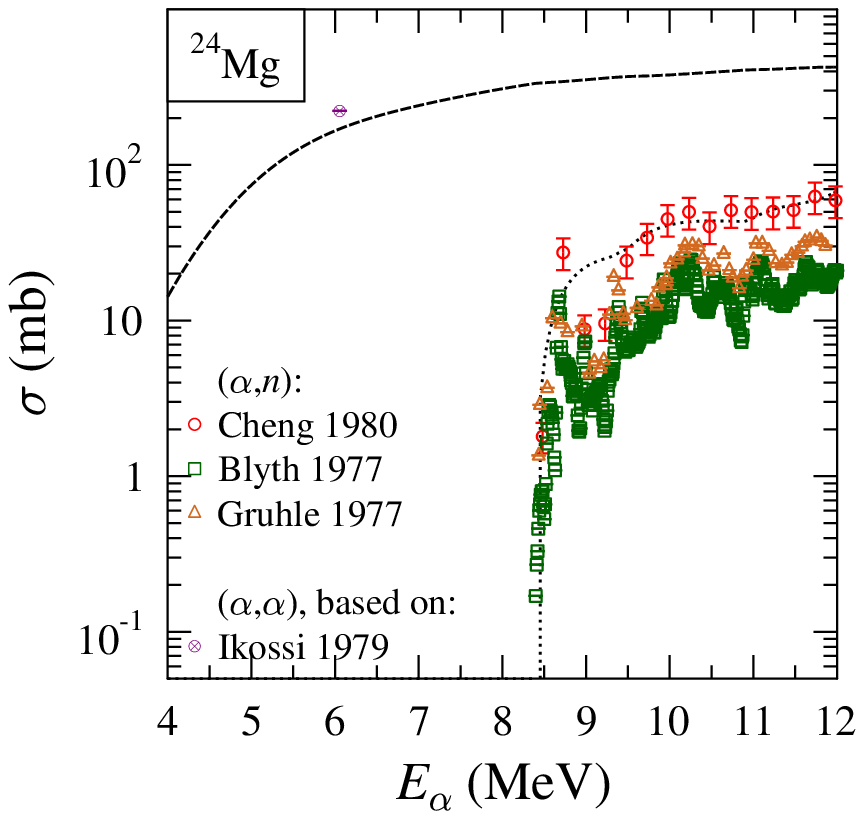}
\caption{
Cross section of the $^{24}$Mg\ran $^{27}$Si and $^{24}$Mg\rap $^{27}$Al
reactions. 
The experimental data have been taken from
\cite{Cheng80,Blyth77,Gruhle77,Ikossi79}.
Further discussion see text.
}
\label{fig:sig_24mg}
\end{figure}
It is difficult to determine the total reaction cross section \sreac\ of
$^{24}$Mg at low energies. The $^{24}$Mg\ran $^{27}$Si reaction has a strongly
negative $Q$-value ($Q = -7.20$\,MeV) and thus does not contribute at low
energies. Also the $^{24}$Mg\rap $^{27}$Al reaction has a slightly negative
$Q$-value ($Q = -1.60$\,MeV), and practically no experimental data can be
found in EXFOR at low energies. At very low energies individual resonances of
the $^{24}$Mg\rag $^{28}$Si reaction dominate.

%
%
\begin{figure}[htb]
  \includegraphics{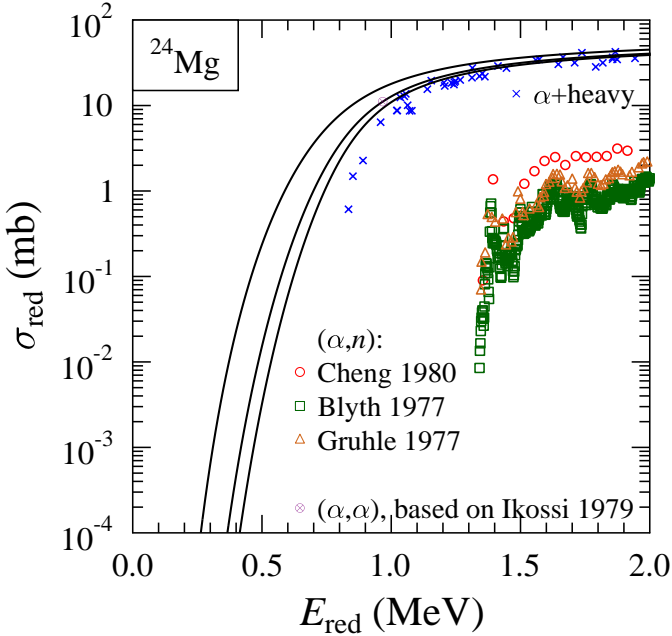}
\caption{
Same as Fig.~\ref{fig:sred_50cr}, but for \al -induced
reactions on $^{24}$Mg. The experimental data have been taken from
\cite{Cheng80,Blyth77,Gruhle77,Ikossi79}.
Further discussion see text.
}
\label{fig:sred_24mg}
\end{figure}
The $^{24}$Mg\ran $^{27}$Si reaction has been studied by Cheng {\it et
  al.}\ \cite{Cheng80}, by Blyth {\it et al.}\ \cite{Blyth77}, and by Gruhle
{\it et al.}\ \cite{Gruhle77}. The data are shown in
Fig.~\ref{fig:sig_24mg}. All experiments used the activation
technique in combination with annihilation spectroscopy. Unfortunately, there
is a disagreement of about a factor of two between the three data
sets. Note that the EXFOR data are based on a numerical table (\cite{Cheng80})
and have been provided by the authors in numerical form (\cite{Blyth77}), and
only the data set of \cite{Gruhle77} had to be re-digitized from a figure;
thus, the above discrepancy cannot be explained by simple digitization
errors. Excellent agreement is found between the StM calculation and the data
by Cheng {\it et al.}\ whereas the data by Blyth {\it et el.}\ and Gruhle {\it
  et al.}\ are overestimated. 

Angular distributions of $^{24}$Mg\raa $^{24}$Mg elastic scattering have been
measured by Ikossi {\it et al.}\ \cite{Ikossi79} at low energies. For one
particular angular distribution at $E_\alpha = 6.055$\,MeV a relatively thick
target (33\,$\mu$g/cm$^2$) was used; this angular distribution is appropriate
for the determination of an average cross section because in addition the
excitation function shows a smooth behavior around this energy. A phase shift
fit leads to \sreac\ $= 222$\,mb, corresponding to \sred\ $= 11.1$\,mb at
\Ered\ $= 0.97$\,MeV.

As the \ran\ cross section contributes only minor to the total reaction cross
section \sreac\ of $^{24}$Mg, the experimental \ran\ data are far below the
expectations for \sred\ as shown in Fig.~\ref{fig:sred_24mg}. However, the data
point from $^{24}$Mg\raa $^{24}$Mg elastic scattering is close to the general
expectation. Therefore, there is no evidence that the \sred\ values for
$^{24}$Mg behave extraordinary.

\subsection{$^{23}$Na}
\label{sec:na23}
The $^{23}$Na\ran $^{26}$Al reaction has significant astrophysical relevance
because it affects the production of the long-lived $^{26}$Al nucleus. The
observation of $\gamma$-rays from the $^{26}$Al decay in our galaxy confirms
ongoing nucleosynthesis \cite{Diehl06}. Very recently, an updated galactic
emission map was derived from the SPI spectrometer data aboard the INTEGRAL
mission \cite{Bouchet15}. Because of a low-lying $0^+$ isomer in
$^{26}$Al which decays directly to $^{26}$Mg and bypasses the $5^+$ ground
state of $^{26}$Al, much efforts have been made to distinguish the
$^{23}$Na\ran $^{26}$Al$_{\rm{g.s.}}$ and $^{23}$Na\ran $^{26{\rm{m}}}$Al
reactions. 

\begin{figure}[htb]
  \includegraphics{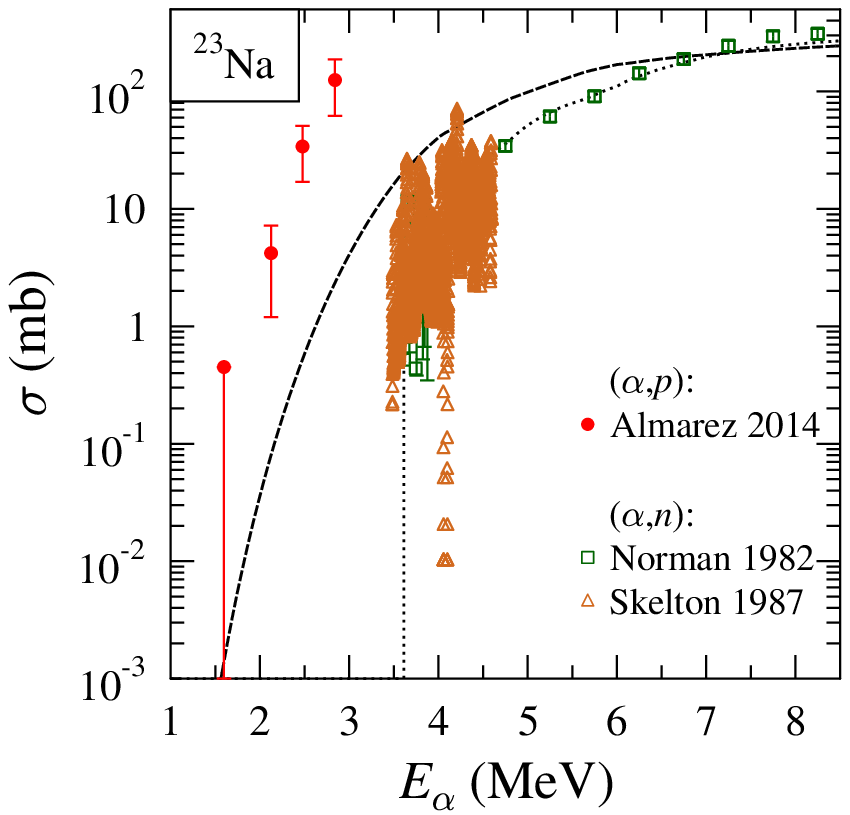}
\caption{
Cross section of the $^{23}$Na\ran $^{26}$Al and $^{23}$Na\rap $^{26}$Mg
reactions. 
The experimental data have been taken from
\cite{Skelton87,Norman82,Alm14}.
Further discussion see text.
}
\label{fig:sig_23na}
\end{figure}
The $^{23}$Na\ran $^{26}$Al reaction has been reviewed in the NACRE
compilation \cite{NACRE}, and the role of the isomer is discussed e.g.\ in
\cite{Ward80}. Because of its negative $Q$-value ($Q = -2.97$\,MeV), this
reaction does not affect the total reaction cross section \sreac\ of $^{23}$Na
at low energies. Therefore, only the total cross section of the $^{23}$Na\ran
$^{26}$Al is studied here.

The NACRE compilation \cite{NACRE} recommends three data sets for the
$^{23}$Na\ran $^{26}$Al reaction: Skelton {\it et al.}\ \cite{Skelton87} and
Norman {\it et al.}\ \cite{Norman82} have measured the total yield which can
be directly converted to the total \ran\ cross section. Doukellis and Rapaport
\cite{Doukellis87} used the time-of-flight technique to resolve the $n_0$,
$n_1$, and $n_2$ neutron groups at six laboratory angles. The data by
Doukellis and Rapaport are not available at EXFOR, and the numerical data at
the NACRE web site seem to be re-digitized because the given energies are
discrepant for the $n_0$, $n_1$, and $n_2$ groups. Consequently, it is
practically impossible to derive the total \ran\ cross section from the data
by Doukellis and Rapaport. Such a determination is further hampered at higher
energies by contributions of higher neutron groups $n_{>2}$. The
recommendation of NACRE that ``The DO87 time of flight experiment is indeed
considered to be more reliable than the NO82 thick target measurements'' is not
well traceable, and the resulting recommendation to scale the Norman {\it et
  al.}\ data by a factor of $1/3$ to adjust to the Doukellis and Rapaport data
is not taken into account in this work. Here the original data of Norman {\it
  et al.}\ \cite{Norman82} and the data by Skelton {\it et
  al.}\ \cite{Skelton87} are shown in Fig.~\ref{fig:sig_23na}. As thin-target
measurements of \cite{Skelton87} show many resonances, the StM calculation is
only able to reproduce the average behavior of the excitation
function. Furthermore, it can be seen from the StM calculations in
Fig.~\ref{fig:sig_23na} that the cross section of the $^{23}$Na\rap $^{26}$Mg
reaction exceeds the $^{23}$Na\ran $^{26}$Al cross section significantly for
energies below about 6\,MeV.

\begin{figure}[hbt]
  \includegraphics{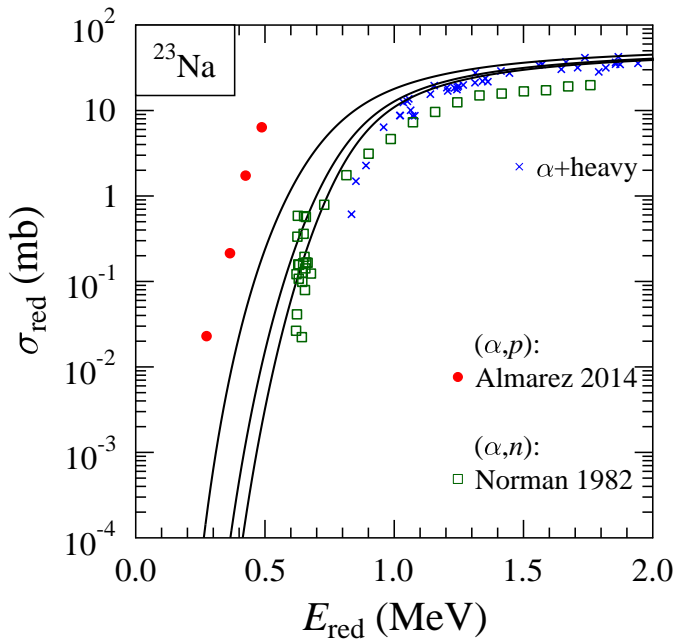}
\caption{
Same as Fig.~\ref{fig:sred_50cr}, but for \al -induced
reactions on $^{23}$Na. The experimental data have been taken from
\cite{Norman82,Alm14}. The lowest data point of Almarez-Calderon {\it et
  al.}\ \cite{Alm14} represents an upper limit only. For better visibility the
data of Skelton {\it et al.}\ \cite{Skelton87} are omitted.
Further discussion see text.
}
\label{fig:sred_23na}
\end{figure}
In a detailed sensitivity study of the production of $^{26}$Al \cite{Ili11} it
has been shown that the $^{23}$Na\rap $^{26}$Mg reaction 
also plays an essential role in the production of $^{26}$Al.
Low-energy data for the $^{23}$Na\rap $^{26}$Mg reaction are available
from Whitmire and Davids \cite{Whitmire74}. However, only resonance strengths
have been determined in \cite{Whitmire74}. Some criticisms to this work have
been reported in \cite{Ili11}, and it was concluded in \cite{Ili11} that StM
calculations should be preferred and that the $^{23}$Na\rap $^{26}$Mg reaction 
is a prime target for future measurements.

In a very recent study new experimental data for the $^{23}$Na\rap $^{26}$Mg
reaction at 
low energies became available. Almarez-Calderon {\it et al.}\ \cite{Alm14}
used a $^{23}$Na beam in inverse kinematics to irradiate a cryogenic $^4$He
gas target, and a silicon strip detector was placed 20\,cm downstream from the
target to detect protons in an angular range from $\vartheta_{\rm{lab}} =
6.8^\circ - 13.5^\circ$. From the observed proton groups $p_0$ and $p_1$ 
average cross sections (averaged over the energy distribution of the beam
which is caused by energy loss and straggling in the entrance window and in
the target gas cell itself) were determined. However, the measured
differential \rap\ cross sections constrain the angular distribution of the
\rap\ cross section only in a very limited angular range. The determination of
angle-integrated cross sections in \cite{Alm14} had to use angular
distributions of the $^{27}$Al\rap $^{30}$Si reaction where similar $J^\pi$ of
the nuclei under study are found. The resulting cross section of the $p_0$ and
$p_1$ groups are finally summed to provide the total $^{23}$Na\rap $^{26}$Mg
cross sections. It can be seen from Fig.~\ref{fig:sig_23na} that the
experimental results are dramatically underestimated by the StM calculation.
The total reaction cross section of $^{23}$Na is well defined by the
$^{23}$Na\rap $^{26}$Mg reaction already below about 6\,MeV, and below the
\ran\ threshold the total reaction cross section \sreac\ is almost entirely
given by the only open particle channel. The results for the reduced cross
section \sred\ are shown in Fig.~\ref{fig:sred_23na}. It is obvious from
Fig.~\ref{fig:sred_23na} that the recent data by Almarez-Calderon {\it et
  al.}\ \cite{Alm14} deviate dramatically from the general behavior which is
otherwise found for nuclei in the $A \approx 20 - 50$ mass region. The new
data lead not only to significantly higher \sred\ values, but also to a
steeper energy dependence than for other nuclei in the $A \approx 20 - 50$
mass range. 

\subsection{$^{22}$Ne}
\label{sec:ne22}
\begin{figure}[htb]
  \includegraphics[clip]{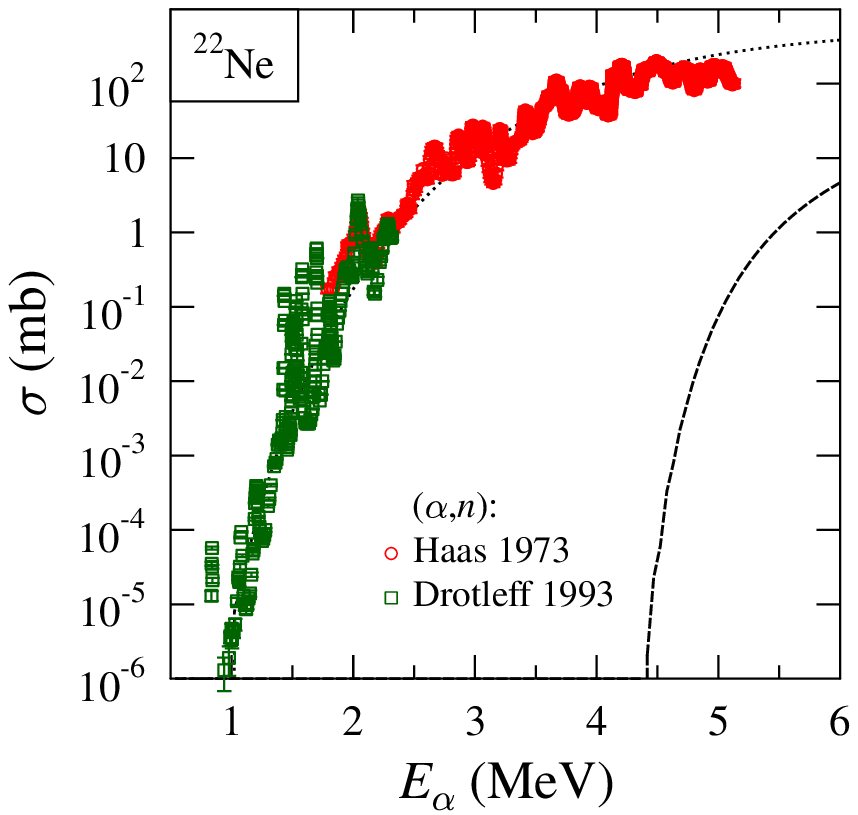}
\caption{
Cross section of the $^{22}$Ne\ran $^{25}$Mg and $^{22}$Ne\rap $^{25}$Na
reactions. 
The experimental data have been taken from
\cite{Haas73,Drotleff93}.
Further discussion see text.
}
\label{fig:sig_22ne}
\end{figure}
Because of the negative $Q$-value of the $^{22}$Ne\rap $^{25}$Na reaction ($Q
= -3.53$\,MeV), at astrophyically relevant energies the $^{22}$Ne\ran
$^{25}$Mg reaction dominates the total reaction cross section of
$^{22}$Ne. This reaction plays a 
major role as neutron source for the astrophysical $s$-process. It is included
in the NACRE compilation \cite{NACRE} where the data of Haas {\it et
  al.}\ \cite{Haas73} and Drotleff {\it et al.}\ \cite{Drotleff91,Drotleff93}
are recommended. These data are shown in Fig.~\ref{fig:sig_22ne} and compared
to a StM calculation. As the cross section is dominated by resonances at low
energies, the StM calculation is only able to reproduce the average properties
of the excitation function. Later data by Jaeger {\it et al.}\ \cite{Jaeger01}
extend the measurements of Drotleff {\it et al.}\ towards lower energies. The
cross section at these very low energies is essentially given by resonant
contributions, and only an experimental yield (but not the cross section) is
presented in \cite{Jaeger01}. Therefore, the data by Jaeger {\it et
  al.}\ \cite{Jaeger01} are not shown in Fig.~\ref{fig:sig_22ne} because there
is no straightforward conversion from the experimental yield to the
\ran\ reaction cross section for extended gas target measurements (see
e.g.\ \cite{Koelle99}). 

\begin{figure}[b]
  \includegraphics{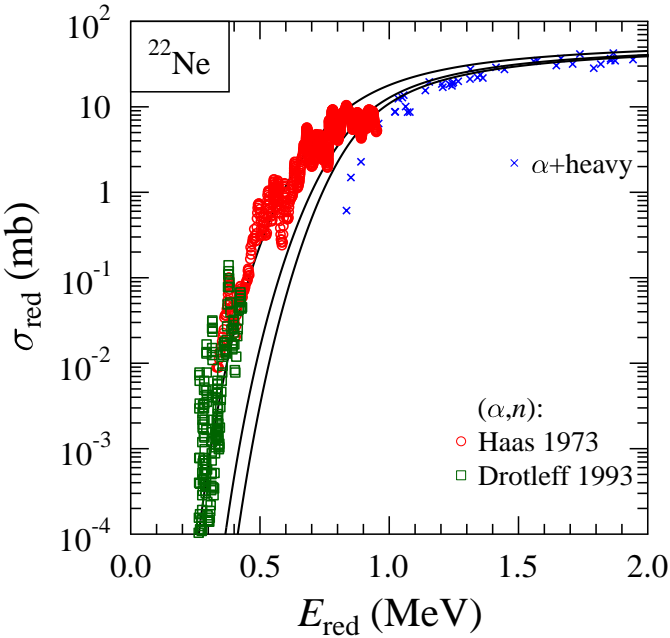}
\caption{
Same as Fig.~\ref{fig:sred_50cr}, but for \al -induced
reactions on $^{22}$Ne. The experimental data have been taken from
\cite{Haas73,Drotleff93}.
Further discussion see text.
}
\label{fig:sred_22ne}
\end{figure}
A full discussion of this reaction and the derived astrophysical reaction rate
\Nsv\ has to include further indirect information (e.g., properties of levels
in the compound $^{26}$Mg nucleus). This is beyond the scope of the present
paper. New results for the $^{22}$Ne\ran $^{25}$Mg reaction after publication
of the first NACRE compilation \cite{NACRE} are e.g.\ summarized in
\cite{Longland12}, and further information is given in
\cite{Longland09,Boer10,Boer14}. 

No data for the $^{22}$Ne\rap $^{25}$Na reaction are listed in the EXFOR
database. Fortunately, this does not affect the determination of the total
reaction cross section \sreac\ of $^{22}$Ne because of the dominating
$^{22}$Ne\ran $^{25}$Mg reaction. The \ran\ cross section is presented as
reduced cross section \sred\ in Fig.~\ref{fig:sred_22ne}. 
Similar to most nuclei under study in this work, the
\sred\ data for $^{22}$Ne do not show a peculiar behavior.

\subsection{$^{21}$Ne}
\label{sec:ne21}
\begin{figure}[bht]
  \includegraphics[clip]{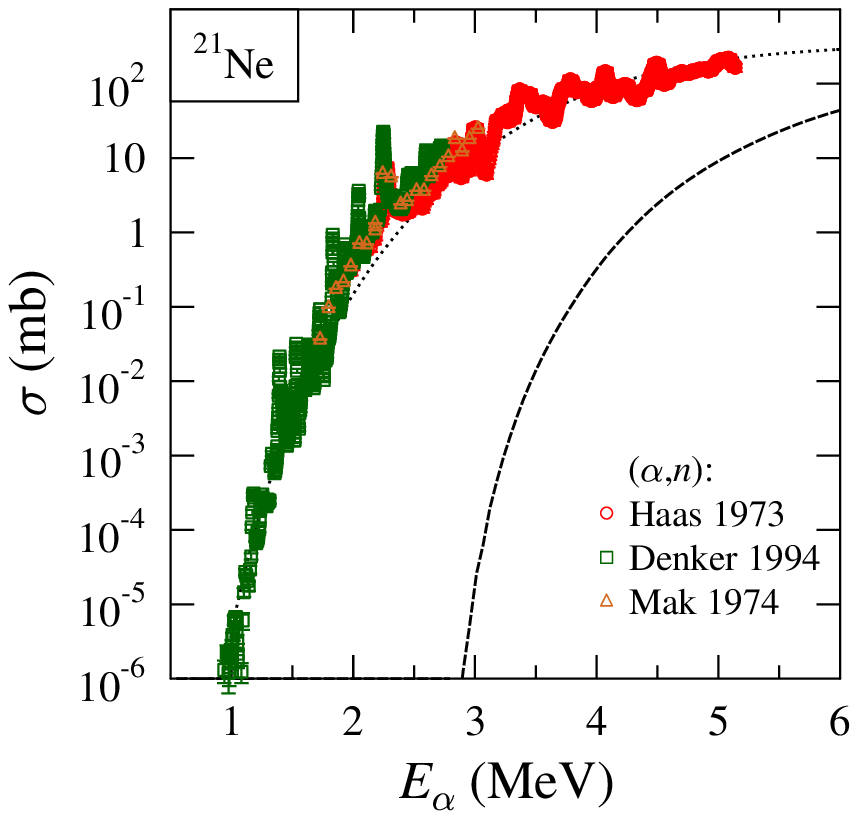}
\caption{
Cross section of the $^{21}$Ne\ran $^{24}$Mg and $^{21}$Ne\rap $^{24}$Na
reactions. 
The experimental data have been taken from
\cite{Haas73,Denker94,Mak74}.
Further discussion see text.
}
\label{fig:sig_21ne}
\end{figure}

Similar to the results for $^{22}$Ne in the previous section, the
$^{21}$Ne\ran $^{24}$Mg cross section is much larger than the $^{21}$Ne\rap
$^{24}$Na cross section. Because of the low natural abundance of $^{21}$Ne,
only very few data exist for this nucleus. The NACRE compilation \cite{NACRE}
recommends the data by Haas {\it et al.}\ \cite{Haas73} and Denker
\cite{Denker94}. Surprisingly, the data of Mak {\it et al.}\ \cite{Mak74} are
not taken into account in NACRE. Mak {\it et al.}\ report average cross
sections (averaged over a about 100\,keV thick neon gas target) which are in
good agreement with the other data which are recommended in NACRE.
In Fig.~\ref{fig:sig_21ne} the experimental data are compared
to a StM calculation which is able to reproduce the average properties of the
$^{21}$Ne\ran $^{24}$Mg cross section. No data are available at EXFOR for the
$^{21}$Ne\rap $^{24}$Na reaction which has a negative $Q$-value of $Q =
-2.18$\,MeV.

\begin{figure}[htb]
  \includegraphics{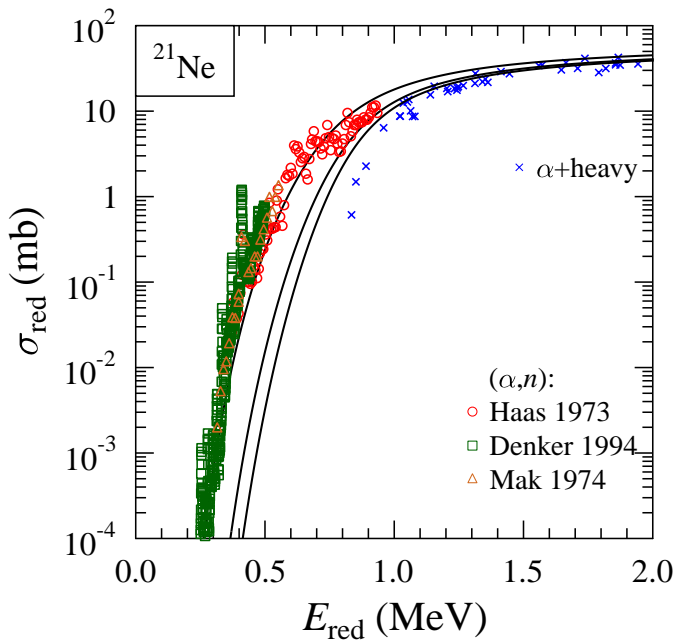}
\caption{
Same as Fig.~\ref{fig:sred_50cr}, but for \al -induced
reactions on $^{21}$Ne. The experimental data have been taken from
\cite{Haas73,Denker94,Mak74}. 
Further discussion see text.
}
\label{fig:sred_21ne}
\end{figure}

The reduced cross sections \sred\ for $^{21}$Ne from the experimental
$^{21}$Ne\ran $^{24}$Mg data of \cite{Haas73,Denker94,Mak74} are shown in
Fig.~\ref{fig:sred_21ne}. 
Similar to most nuclei under study in this work, the
\sred\ data for $^{21}$Ne do not show a peculiar behavior.

\subsection{$^{20}$Ne}
\label{sec:ne20}
Because of the negative $Q$-values of the $^{20}$Ne\ran $^{23}$Mg ($Q =
-7.22$\,MeV) and $^{20}$Ne\rap $^{23}$Na ($Q = -2.38$\,MeV) reactions, it is
not possible to determine the total cross section of $^{20}$Ne at low energies
from \rap\ and \ran\ data. The $^{20}$Ne\rag $^{24}$Mg reaction at low
energies is dominated by isolated resonances (e.g.,
\cite{Schmalbrock83,Koelle99}), and the experimental yield in these gas target
measurements is dominated over broad energy ranges by the tails of strong
resonances. 

\begin{figure}[htb]
  \includegraphics[clip]{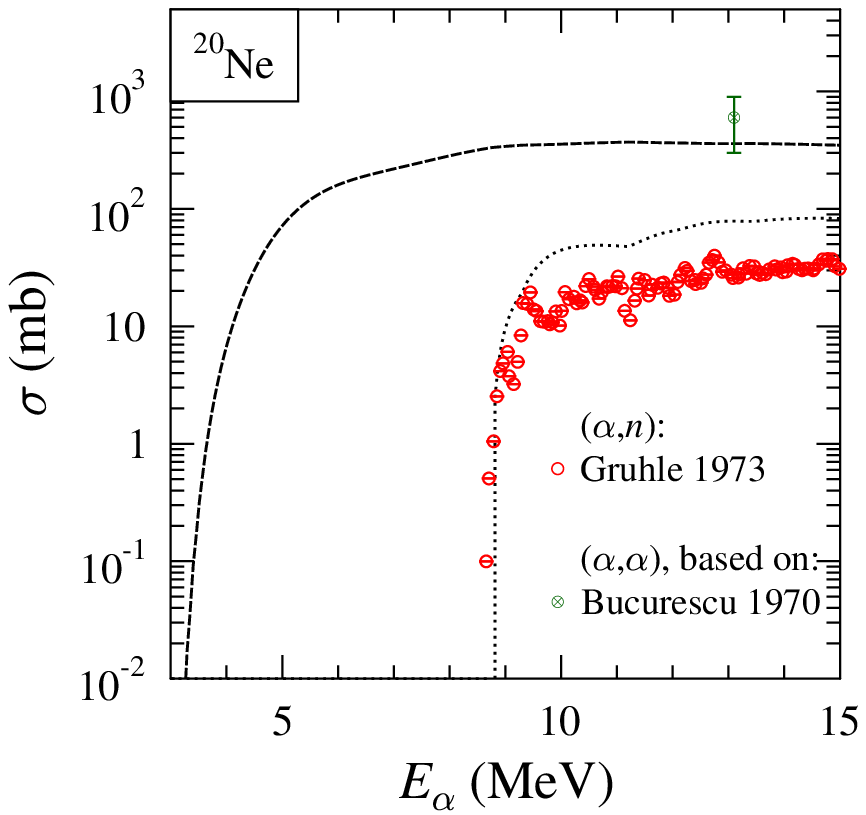}
\caption{
Cross section of the $^{20}$Ne\ran $^{23}$Mg and $^{20}$Ne\rap $^{23}$Na
reactions. 
The experimental data have been taken from
\cite{Gruhle73,Bucurescu70}.
Further discussion see text.
}
\label{fig:sig_20ne}
\end{figure}
Fig.~\ref{fig:sig_20ne} shows the available $^{20}$Ne\ran $^{23}$Mg data at
higher energies, i.e.\ above the \ran\ threshold, in comparison to a StM
calculation. As the cross section of the $^{20}$Ne\ran $^{23}$Mg reaction is
much smaller than the cross section of the $^{20}$Ne\rap $^{23}$Na reaction,
an attempt was made to estimate the total reaction cross section \sreac\ from
$^{20}$Ne\raa $^{20}$Ne elastic scattering at low energies. However, a phase
shift analysis to the data of \cite{Bucurescu70} at 13.1\,MeV is only able to
constrain \sreac\ in a relatively wide range of \sreac\ $= 600 \pm 300$\,mb
because of the limited number of data points in \cite{Bucurescu70}; this
corresponds to \sred\ $= 32$\,mb at \Ered $= 2.35$\,MeV. Somewhat higher
\sred\ values between 58 and 78\,mb were found from the analysis of full
angular distributions from \cite{Frickey71} at slightly higher energies from
15.8 to 17.8\,MeV, corresponding to \Ered\ between 2.83 and 3.19\,MeV.

\begin{figure}[b]
  \includegraphics{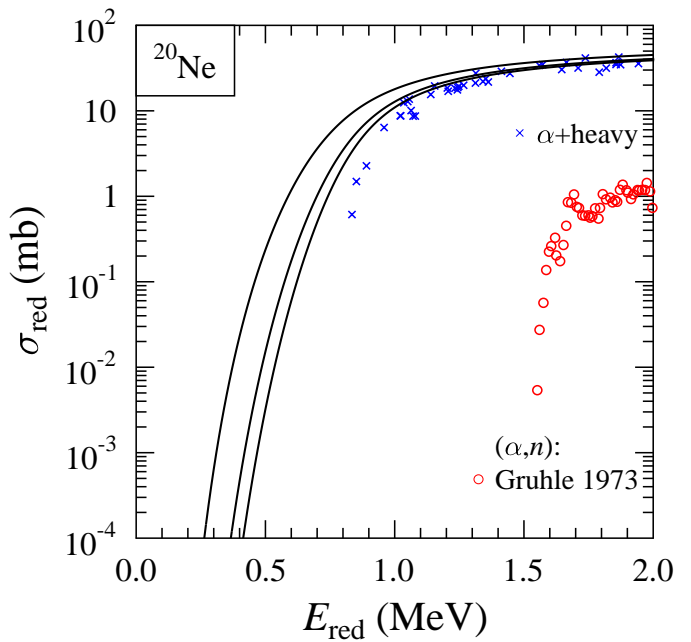}
\caption{
Same as Fig.~\ref{fig:sred_50cr}, but for \al -induced
reactions on $^{20}$Ne. The experimental data have been taken from
\cite{Gruhle73}. 
Further discussion see text.
}
\label{fig:sred_20ne}
\end{figure}
Unfortunately, no data for the $^{20}$Ne\rap $^{23}$Na reaction can be found
in the EXFOR database. $\gamma$-ray yields after bombardment of $^{20}$Ne with
\al -particles have been reported in \cite{Seamster84}. In principle, these
yields should allow to constrain the $^{20}$Ne\rap $^{23}$Na cross
section. However, the strong $\gamma$ transition in $^{23}$Na at 1637\,keV
($7/2^+$;\,2076\,keV $\rightarrow$ $5/2^+$;\,440\,keV) almost coincides with
the first $2^+$ in $^{20}$Ne at 1634\,keV which is excited by inelastic
scattering, and therefore it was not possible to distinguish between the
\rap\ reaction and inelastic scattering in \cite{Seamster84}. No information
is given in \cite{Seamster84} on the 440\,keV transition from the first
excited state in $^{23}$Na to the ground state.

It is clear that the $^{20}$Ne\ran $^{23}$Mg data are much lower than the
expected values for reduced cross sections \sred\ (see
Fig.~\ref{fig:sred_20ne}) because of the dominating 
$^{20}$Ne\rap $^{23}$Na reaction. Nevertheless, together with the additional data
from $^{20}$Ne\raa $^{20}$Ne elastic scattering at higher energies (above the
shown range in Fig.~\ref{fig:sred_20ne}) it can be concluded that there
is at least no evidence for an unexpected behavior of \sred\ of $^{20}$Ne.

\subsection{$^{18}$Ne}
\label{sec:ne18}
The experimental situation for the unstable $^{18}$Ne nucleus is completely
different from all above examples. Only indirect data are available to
constrain the $^{18}$Ne\rap $^{21}$Na cross section, and the $^{18}$Ne\ran
$^{21}$Mg reaction does not contribute to the total reaction cross section
\sreac\ at low energies because of the strongly negative $Q$-value ($Q =
-11.24$\,MeV). The available experimental information for the $^{18}$Ne\rap
$^{21}$Na reaction has been summarized recently in \cite{Mohr14b,Mohr13c}.
Although the $^{18}$Ne\rap $^{21}$Na cross section is dominated by individual
resonances, it has been shown in \cite{Mohr14b} that a StM calculation is
roughly able to reproduce the average properties of the excitation function.
\begin{figure}[b]
  \includegraphics{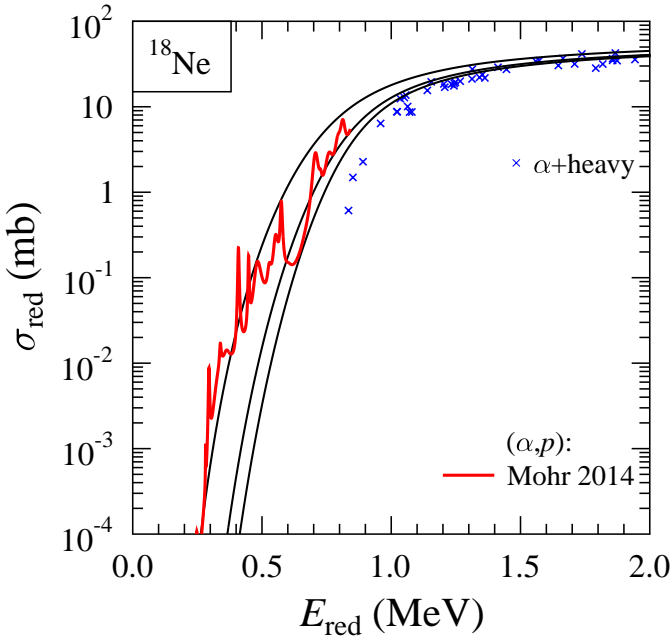}
\caption{
Same as Fig.~\ref{fig:sred_50cr}, but for \al -induced
reactions on $^{18}$Ne. The estimate for the experimental cross section is
taken from \cite{Mohr14b}.
Further discussion see text.
}
\label{fig:sred_18ne}
\end{figure}

The latest result of \cite{Mohr14b} for the $^{18}$Ne\rap $^{21}$Na cross
section has been converted to the reduced cross section \sred ; the result is
shown in Fig.~\ref{fig:sred_18ne}.
Similar to most nuclei under study in this work, the
\sred\ data for $^{18}$Ne do not show a peculiar behavior.

\subsection{$^{19}$F}
\label{sec:f19}
Because of its positive $Q$-value, the $^{19}$F\rap $^{22}$Ne reaction
dominates at low energies. Unfortunately, no total cross section data are
available at EXFOR. The differential cross sections measured by Ugalde {\it et
  al.}\ \cite{Ugalde08} are fitted by an R-matrix calculation, and the
R-matrix result is directly converted to the stellar reaction rate in
\cite{Ugalde08}. The adopted rate of \cite{Ugalde08} is well reproduced by StM
calculations using the \al -nucleus potential by McFadden and Satchler
\cite{McF66} (see Fig.~9 of \cite{Ugalde08}).
\begin{figure}[htb]
  \includegraphics[clip]{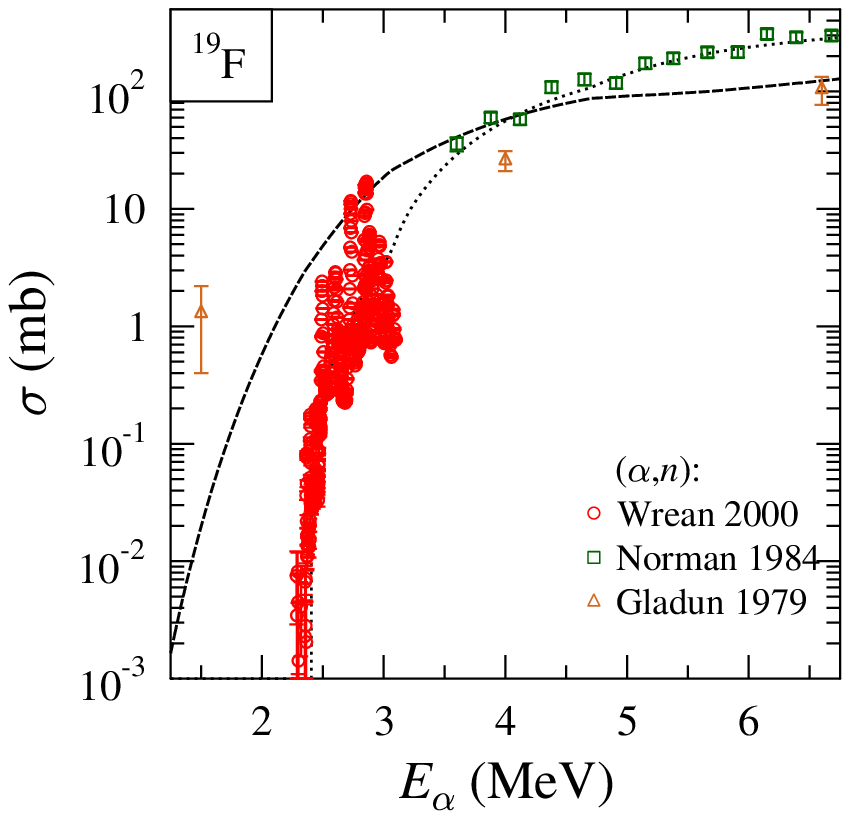}
\caption{
Cross section of the $^{19}$F\ran $^{22}$Na and $^{19}$F\rap $^{22}$Ne
reactions. 
The experimental data have been taken from
\cite{Wrean00,Norman84,Gladun79}.
Further discussion see text.
}
\label{fig:sig_19f}
\end{figure}

Several data sets are available for the $^{19}$F\ran $^{22}$Na reaction. Wrean
and Kavanagh \cite{Wrean00} have used thin targets and direct neutron
detection for their measurement at low energies below about 3.5\,MeV. At
higher energies Norman {\it et al.}\ \cite{Norman84} have measured
thick-target neutron yields which were converted to cross sections by
differentiation. The earlier data by Gladun and Chursin \cite{Gladun79} have
huge uncertainties in energy and deviate from the other experimental results
\cite{Wrean00,Norman84}. The thick-target data at higher energies
\cite{Norman84} are well reproduced by the StM, and as expected, at lower
energies the StM is only able to reproduce the average energy dependence of
the experimental data of \cite{Wrean00}. Earlier data by Balakrishnan {\it et
  al.}\ \cite{Bala78} are omitted because of problems with background from
$^{13}$C (see discussion in \cite{Wrean00} and similar problems of these
authors for $^{29}$Si, see Sec.~\ref{sec:si29}).

\begin{figure}[htb]
  \includegraphics{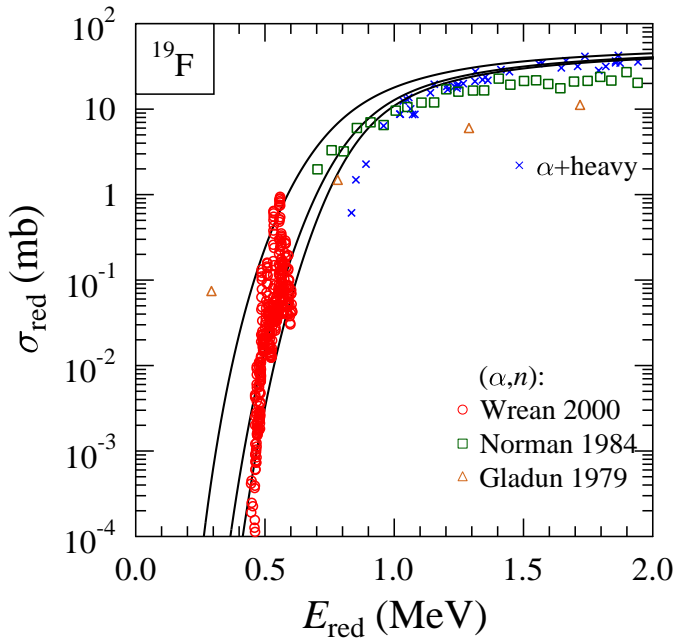}
\caption{
Same as Fig.~\ref{fig:sred_50cr}, but for \al -induced
reactions on $^{19}$F. The experimental data have been taken from
\cite{Wrean00,Norman84,Gladun79}.
Further discussion see text.
}
\label{fig:sred_19f}
\end{figure}
As the $^{19}$F\rap $^{22}$Ne reaction dominates at low energies, it is not
surprising that the reduced cross sections \sred\ from the $^{19}$F\ran
$^{22}$Na data \cite{Wrean00,Norman84,Gladun79} are somewhat lower than the
general trend of reduced cross sections. Nevertheless, from the excellent
agreement of the experimental data and the StM calculations for the
$^{19}$F\rap $^{22}$Ne and $^{19}$F\ran $^{22}$Na reactions it can be
concluded that there is no evidence for a peculiar behavior of $^{19}$F.

\section{Discussion}
\label{sec:disc}
A first finding of the above presentation is that many experimental data for
\al -induced reactions in the $A \approx 20 - 50$ mass range show reasonable
agreement. Major discrepancies between individual data sets have been
discussed in the corresponding sections above. Nevertheless it should be kept
in mind that in several cases absolute normalizations of the experimental data
are stated with additional uncertainties of $10 - 20$\,\% which are not
included in the shown error bars. New experimental data should provide
absolute cross sections with small uncertainties, as has been done e.g.\ by
Vonach {\it et al.}\ \cite{Vonach83}. Additionally, for some reactions only
one or even no data set is available. Here of course new data should provide
excitation functions in small energy steps.

The general agreement between the experimental data and calculations in the StM
is very good. The basic ingredient for the StM calculations is the \al
-nucleus potential which essentially defines the total reaction cross section
\sreac\ for \al -induced reactions. As long as either the \rap\ or \ran\ cross
section is the dominant (say greater than about $70 - 80$\,\%) contribution to
\sreac , the StM calculation is practically insensitive to all other
ingredients of the StM. Fortunately, this is the case for many nuclei under
study in the $A \approx 20 - 50$ mass region which allows are careful test of
the \al -nucleus potential.

In the present study the simple energy-independent 4-parameter potential by
McFadden and Satchler \cite{McF66} has been chosen. As this potential leads to
overestimation of \al -induced cross sections for a wide range of heavy target
nuclei (above $A \approx 60$) at low energies, the very good agreement between
the experimental data and the StM calculation in the $A \approx 20 - 50$ mass
region is somewhat surprising. For heavy nuclei it has been suggested to add
an energy dependence to the imaginary part of the McFadden/Satchler potential
to avoid this typical overestimation of cross sections (e.g.,
\cite{Som98,Sau11}). Such an energy dependence is also expected from
theoretical side, and various parametrizations have been suggested like a
Fermi-type function \cite{Som98,Sau11}, the Brown-Rho parametrization
\cite{Brown81}, or a resonance-like parametrization \cite{Mohr13}. All these
parameterizations have two features in common: ($i$) They start from very
small imaginary parts at very low energies and end at a saturation value at
higher energies significantly above the Coulomb barrier; ($ii$) the increase
is characterized by an energy where e.~g.\ half of the saturation value is
reached, and by a slope parameter. Such an energy dependence of the imaginary
part reduces the total reaction cross section \sreac\ towards lower
energies, compared to the original McFadden/Satchler potential. 

Obviously, such a reduction is not needed for the nuclei in the $A \approx 20
- 50$ mass range. Following the discussion in McFadden and Sachler
\cite{McF66}, it is stated that ``There is a tendency for the heavier nuclei
to favour smaller $r_0$, and for the lighter ones to favor larger $r_0$''. A
larger radius parameter $r_0$ leads to increased total reaction cross sections
\sreac . Thus, strictly speaking, the 4-parameter McFadden/Satchler potential
with the fixed radius parameter $r_0 = 1.4$\,fm has two shortcomings. First,
the calculations should overestimate the experimental reaction cross sections
\sreac\ towards lower energies because of the missing energy dependence of the
imaginary part. Second, the calculations should underestimate \sreac\ because
the adjustment of the potential to elastic scattering data in \cite{McF66}
requires a larger radius parameter than the fixed average value of $r_0 =
1.4$\,fm which is adopted by McFadden and Satchler. Therefore, the very good
agreement between experimental data and the 
StM calculation using the McFadden/Satchler potential may even be considered
as somewhat accidental because the missing energy dependence of the imaginary
part may partly compensate the missing $A$ dependence of the radius parameter
$r_0$. 

For some nuclei under study a good agreement between the StM calculation and
the experimental data is found for the dominating channel whereas the StM
calculation deviates strongly from the experimental data for the weak
channel. The most prominent example for such a behavior is $^{26}$Mg
where the dominating $^{26}$Mg\ran $^{29}$Si cross section is well reproduced
by the StM, but the $^{26}$Mg\rap $^{29}$Al cross section is overestimated by
about one order of magnitude. Such a behavior points to a deficiency in the
theoretical treatment of the $^{29}$Al + $p$ channel, as the \al -nucleus
potential is confirmed by the reproduction of the \ran\ channel. However,
although two independent data sets \cite{Minai90,Probst76} are available for
the $^{26}$Mg\rap $^{29}$Al reaction (see Sec.~\ref{sec:mg26}), both data sets
have been obtained from the stacked-foil activation technique which has turned
out to be not very reliable for low energies (see e.g.\ the huge scatter of
such data for $^{51}$V in Fig.~\ref{fig:sig_51v} in Sec.~\ref{sec:v51}). A
quite similar deviation for the \rap\ channel can be seen for the neighboring
$^{25}$Mg nucleus; however, here the experimental data represent only a
partial cross section of the \rap\ reaction (see Fig.~\ref{fig:sig_25mg} in
Sec.~\ref{sec:mg25}). Interestingly, for both cases $^{25}$Mg and $^{26}$Mg
the estimated \rap\ cross sections from the thick-target yields in
\cite{Roughton83} show much better agreement with the StM calculations. Thus,
it is not fully clear whether there is really a deficieny in the StM
calculations or an experimental problem in the \rap\ data of
\cite{Minai90,Probst76}. 

The agreement between experimental data and the StM calculation is limited to
cases where the experimental cross section is averaged over a sufficient
number of resonances in the compound nucleus. This sufficiently high level
density is achieved for nuclei at the upper end of the mass range $A \approx
20 - 50$ under study. For the lighter nuclei individual resonances become more
and more visible. This obviously depends crucially on the experimental
conditions like the energy spread of the beam and in particular on the target
thickness. A nice example has been given for $^{27}$Al where thin-target data
show many individual resonances whereas thick-target data from the same
experiment show a smooth (i.e., non-resonant) energy dependence (see data from
\cite{Flynn78} in Fig.~\ref{fig:sig_27al} in Sec.~\ref{sec:al27}). As soon as
the level density in the compound nucleus is not high enough, the StM
calculation is only able to reproduce the general trend of the energy
dependence of the excitation function, but not all the individual resonances. 

The present study attempts to provide a comparison of reaction cross sections
for various target nuclei at energies below and above the Coulomb barrier. For
this purpose the method of reduced energies \Ered\ and reduced cross sections
\sred\ was used \cite{Gom05}. It is found that the data for \al -induced cross
sections in the $A \approx 20 - 50$ mass range are slightly higher than the
general results for \al -induced reactions on heavy (above $A \approx 90$)
targets \cite{Mohr10,Mohr13}. The \sred\ values increase relatively smoothly
with decreasing target mass $A$. The expected range of \sred\ values is
indicated by three lines in all figures with reduced cross sections \sred ;
these lines correspond to the theoretical predictions for $^{51}$V, $^{36}$Ar,
and $^{21}$Ne (i.e., covering the mass range under study). An expected
exception is the doubly-magic ($Z = N = 20$) $^{40}$Ca nucleus which shows
slightly smaller \sred\ compared to its neighboring nuclei (see
Fig.~\ref{fig:sred_40ca} in Sec.~\ref{sec:ca40}). Surprisingly, not much
differences are seen for even-even, even-odd, odd-even, and odd-odd
nuclei. Unfortunately, experimental data for odd-odd nuclei are only scarcely
available.

Four more significant exceptions from this generally smooth behavior of
\sred\  have
been found. The results for $^{40}$Ar and $^{36}$Ar are far below the expected
range of \sred\ values. However, there are only very few data points which are
based on one particular very old experiment \cite{Schwartz56}. New data for
$^{40}$Ar and $^{36}$Ar are highly desirable to illustrate this behavior. The
recent results from the $^{33}$S\rap $^{36}$Cl reaction \cite{Bowers13} are
slightly above the expected range; this discrepancy sharpens dramatically as
soon as the additional cross section of the $^{33}$S\ran $^{36}$Ar reaction
(estimated from the theoretical ratio between \rap\ and \ran\ channel, see
\cite{Mohr14}) is
taken into account. The summed \rap\ and \ran\ cross sections are far above
the expected range for \sred , and the energy depencence is much flatter than
expected (see Fig.~\ref{fig:sred_33s} in Sec.~\ref{sec:s33} and detailed
discussion in \cite{Mohr14}). Finally, the recent $^{23}$Na\rap $^{26}$Mg data
\cite{Alm14} are very far above the expected range, and they show a much
steeper energy dependence than expected (see Fig.~\ref{fig:sred_23na} in
Sec.~\ref{sec:na23}). As for both $^{33}$S and $^{23}$Na only one data set is
available in the relevant energy range, new experimental data would be very
helpful to confirm the unexpected behavior of these two nuclei. 

The \sred\ vs.\ \Ered\ reduction scheme is very simple, and  also other
reduction schemes have been suggested. For $^{33}$S
it was stated \cite{Mohr14} that also the reduction scheme from
\cite{Wolski13} leads to similar conclusions.

The strong deviation between expected reduced cross sections \sred\ and
experimental results for $^{40}$Ar, $^{36}$Ar, $^{33}$S, and $^{23}$Na is
correlated with a poor agreement between the experimental data and the StM
calculations. From the otherwise smooth behavior of \sred\ values for nuclei
with $A \approx 20 - 50$ it is obvious that it is not possible to find an
\al -nucleus potential with smoothly varying parameters which is able to
reproduce the general trend of \sred\ and the lower outliers $^{40}$Ar and
$^{36}$Ar and the upper outliers $^{33}$S and $^{23}$Na simultaneously.

\section{Summary and conclusions}
\label{sec:conc}
Reduced cross sections \sred\ were derived for \al -induced reactions on
nuclei in the $A \approx 20 - 50$ mass range. This simple reduction scheme
(reduced cross section \sred\ versus reduced energy \Ered\ as suggested in
\cite{Gom05}) shows a very similar behavior for heavy-ion induced reactions in
a broad energy range. It has been found earlier \cite{Mohr10,Mohr13} that this
reduction scheme works also well for \al -induced reactions on heavy
nuclei. The present study shows that \al -induced reactions in the $A \approx
20 - 50$ mass range show a trend of slightly larger reduced cross sections at
low energies (below \Ered\ $\approx 1$\,MeV) with decreasing target
mass. However, this trend is weak and relatively smooth. Four outliers are
identified: $^{36}$Ar and $^{40}$Ar with smaller \sred\ values (based on early
experimental data of \cite{Schwartz56}) and $^{23}$Na (based on \cite{Alm14})
and $^{33}$S (based on \cite{Bowers13}) with significantly increased
\sred\ values. 

In general, the calculation of \Ered\ and \sred\ allows for a quick and simple
test whether the cross section of an \al -induced reaction for a particular
nucleus behaves regularly or extraordinary. The present study provides the
basis for such a comparison. From my point of view, such a test is strongly
recommended for any new data on \al -induced reactions. 

As a byproduct of the
present study it was found that the reduced energy \Ered\ has a simple
approximate relation to the most effective energy for astrophysical reaction
rates (the so-called Gamow window): e.g., for $T_9 = 2$ the Gamow window
appears around \Ered\ $\approx 0.45$\,MeV for all nuclei under study in this
work.

The experimental cross sections of \al -induced reactions in the $A \approx 20
- 50$ mass range are compared to calculations in the statistical model. Here
the cross section factorizes into a production cross section of the compound
nucleus which depends on the chosen \al -nucleus potential, and into a decay
branching of the compound state which depends on the other ingredients of the
statistical model, but is almost independent of the \al -nucleus
potential. Fortunately, for most of the nuclei under study, one particular
reaction channel -- \rap\ or \ran\ -- is dominating which allows a strict test
of the chosen \al -nucleus potential by comparing only the cross section of
the dominating \rap\ or \ran\ channel; this test is only weakly affected by
the other ingredients of the statistical model. Surprisingly it is found that
the old and very simple 4-parameter potential by McFadden and Satchler
\cite{McF66} leads to very good agreement with most of the experimental data;
i.e., the smooth energy dependence of excitation functions for nuclei in the
upper half of the mass range $A \approx 20 - 50$ is reproduced, and for the
lighter nuclei under study the statistical model reproduces only the average
energy dependence of the experimental excitation function which is governed by
individual resonances. For the four outliers in the \sred\ reduction scheme
($^{40}$Ar, $^{36}$Ar, $^{33}$S, $^{23}$Na) it is found that these nuclei also
show poor agreement between the experimental data and the statistical model
calculations for \al -induced reactions.

\section*{Acknowledgments}
I thank A.\ Laird for the encouragement to tackle this broad study and
Z.\ M\'at\'e for providing the numerical data for $^{34}$S\raa $^{34}$S and
$^{50}$Cr\raa $^{50}$Cr. The pleasant long-term collaboration on the \al
-nucleus interaction with Zs.\ F\"ul\"op, Gy.\ Gy\"urky, G.\ G.\ Kiss, and
E.\ Somorjai is gratefully acknowledged. The present study benefits from the
availability of experimental data in the EXFOR database - special thanks to
N.\ Otsuka and coworkers for maintaining EXFOR. This work was supported by
OTKA (K101328 and K108459). 

\bigskip

{\noindent\underline{Note added in Proof:}}\\
According to a private communcation with Alison Laird, new experiments on the
$^{23}$Na\rap $^{26}$Mg reaction have been done very recently. She points
out that the ``cross sections agree with NON-SMOKER, apart from at the lowest
energies (below $E_{\rm{c.m.}} = 1.4$\,MeV)'' where the new data are even lower
than the theoretical prediction. As the calculated $^{23}$Na\rap $^{26}$Mg
cross section in the StM depends essentially only on the chosen \al -nucleus
potential, the agreement between the latest experimental data and theory also
holds for the TALYS calculations in Fig.~\ref{fig:sig_23na} (see
Sec.~\ref{sec:na23}). Thus, the status of $^{23}$Na may change from
``outlier'' to ``regular behavior''.


\end{document}